\documentclass[11pt]{article}  
\usepackage{graphicx}
\usepackage{a41}                
\usepackage{color}
\usepackage{cite}
\usepackage[rflt]{floatflt}
\usepackage{float}
\newcommand{\Ctil}{\tilde{C}}
\newcommand{\lsim}{\raisebox{-0.07cm   }
{$\, \stackrel{<}{{\scriptstyle\sim}}\, $}}
\newcommand{\gsim}{\raisebox{-0.07cm   }
{$\, \stackrel{>}{{\scriptstyle\sim}}\, $}}

\usepackage{array}

\usepackage[english]{babel}
\usepackage[latin1]{inputenc}
\usepackage[T1]{fontenc}
\usepackage{ae}

\usepackage{url}

\usepackage{amsmath, amsthm, amssymb}
\newtheorem{thm}{Theorem}[section]

\newtheorem{definition}[thm]{Definition}

\newcommand{\GeV}{\rm GeV}
\newcommand{\MS}{\overline{{\sf MS}}}

\newcommand\NN{\nonumber}

\newcommand{\Li}{{\rm Li}}

\newcommand{\ep}{\varepsilon}

\usepackage{rotating}

\usepackage{graphicx}

\newcounter{mmacnt}
\def\restartmma{\setcounter{mmacnt}{0}}
\restartmma \catcode`|=\active
\def|#1|{\mathrm{#1}}
\catcode`|=12
\newenvironment{mma}{
 \par\smallskip
 \catcode`|=\active
 \parskip=0pt\parindent=0pt 
 \small
 \def\In##1\\{%
   \def\linebreak{\hfill\break\null\qquad}%
   \refstepcounter{mmacnt}
   \hangindent=2.5em\hangafter=0
   \leavevmode
   \llap{\tiny\sffamily In[\arabic{mmacnt}]:=\kern.5em}%
   \mathversion{bold}\footnotesize$\displaystyle##1$\normalsize
   \mathversion{normal}\par
 }%
 \def\Print##1\\{%
   \def\linebreak{\hfill\break}%
   \hangindent=2.5em\hangafter=0
   \leavevmode ##1\par}%
 \def\Out##1\\{%
   \def\linebreak{$\hfill\break\null\hfill$}%
   \kern\abovedisplayskip\par
   \hangindent=2.5em\hangafter=0
   \leavevmode
   \llap{\tiny\sffamily Out[\arabic{mmacnt}]=\kern.5em}
   \footnotesize$\displaystyle##1$\normalsize\hfill\null\par
   \kern\belowdisplayskip
 }%
 \def\Warning##1##2\\{%
   \def\linebreak{\hfill\break}%
   \hangindent=2.5em\hangafter=0
   \leavevmode
   {\scriptsize##1 : ##2}\par}%
}{%
 \par\smallskip
}

\usepackage{color}

\newenvironment{fshaded}{%
\MakeFramed {\FrameRestore}
}%
{\endMakeFramed}

\usepackage{tikz}
\usetikzlibrary{matrix}

\allowdisplaybreaks[4]

\begin{document}
\setlength{\baselineskip}{0.515cm}
\sloppy
\thispagestyle{empty}
\begin{flushleft}
DESY 13--232
\\
DO--TH 13/32\\
SFB/CPP-14-63\\
MITP/14--057\\
LPN 14--103\\
August 2014\\
\end{flushleft}

\mbox{}
\vspace*{\fill}
\begin{center}

{\LARGE\bf The 3-Loop Pure Singlet Heavy Flavor}

\vspace*{2mm}
{\LARGE\bf Contributions to the Structure Function \boldmath $F_2(x,Q^2)$}

\vspace*{2mm}
{\LARGE\bf and the Anomalous Dimension}

\vspace{4cm}
{\large
J.~Ablinger$^a$,
A.~Behring$^b$,
J.~Bl\"umlein$^b$,
A.~De Freitas$^b$,
A.~von~Manteuffel$^c$, \\
and C.~Schneider$^a$}

\vspace*{5mm}
{\it $^a$~Research Institute for Symbolic Computation (RISC),\\
                          Johannes Kepler University, Altenbergerstra\ss{}e 69,
                          A--4040, Linz, Austria}\\

\vspace*{3mm}
{\it  $^b$ Deutsches Elektronen--Synchrotron, DESY,}\\
{\it  Platanenallee 6, D-15738 Zeuthen, Germany}

\vspace*{3mm}
{\it  $^c$ PRISMA Cluster of Excellence,  Institute of Physics, J. Gutenberg 
University,}\\
{\it D-55099 Mainz, Germany.}
\\

\end{center}
\normalsize
\vspace{\fill}
\begin{abstract}
\noindent 
The pure singlet asymptotic heavy flavor corrections to 3--loop order for the deep--inelastic scattering 
structure function $F_2(x,Q^2)$ and the corresponding transition matrix element $A_{Qq}^{(3), \sf PS}$ 
in the variable flavor number scheme are computed. In Mellin-$N$ space these inclusive quantities  depend 
on generalized harmonic sums. We also recalculate the complete 3-loop pure singlet anomalous dimension for 
the first time. Numerical results for the Wilson coefficients, the operator matrix element and the contribution 
to the structure function $F_2(x,Q^2)$ are presented.
\end{abstract}

\vspace*{\fill}
\noindent
\numberwithin{equation}{section}
\newpage
\section{Introduction}
\label{sec:1}

\vspace{1mm}
\noindent
The present precision of deep-inelastic scattering data \cite{DIS} allows for the measurement of  the strong coupling constant 
$\alpha_s(M_Z^2)$
at an accuracy of 1\% \cite{Bethke:2011tr} and for a precision determination of the parton distribution functions 
\cite{PDF,Alekhin:2013nda} and the mass of the charm quark $m_c$ \cite{Alekhin:2012vu}. In the future, new dedicated deep-inelastic 
experiments may be carried out at high luminosities at the EIC \cite{EIC} and at even higher energies than those at HERA \cite{HERA} as 
planned for the
LHeC~\cite{AbelleiraFernandez:2012cc}. At these facilities the experimental resolution will be even higher. The corresponding 
analyses require 3-loop accuracy, including the heavy flavor Wilson coefficients. At present the heavy flavor corrections to 
deep-inelastic scattering are known to 2-loop order in semi-analytic form \cite{NLOH}\footnote{For a precise implementation in 
Mellin space see \cite{Alekhin:2003ev}.}. As has been shown in Ref.~\cite{Buza:1995ie} for scales $Q^2 \gsim 10~m^2$, 
with $m$ the heavy quark mass, 
the heavy flavor contributions to the structure function $F_2(x,Q^2)$ can be calculated to the 1\% level employing
a factorization of the scattering cross section into massive operator matrix elements (OMEs) and massless Wilson coefficients
\cite{Buza:1995ie}. This enables us to calculate the higher order corrections in Quantum Chromodynamics (QCD) in analytic form.

At 3-loop order this calculation has been performed for a series of Mellin moments in Ref.~\cite{Bierenbaum:2009mv} in 2009. The calculation
of the corresponding results for general values of the Mellin variable $N$ requires by far different techniques than those 
having been used in \cite{Bierenbaum:2009mv}. To a wide extent, they were not previously available and had to be newly developed in course 
of the present 
calculation.
In the past we have recalculated and corrected the 2-loop results \cite{Buza:1995ie,Buza:1996xr,Buza:1996wv,Buza:1997mg} using 
more systematic summation and integration 
methods in Refs.~\cite{Bierenbaum:2007qe,Bierenbaum:2007pn,Bierenbaum:2009zt,Bierenbaum:2008yu,Blumlein:2014fqa}.
Furthermore, we calculated the asymptotic heavy flavor corrections to the structure function $F_L(x,Q^2)$  
\cite{Blumlein:2006mh,Behring:2014eya}. Very recently, we presented results on the operator matrix element $A_{gq}^{(3)}$ 
\cite{Ablinger:2014lka} and the flavor non-singlet OMEs and Wilson coefficients \cite{Ablinger:2014vwa}. 
Furthermore, the 3-loop contributions of $O(N_F T_F^2)$ have been computed completely \cite{Ablinger:2010ty,Blumlein:2012vq}, as well as 
the contributions $O(T_F^2)$ to the OMEs $A_{gq}$ and $A_{gg}$ \cite{Ablinger:2014uka} stemming from graphs with two internal fermion lines 
carrying the same mass\footnote{For contributions of graphs with two internal fermion lines of different heavy quark masses see 
\cite{BW1}.}.
Technical aspects of these calculations have been presented in Refs.~\cite{Ablinger:2012qm,Ablinger:2014yaa}. 
In all these calculations the respective contributions to the 3--loop anomalous dimension are obtained as a by-product. 

In the present paper we calculate the pure singlet contributions to the heavy flavor Wilson coefficient $H_{2,q}^{\rm PS}$ at 
3-loop order 
in the asymptotic region and present the operator matrix element $A_{Qq}^{(3),\rm PS}$, which also appears as one of the matching 
coefficients
in the variable flavor number scheme (VFNS). As in previous calculations \cite{Ablinger:2014uka,Ablinger:2014bra}, new mathematical 
structures emerge in intermediary steps. In $x$-space they appear as generalized harmonic polylogarithms \cite{Ablinger:2013cf}. 
In the physical result they can be mapped back to the usual harmonic polylogarithms \cite{Remiddi:1999ew} at the arguments $x$ 
and a new one at $y = 1 - 2x$. In Mellin-$N$ space, generalized harmonic sums contribute to the result \cite{Moch:2001zr}.

The paper is organized as follows. In Section~\ref{sec:2} we briefly describe the basic formalism. Some technical aspects of the 
calculation of the massive OME are outlined in Section~\ref{sec:3}. This concerns the reduction of the Feynman diagrams to 
master integrals and the different methods we have applied for their calculation. The unpolarized pure singlet anomalous dimensions 
to 3-loop order is presented in Section~\ref{sec:4} and compared with results in the literature. The massive OME 
$A_{Qq}^{(3),\rm PS}$ is given in Section~\ref{sec:5}. Here we also discuss asymptotic expansions for small and large 
values of the momentum fraction $x$. The asymptotic heavy flavor Wilson coefficient $H_{2,q}^{(3),\rm PS}$ is presented in 
Section~\ref{sec:6} and numerical illustrations are given for the pure singlet contribution to the structure function 
$F_2(x,Q^2)$ due to charm and bottom quarks. Section~\ref{sec:7} contains the conclusions. In Appendix~A we discuss aspects
of the contributing integral families. Mellin representations of the newly contributing generalized harmonic sums are
given in Appendix~B. The expression for the operator matrix element $A_{Qq}^{\rm PS}$ and the asymptotic massive
Wilson coefficient $H_{2,q}^{\rm PS}$ in $x$-space are given in Appendix~C.
\section{Basic Formalism}
\label{sec:2}

\vspace{1mm}
\noindent
The renormalized pure singlet OME in the ${\MS}$--scheme for the coupling constant to 3-loop order 
\cite{Bierenbaum:2009mv} is given by
\begin{eqnarray}
A_{Qq}^{\rm PS, \MS}&=& 
  a_s^2 A_{Qq}^{(2),\rm  PS, \MS}
+ a_s^3 A_{Qq}^{(3),\rm PS, \MS}~.
\end{eqnarray}
It describes the transition between massless on-shell quark states $\langle q|$, characterized by
a local quark operator in the light-cone expansion \cite{LCE}, which is located on the heavy quark line.
The corresponding pure singlet contributions in case the operator is located on an internal light quark line
has been dealt with in Refs.~\cite{Ablinger:2010ty,Behring:2014eya}. The OMEs at 2- and 3-loop order are given by
\begin{eqnarray}
A_{Qq}^{(2),\rm PS, \MS}&=&
                -\frac{\hat{\gamma}_{qg}^{(0)}
                             \gamma_{gq}^{(0)}}{8}
                   \ln^2 \left(\frac{m^2}{\mu^2}\right)
                +\frac{\hat{\gamma}_{qq}^{(1), {\rm PS}}}{2}
                   \ln \left(\frac{m^2}{\mu^2}\right)
                +a_{Qq}^{(2),{\rm PS}}
                +\frac{\hat{\gamma}_{qg}^{(0)}
                             \gamma_{gq}^{(0)}}{8}\zeta_2~,
\label{AQq2PSMSren} 
\\
A_{Qq}^{(3),{\rm PS}, \MS}&=&
      \frac{\hat{\gamma}_{qg}^{(0)}\gamma_{gq}^{(0)}}{48}
                  \Biggl\{
                         \gamma_{gg}^{(0)}
                        -\gamma_{qq}^{(0)}
                        +6\beta_0
                        +16\beta_{0,Q}
                  \Biggr\}
              \ln^3 \left(\frac{m^2}{\mu^2}\right)
  +    \frac{1}{8}\Biggl\{
                         -4\hat{\gamma}_{qq}^{(1),{\rm PS}}
                               \Bigl(
                                 \beta_0
                                +\beta_{0,Q}
                               \Bigr)
\NN\\ &&
                        +\hat{\gamma}_{qg}^{(0)}
                               \Bigl(
                                 \hat{\gamma}_{gq}^{(1)}
                                -\gamma_{gq}^{(1)}
                               \Bigr)
                        -\gamma_{gq}^{(0)}\hat{\gamma}_{qg}^{(1)}
                  \Biggr\}
              \ln^2 \left(\frac{m^2}{\mu^2}\right)
  +   \frac{1}{16}\Biggl\{
                         8\hat{\gamma}_{qq}^{(2),{\rm PS}}
                        -8N_F\hat{\tilde{\gamma}}_{qq}^{(2),{\rm PS}}
\NN\\ &&
                        -32a_{Qq}^{(2),{\rm PS}}(\beta_0+\beta_{0,Q})
                        +8\hat{\gamma}_{qg}^{(0)}a_{gq,Q}^{(2)}
                        -8\gamma_{gq}^{(0)}a_{Qg}^{(2)}
                        -\hat{\gamma}_{qg}^{(0)}\gamma_{gq}^{(0)}\zeta_2\
                          \Bigl(
                                 \gamma_{gg}^{(0)}
                                -\gamma_{qq}^{(0)}
\NN\\ &&
                                +6\beta_0
                                +8\beta_{0,Q}
                          \Bigr)
                  \Biggr\}
              \ln \left(\frac{m^2}{\mu^2}\right)
    +4(\beta_0+\beta_{0,Q})\overline{a}_{Qq}^{(2),{\rm PS}}
    +\gamma_{gq}^{(0)}\overline{a}_{Qg}^{(2)}
    -\hat{\gamma}_{qg}^{(0)}\overline{a}_{gq,Q}^{(2)}
\NN\\ &&
    +\frac{\gamma_{gq}^{(0)}\hat{\gamma}_{qg}^{(0)}\zeta_3}{48}
                  \Bigl(
                         \gamma_{gg}^{(0)}
                        -\gamma_{qq}^{(0)}
                        +6\beta_0
                  \Bigr)
    +\frac{\hat{\gamma}_{qg}^{(0)}\gamma_{gq}^{(1)}\zeta_2}{16}
    -\delta m_1^{(1)} \hat{\gamma}_{qg}^{(0)}
                        \gamma_{gq}^{(0)}
    +\delta m_1^{(0)} \hat{\gamma}_{qq}^{(1),{\rm PS}}
\NN\\ &&
    +2 \delta m_1^{(-1)} a_{Qq}^{(2),{\rm PS}} 
    + {a_{Qq}^{(3),{\rm PS}}}~.     
\label{AQq3PSMSren} 
\end{eqnarray}
Here $\gamma_{ij}^{(k)},~~k = 0,1,2$ denote the anomalous dimensions, $a_{ij}^{(k)},~~k = 1,2,3$
is the constant part of the unrenormalized OME at $O(a_s^k)$, with the strong coupling constant
$g_s$ expressed as $a_s = g_s^2/(4\pi)^2 \equiv \alpha_s/(4\pi)$, $\bar{a}_{ij}^{(k)},~~k = 1,2$
denotes the part $\propto \ep$ of the unrenormalized OME at $O(a_s^k)$, with $\ep = D - 4$ the dimensional parameter,
$\beta_k$ and $\beta_{Q,k}$ are the expansion coefficients of the QCD $\beta$-function in the 
$\overline{\rm MS}$--scheme and for massive contributions, $\delta m_k^{(l)}$ are the expansion coefficients
of the renormalized quark mass $m$, $\mu$ is the renormalization scale, $N_F$ denotes the number of light quark flavors,
and $\zeta_k = \sum_{l=1}^\infty (1/l^k),~~k \in \mathbb{N}, k \geq 2$ denotes the Riemann $\zeta$-function at integer
argument. For details of the notation see Ref.~\cite{Bierenbaum:2009mv}.
Here and in the following we also use the shorthand notations
\begin{eqnarray}
\hat{f}(x,N_F)   &\equiv& f(x,N_F+1) - f(x,N_F)\\
\tilde{f}(x,N_F) &\equiv& \frac{f(x,N_F)}{N_F}~.
\end{eqnarray}
In the asymptotic region $Q^2 \gg m^2$ the pure singlet heavy flavor Wilson coefficient is given by 
\cite{Bierenbaum:2009mv}
\begin{eqnarray}
H_{2,q}^{\rm PS}(N_F)
&=& a_s^2 \left[~A_{Qq}^{{\rm PS}, (2)}(N_F)
+~\Ctil_{2,q}^{{\rm PS}, (2)}(N_F+1)\right]   
\nonumber\\   
&+& a_s^3 \Bigl[~A_{Qq}^{{\rm PS}, (3)}(N_F)
+~\Ctil_{2,q}^{{\rm PS}, (3)}(N_F+1)
+ A_{gq,Q}^{(2)}(N_F)~\Ctil_{2,g}^{(1)}(N_F+1)
\nonumber\\ && \hspace*{5mm}  
+ A_{Qq}^{{\rm PS},(2)}(N_F)~C_{2,q}^{{\rm NS}, (1)}(N_F+1)
\Bigr].
\label{eq:WILS}
\end{eqnarray}
Here $C_{2,j}^{(l)}$, with $j = q,g,~~l = 1,2,3$ denote the corresponding light flavor Wilson coefficients
\cite{Furmanski:1981cw,ZN,Moch:1999eb,Vermaseren:2005qc}.
The OME $A_{Qq}^{(2),\rm PS}$ has been calculated in 
\cite{Buza:1995ie,Bierenbaum:2007qe} 
and $A_{gq,Q}^{(2)}$ in \cite{Buza:1996wv,Bierenbaum:2009zt}.

The heavy flavor pure singlet contribution to the structure function $F_2(x,Q^2)$ in case of the coupling of the exchanged
virtual photon of virtuality $Q^2$ to the heavy quark line of charge $e_Q$ is obtained by \cite{Behring:2014eya}
\begin{eqnarray}
F_2^{Q\bar{Q},\sf PS}(x,Q^2) = e_Q^2 x \int_x^1 \frac{dy}{y} H_{2,q}^{\rm PS}\left(y,\frac{Q^2}{\mu^2}, \frac{m^2}{\mu^2}\right) 
\Sigma\left(\frac{x}{y}, \mu^2\right)~.
\end{eqnarray}
The quark-singlet distribution is given by 
\begin{eqnarray}
\Sigma(x,\mu^2) = \sum_{k=1}^{N_F} \left[q_k(x,\mu^2) + \bar{q}_k(x,\mu^2)\right]
\end{eqnarray}
and $q(x) (\bar{q}(x))$ denote the light flavor quark and anti-quark number densities, respectively.

Before we present the physical results on the pure singlet 3-loop anomalous dimension, the massive OME and Wilson coefficient,
and numerical results on the pure singlet contribution to the structure function $F_2(x,Q^2)$, we discuss a series of technical 
details of the present calculation. 
\section{Details of the Calculation}
\label{sec:3}

\begin{figure}[H]
\begin{minipage}[c]{0.23\linewidth}
     \includegraphics[width=1\textwidth]{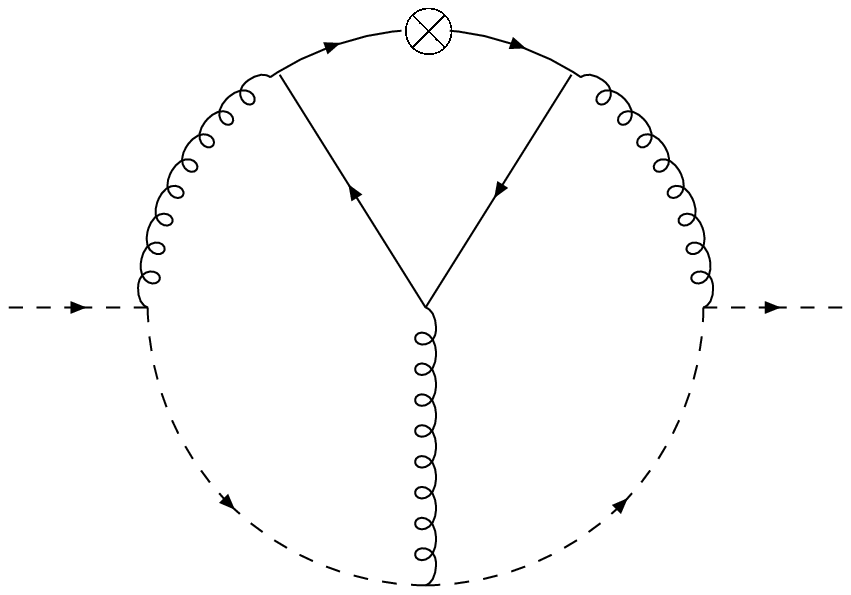}
\vspace*{-8mm}
\begin{center}
{\footnotesize (a)}
\end{center}
\end{minipage}
\hspace*{1mm}
\begin{minipage}[c]{0.23\linewidth}
     \includegraphics[width=1\textwidth]{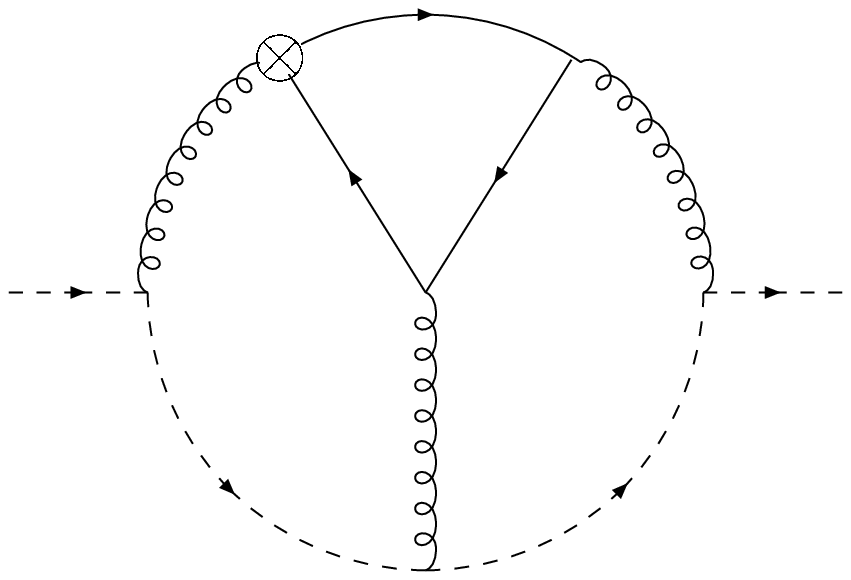}
\vspace*{-8mm}
\begin{center}
{\footnotesize (b)}
\end{center}
\end{minipage}
\hspace*{1mm}
\begin{minipage}[c]{0.23\linewidth}
     \includegraphics[width=1\textwidth]{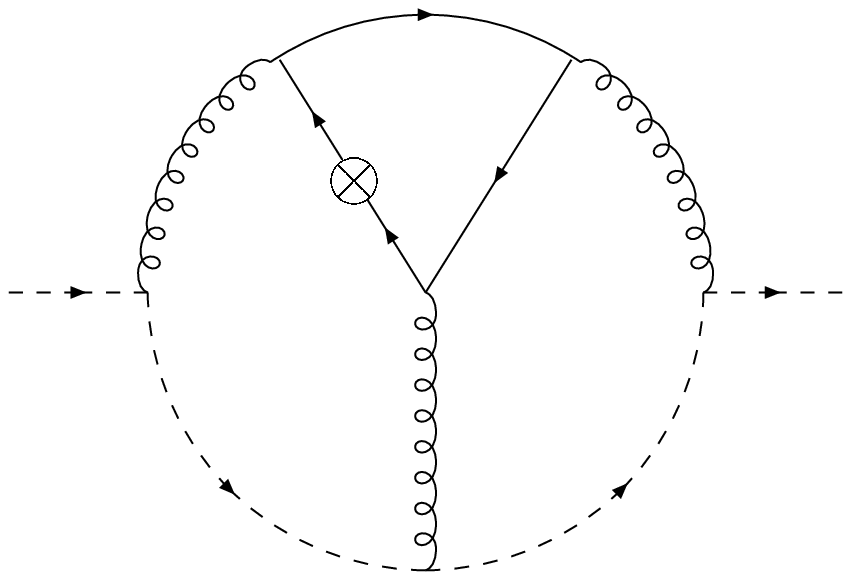}
\vspace*{-8mm}
\begin{center}
{\footnotesize (c)}
\end{center}
\end{minipage}
\hspace*{1mm}
\begin{minipage}[c]{0.23\linewidth}
     \includegraphics[width=1\textwidth]{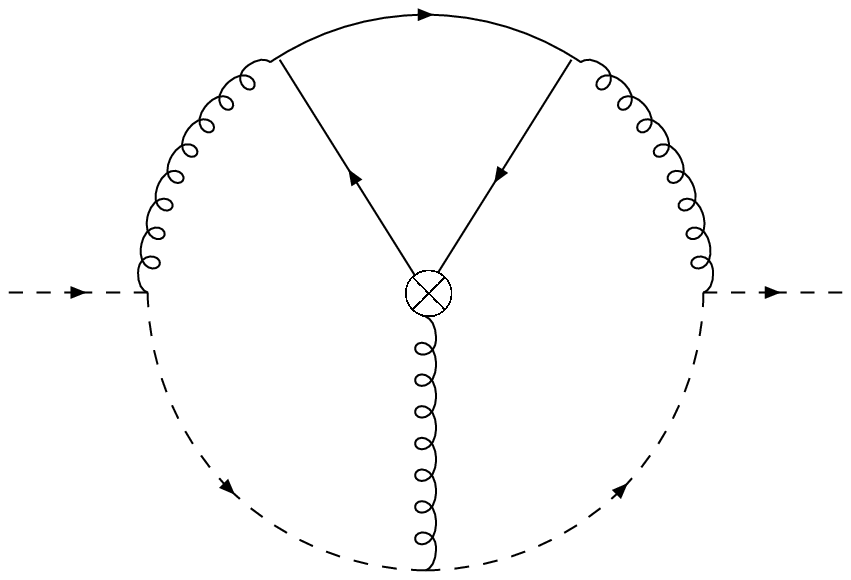}
\vspace*{-8mm}
\begin{center}
{\footnotesize (d)}
\end{center}
\end{minipage}

\vspace*{4mm}
\begin{minipage}[c]{0.23\linewidth}
     \includegraphics[width=1\textwidth]{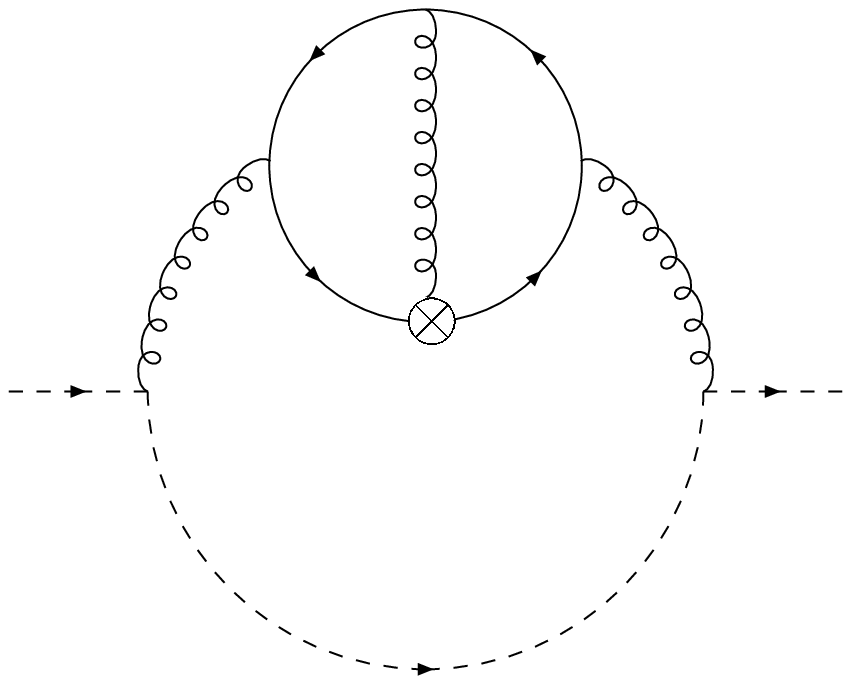}
\vspace*{-8mm}
\begin{center}
{\footnotesize (e)}
\end{center}
\end{minipage}
\hspace*{1mm}
\begin{minipage}[c]{0.23\linewidth}
     \includegraphics[width=1\textwidth]{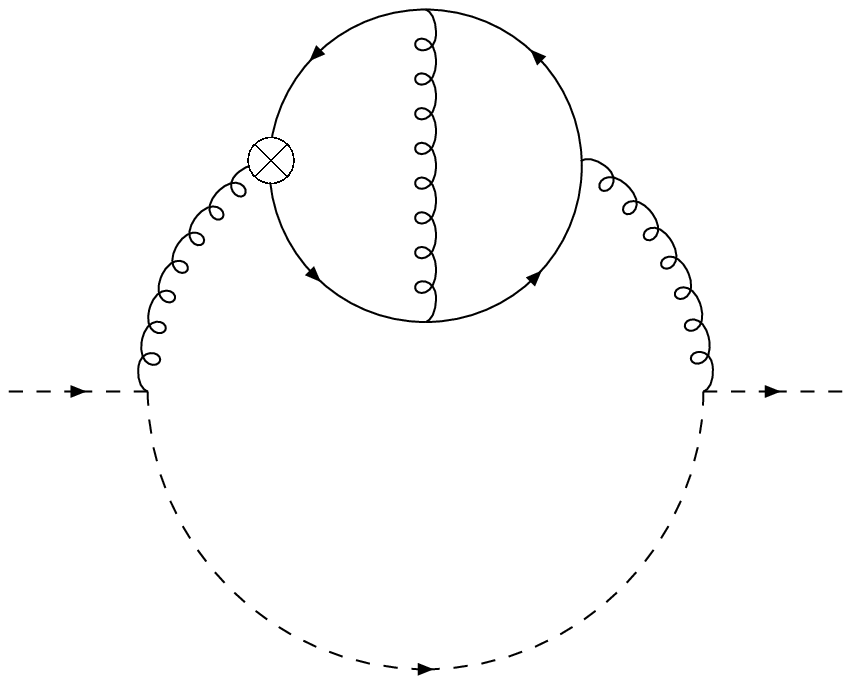}
\vspace*{-8mm}
\begin{center}
{\footnotesize (f)}
\end{center}
\end{minipage}
\hspace*{1mm}
\begin{minipage}[c]{0.23\linewidth}
     \includegraphics[width=1\textwidth]{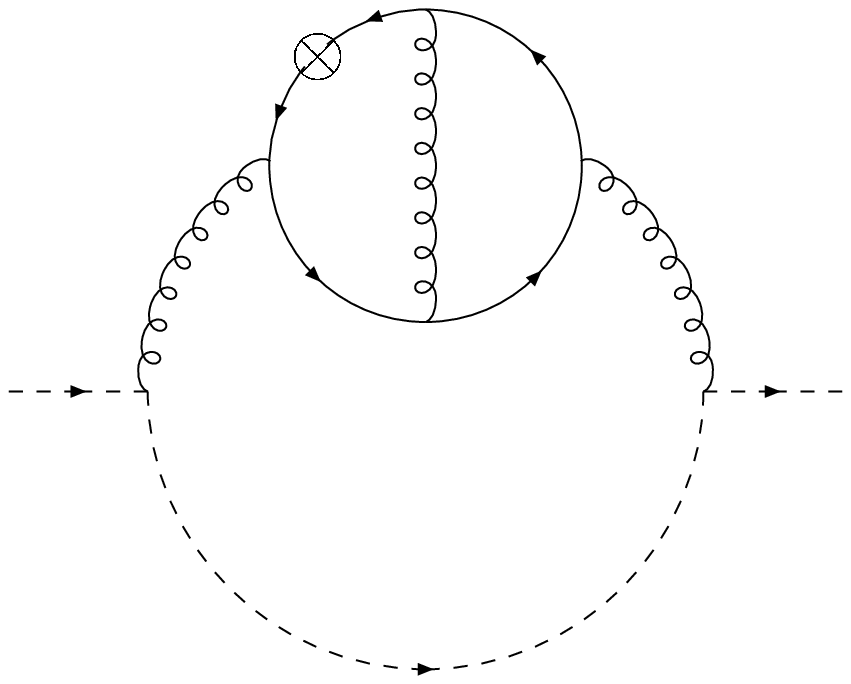}
\vspace*{-8mm}
\begin{center}
{\footnotesize (g)}
\end{center}
\end{minipage}
\hspace*{1mm}
\begin{minipage}[c]{0.23\linewidth}
     \includegraphics[width=1\textwidth]{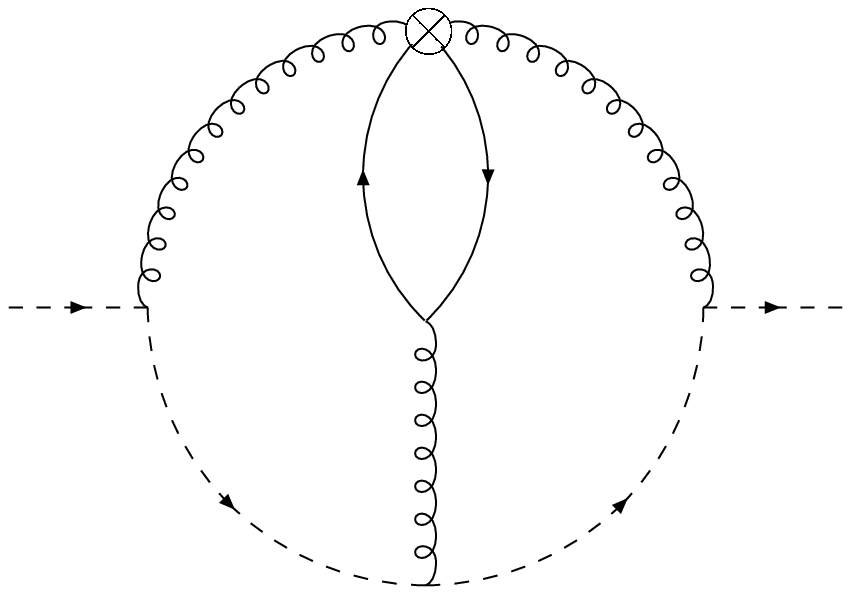}
\vspace*{-8mm}
\begin{center}
{\footnotesize (h)}
\end{center}
\end{minipage}

\vspace*{4mm}
\begin{minipage}[c]{0.23\linewidth}
     \includegraphics[width=1\textwidth]{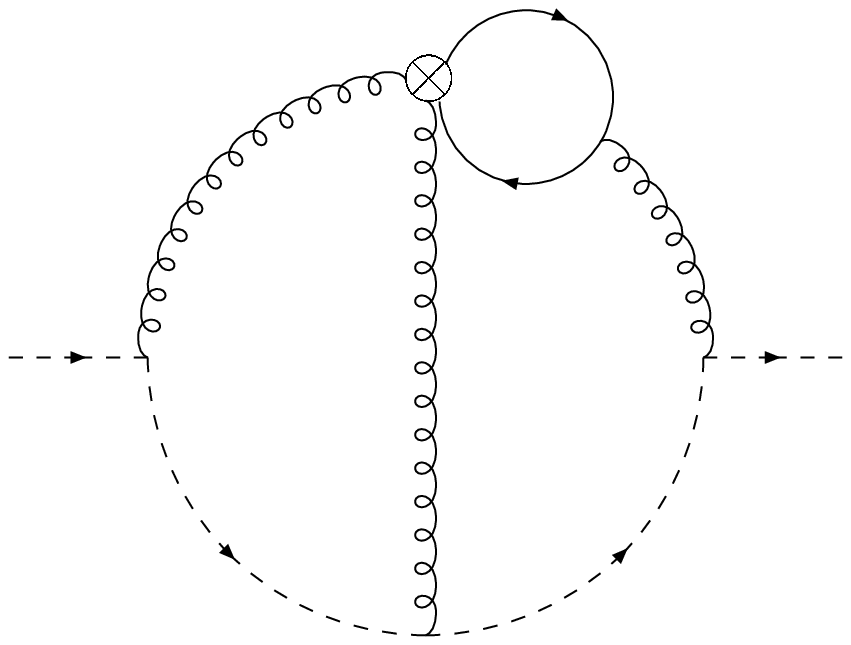} 
\vspace*{-8mm}
\begin{center}
{\footnotesize (i)}
\end{center}
\end{minipage}
\hspace*{1mm}
\begin{minipage}[c]{0.23\linewidth}
     \includegraphics[width=1\textwidth]{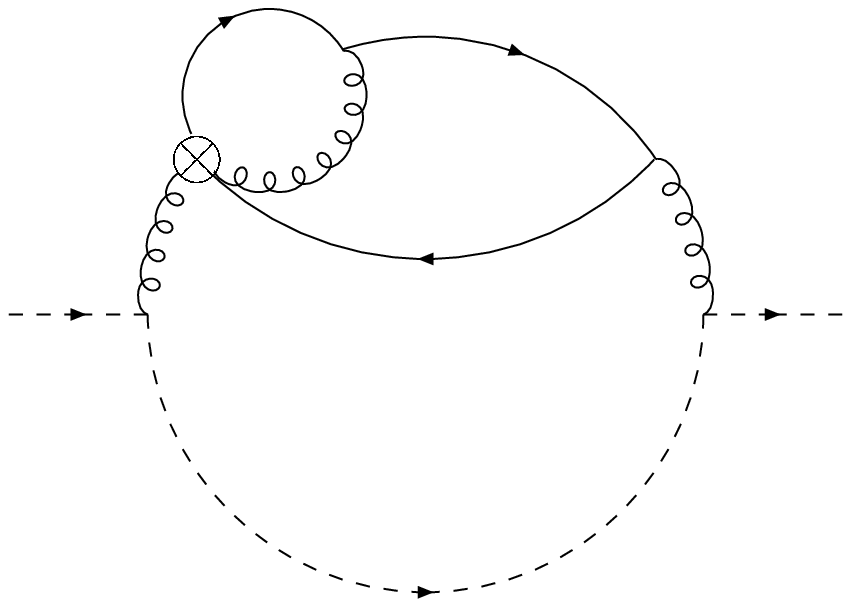} 
\vspace*{-8mm}
\begin{center}
{\footnotesize (j)}
\end{center}
\end{minipage}
\hspace*{1mm}
\begin{minipage}[c]{0.23\linewidth}
     \includegraphics[width=1\textwidth]{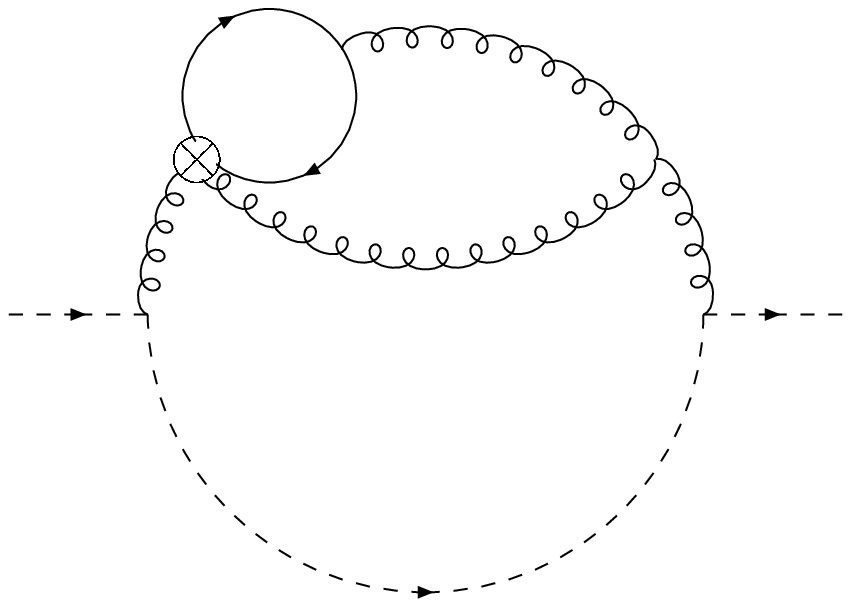} 
\vspace*{-8mm}
\begin{center}
{\footnotesize (k)}
\end{center}
\end{minipage}
\hspace*{1mm}
\begin{minipage}[c]{0.23\linewidth}
     \includegraphics[width=1\textwidth]{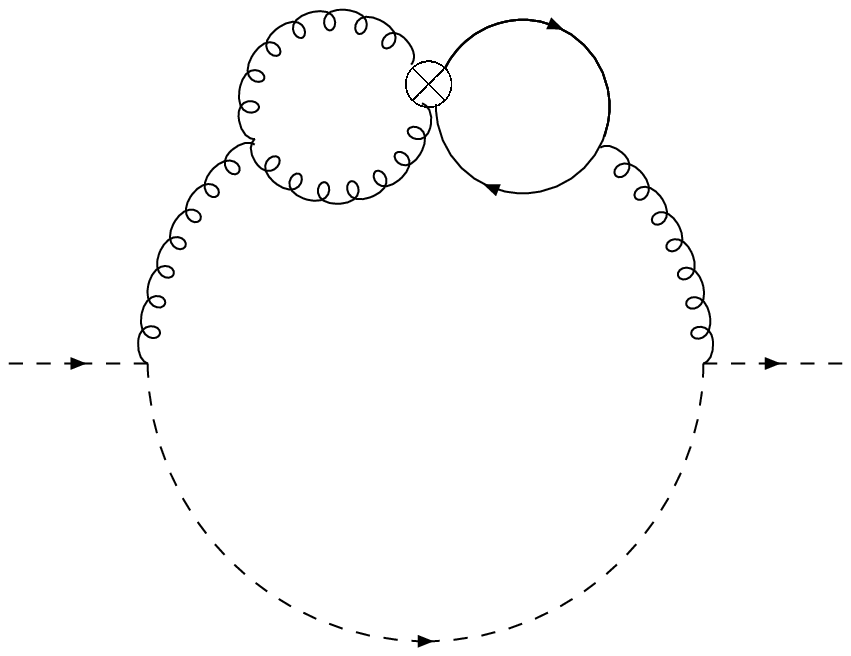}
\vspace*{-8mm}
\begin{center}
{\footnotesize (l)}
\end{center}
\end{minipage}

\vspace*{4mm}
\begin{minipage}[c]{0.23\linewidth}
     \includegraphics[width=1\textwidth]{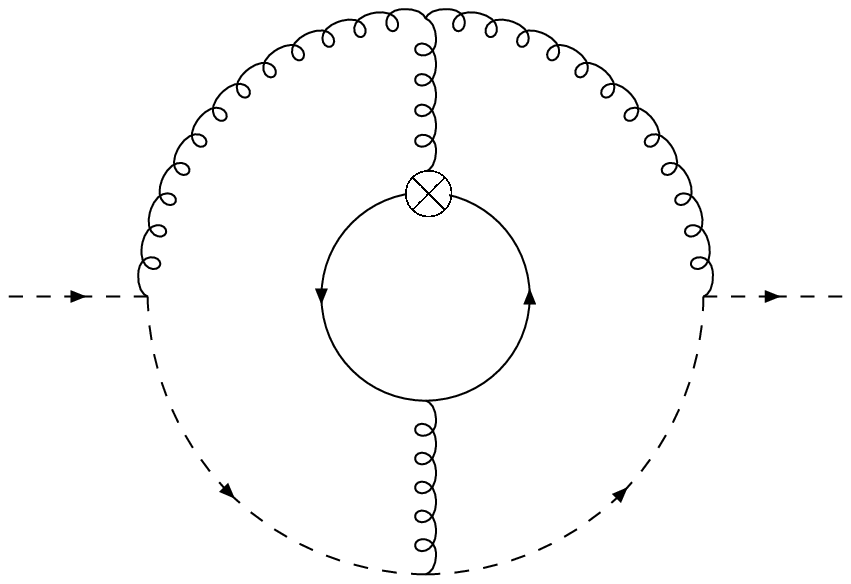}
\vspace*{-8mm}
\begin{center}
{\footnotesize (m)}
\end{center}
\end{minipage}
\hspace*{1mm}
\begin{minipage}[c]{0.23\linewidth}
     \includegraphics[width=1\textwidth]{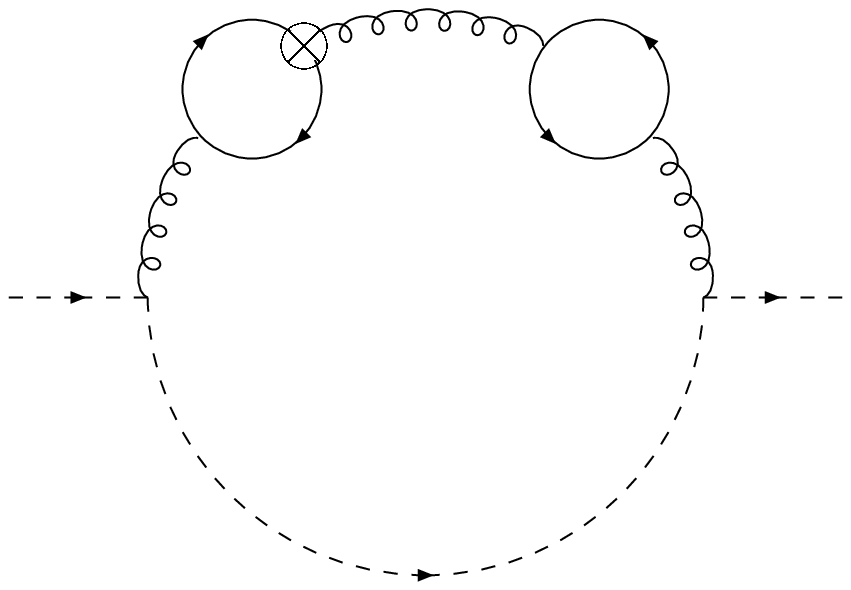}
\vspace*{-8mm}
\begin{center}
{\footnotesize (n)}
\end{center}
\end{minipage}
\hspace*{1mm}
\begin{minipage}[c]{0.23\linewidth}
     \includegraphics[width=1\textwidth]{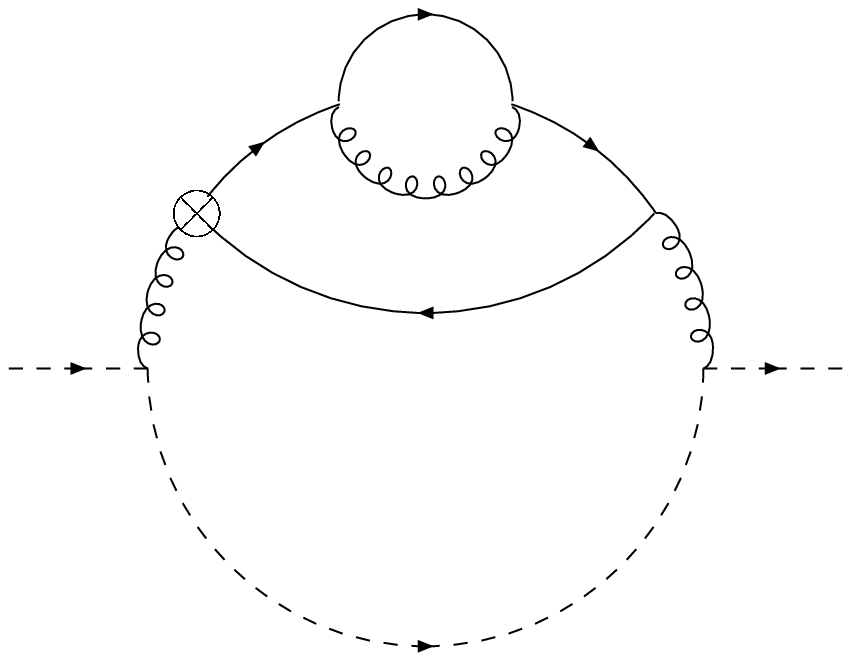}
\vspace*{-8mm}
\begin{center}
{\footnotesize (o)}
\end{center}
\end{minipage}
\hspace*{1mm}
\begin{minipage}[c]{0.23\linewidth}
     \includegraphics[width=1\textwidth]{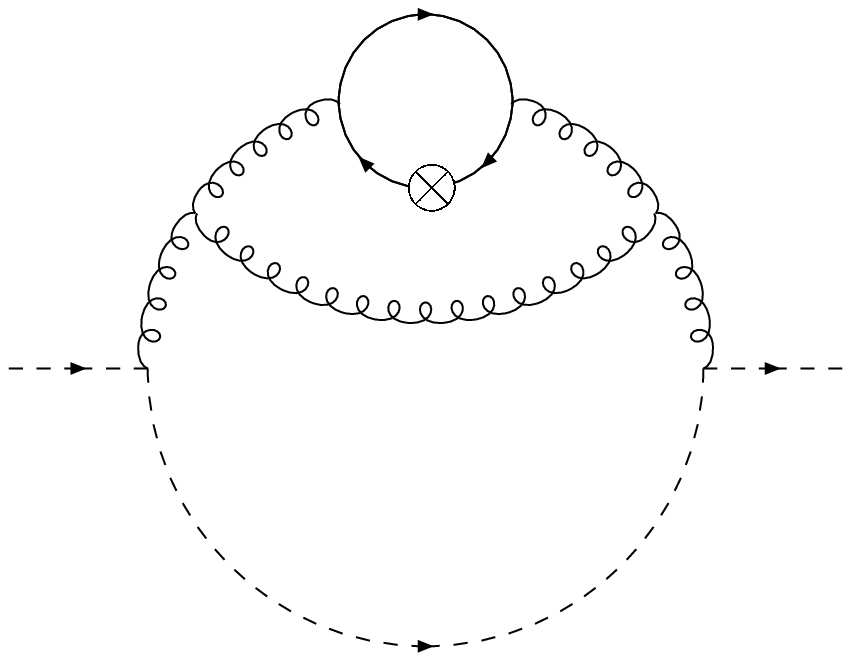}
\vspace*{-8mm}
\begin{center}
{\footnotesize (p)}
\end{center}
\end{minipage}

\vspace*{4mm}
\begin{minipage}[c]{0.23\linewidth}
     \includegraphics[width=1\textwidth]{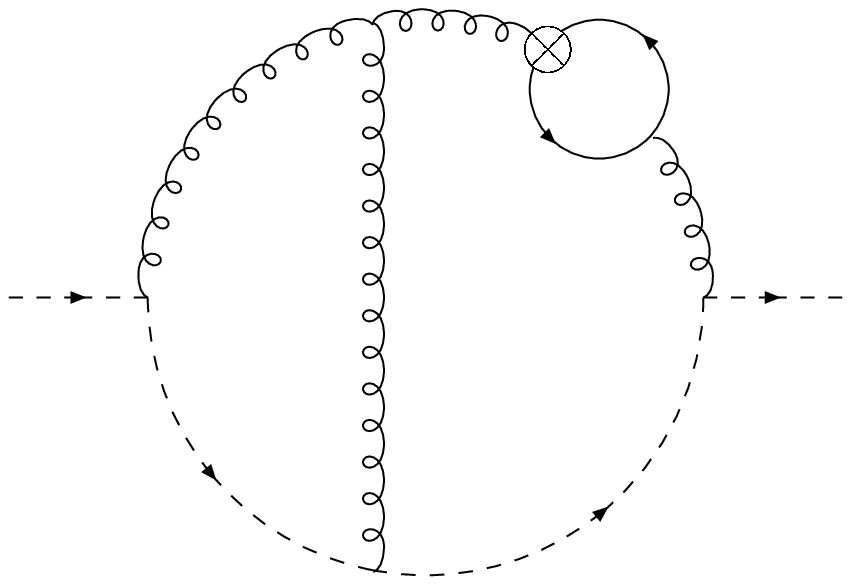}
\vspace*{-8mm}
\begin{center}
{\footnotesize (q)}
\end{center}
\end{minipage}
\hspace*{1mm}
\begin{minipage}[c]{0.23\linewidth}
     \includegraphics[width=1\textwidth]{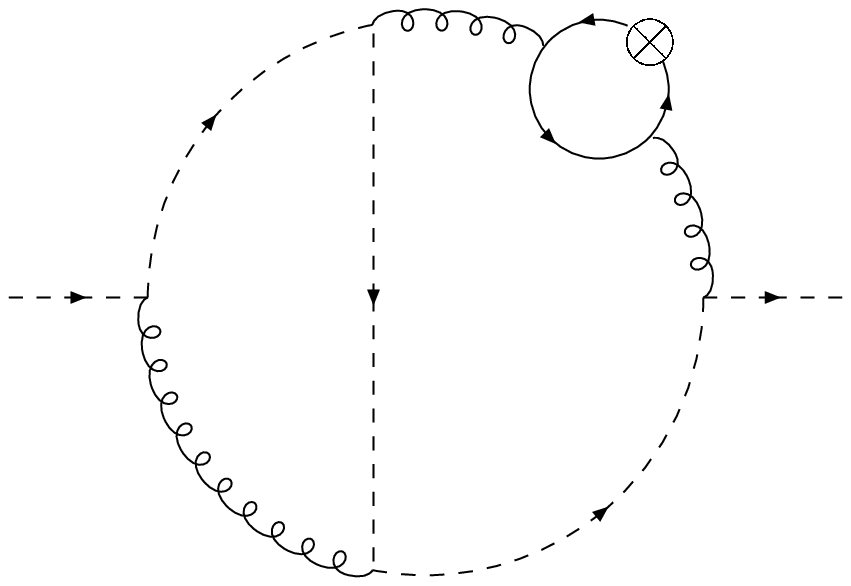}
\vspace*{-8mm}
\begin{center}
{\footnotesize (r)}
\end{center}
\end{minipage}
\hspace*{1mm}
\begin{minipage}[c]{0.23\linewidth}
     \includegraphics[width=1\textwidth]{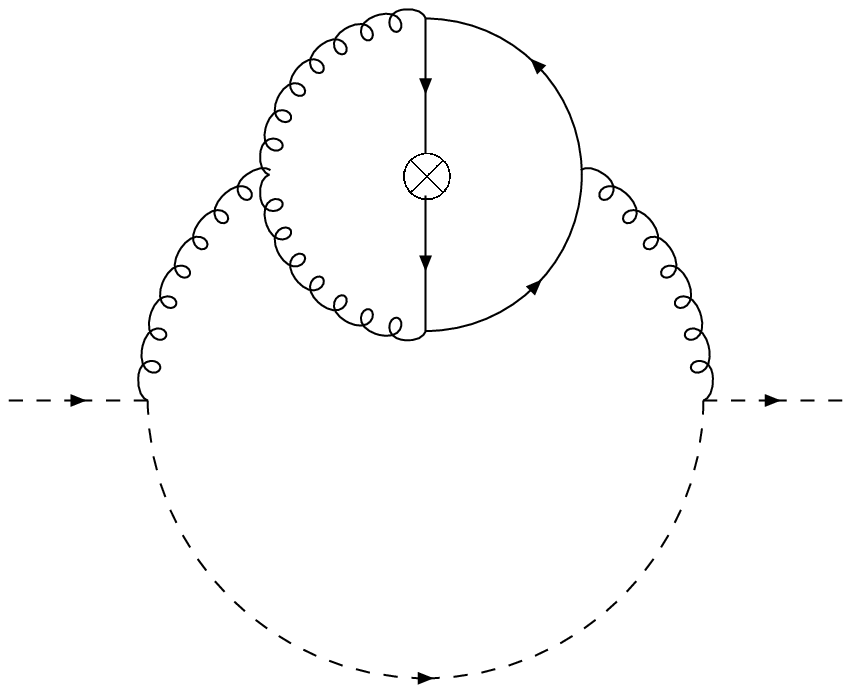}
\vspace*{-8mm}
\begin{center}
{\footnotesize (s)}
\end{center}
\end{minipage}
\hspace*{1mm}
\begin{minipage}[c]{0.23\linewidth}
     \includegraphics[width=1\textwidth]{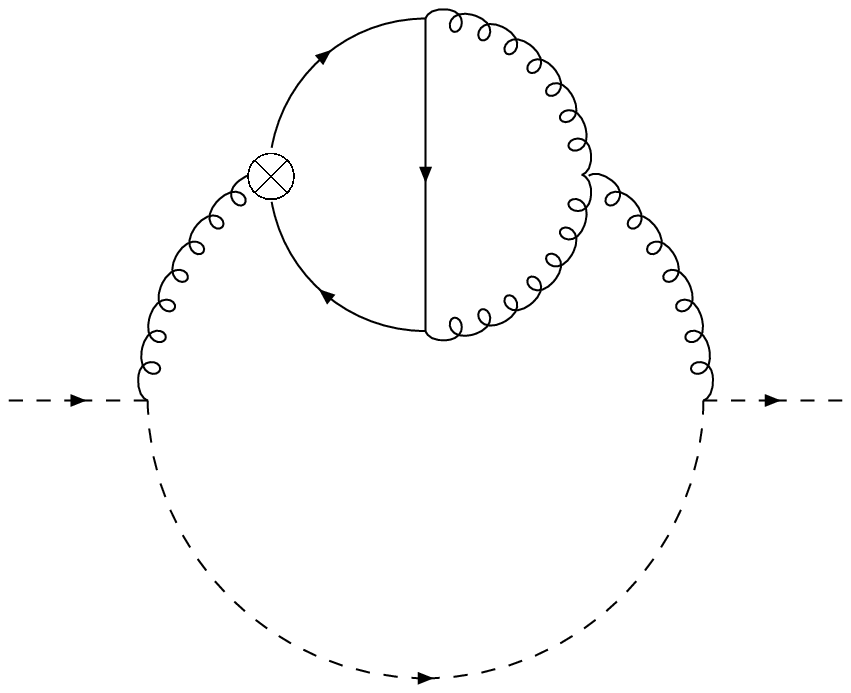}
\vspace*{-8mm}
\begin{center}
{\footnotesize (t)}
\end{center}
\end{minipage}
\caption{\sf \small Sample of diagrams for $A_{Qq}^{(3), \rm PS}$. The dashed arrow lines represent
massless quarks, while the solid arrow lines represent massive quarks, and curly lines are gluons. 
In terms of Feynman integrals, diagrams (a--j) represent the
main topologies, in other words, other diagrams, such as diagrams (k)-(t) can be seen as sub-topologies and/or
are related to diagrams (a--j) by symmetry. The symbol $\otimes$ denotes the local operator insertion, see Ref.~\cite{Bierenbaum:2009mv}.
}
\label{samplediagrams}
\end{figure}

\vspace{1mm}
\noindent
The massive OME $A_{Qq}^{(3), \rm PS}$ is represented by 125 Feynman diagrams, a sample of which is shown in Figure~\ref{samplediagrams}. 
The diagrams are generated using {\tt QGRAF} \cite{Nogueira:1991ex}. Here the  operator insertions are realized in 
terms of 
vertices with non-propagating scalar particles attached to them, cf. \cite{Bierenbaum:2009mv}. The propagators, vertices and operator 
insertions from the output of {\tt QGRAF} are then replaced by the corresponding Feynman rules using a {\tt FORM} \cite{Tentyukov:2007mu} 
program \cite{Bierenbaum:2009mv}, which also allows us to introduce the corresponding projector for the Green function under 
consideration and perform the Dirac-matrix algebra in the numerator of the Feynman integrals. After this, the diagrams end 
up being expressed as linear combinations of scalar integrals. In the case of the contributing bubble topologies we 
used hypergeometric
techniques \cite{Bierenbaum:2007qe,Bierenbaum:2007pn,Bierenbaum:2009zt,Bierenbaum:2008yu,Blumlein:2014fqa,GHYP,Slater,Appell} 
and calculated the corresponding graphs directly, cf.~\cite{Ablinger:2010ty,Behring:2013dga}. 
The packages {\tt Sigma} \cite{SIG1,SIG2}, {\tt EvaluateMultiSums, SumProduction} \cite{EMSSP}, {\tt $\rho$sum} \cite{RHOSUM}, 
{\tt HarmonicSums} \cite{HARMONICSUMS,Ablinger:2013cf} and {\tt  OreSys} \cite{ORESYS} have been used extensively. 
There are nine types of 3-loop integrals involved in the calculation of 
$A_{Qq}^{(3), \rm PS}$. The first five are
\begin{eqnarray}
K_1(\{\nu_i\};a,b,c;N) &=& \int dk  \,\, \frac{(\Delta.k_1)^a  (\Delta.k_2)^b (\Delta.k_3)^c}{D_1^{\nu_1} \cdots D_9^{\nu_9}} (\Delta.k_3)^{N-1}, 
\label{K1} \\
K_2(\{\nu_i\};a,b,c;N) &=& \int dk  \,\, \frac{(\Delta.k_1)^a  (\Delta.k_2)^b (\Delta.k_3)^c}{D_1^{\nu_1} \cdots D_9^{\nu_9}} (\Delta.k_3 - \Delta.k_1)^{N-1}, 
\label{K2} \\
K_3(\{\nu_i\};a,b,c;N) &=& \int dk  \,\, \frac{(\Delta.k_1)^a  (\Delta.k_2)^b (\Delta.k_3)^c}{D_1^{\nu_1} \cdots D_9^{\nu_9}} \sum_{j=0}^{N-2} (\Delta.k_3)^j (\Delta.k_3 - \Delta.k_1)^{N-j-2}, 
\label{K3} \\
K_4 (\{\nu_i\};a,b,c;N)&=& \int dk  \,\, \frac{(\Delta.k_1)^a  (\Delta.k_2)^b (\Delta.k_3)^c}{D_1^{\nu_1} \cdots D_9^{\nu_9}} \nonumber \\ &&
                                                 \,\,   \times \sum_{j=0}^{N-2} (\Delta.k_3 - \Delta.k_1)^j (\Delta.k_3 - \Delta.k_2)^{N-j-2}, 
\label{K4} \\
K_5(\{\nu_i\};a,b,c;N) &=& \int dk  \,\, \frac{(\Delta.k_1)^a  (\Delta.k_2)^b (\Delta.k_3)^c}{D_1^{\nu_1} \cdots D_9^{\nu_9}} \nonumber \\ &&
                                                  \,\,   \times \sum_{j=0}^{N-3} \sum_{l=j+1}^{N-2} (\Delta.k_3)^j 
(\Delta.k_3-\Delta.k_1)^{N-l-2} (\Delta.k_3-\Delta.k_2)^{l-j-1}.
\label{K5}
\end{eqnarray}
Here $\Delta$ denotes a general light-like vector. The Feynman rules, including those for the local operator insertions, are given in 
Ref.~\cite{Bierenbaum:2009mv}.
\noindent
In Eqs.~(\ref{K1}--\ref{K5}),
$\nu_1, \ldots, \nu_9$, $a$, $b$ and $c$ are integers, and we use the shorthand notation
\begin{equation}
\int dk \rightarrow \int \frac{d^D k_1}{(2 \pi)^D} \int \frac{d^D k_2}{(2 \pi)^D} \int \frac{d^D k_3}{(2 \pi)^D}. 
\end{equation}
The (inverse) propagators $D_1, \ldots, D_9$ are given by
\begin{eqnarray}
& D_1 = k_1^2, \quad D_2 = (k_1-p)^2, \quad D_3 = k_2^2, \quad D_4 = (k_2-p)^2,  \nonumber \\
& D_5 = k_3^2-m^2, \quad D_6 = (k_1-k_3)^2-m^2, \quad D_7 = (k_2-k_3)^2-m^2, \nonumber \\
& D_8 = (k_1-k_2)^2 \quad {\rm and} \quad D_9 = (k_3-p)^2-m^2\, , \label{propsB1a} & 
\end{eqnarray}
where $m$ is the mass of the heavy quark and $p$ is the momentum of the external massless quark, which is taken on-shell ($p^2=0$).
For example, the diagram in Figure~\ref{samplediagrams}a can be written as a linear combination of the $K_1$-type integrals defined 
in Eq.~(\ref{K1}), and the diagram in Figure~\ref{samplediagrams}b can be written 
in terms of the $K_3$-type integrals defined in 
Eq.~(\ref{K3}). 

The last four types of integrals have a different propagator structure and are given by
\begin{eqnarray}
K_6(\{\nu_i\};a,b,c;N) &=& \int dk  \,\, \frac{(\Delta.k_1)^a  (\Delta.k_2)^b (\Delta.k_3)^c}{{D'}_1^{\nu_1} \cdots {D'}_9^{\nu_9}} \sum_{j=0}^{N-2} (\Delta.k_1)^j (\Delta.k_2)^{N-j-2}, 
\label{K6} \\
K_7(\{\nu_i\};a,b,c;N) &=& \int dk  \,\, \frac{(\Delta.k_1)^a  (\Delta.k_2)^b (\Delta.k_3)^c}{{D'}_1^{\nu_1} \cdots {D'}_9^{\nu_9}} (\Delta.k_1)^{N-1},  
\label{K7} \\
K_8(\{\nu_i\};a,b,c;N) &=& \int dk  \,\, \frac{(\Delta.k_1)^a  (\Delta.k_2)^b (\Delta.k_3)^c}{{D'}_1^{\nu_1} \cdots {D'}_9^{\nu_9}} \sum_{j=0}^{N-2} (\Delta.k_1)^j (\Delta.k_1-\Delta.k_3)^{N-j-2}, 
\label{K8} 
\nonumber\\
\\
K_9(\{\nu_i\};a,b,c;N) &=& \int dk  \,\, \frac{(\Delta.k_1)^a  (\Delta.k_2)^b (\Delta.k_3)^c}{{D'}_1^{\nu_1} \cdots {D'}_9^{\nu_9}} \nonumber \\ &&
                                                \,\,  \times \sum_{j=0}^{N-3} \sum_{l=j+1}^{N-2} (\Delta.k_1)^j 
(\Delta.k_1-\Delta.k_3)^{N-l-2} (\Delta.k_2-\Delta.k_3)^{l-j-1},
\label{K9}
\end{eqnarray}
where
\begin{eqnarray}
& {D'}_1 = k_1^2-m^2, \quad {D'}_2 = (k_1-p)^2-m^2, \quad {D'}_3 = k_2^2-m^2, \quad {D'}_4 = (k_2-p)^2-m^2,  \nonumber \\
& {D'}_5 = k_3^2, \quad {D'}_6 = (k_1-k_3)^2-m^2, \quad {D'}_7 = (k_2-k_3)^2-m^2, \nonumber \\
& {D'}_8 = (k_1-k_2)^2 \quad {\rm and} \quad {D'}_9 = (k_3-p)^2\, . & 
\end{eqnarray}
For example, the diagrams in Figures~\ref{samplediagrams}e, \ref{samplediagrams}f and \ref{samplediagrams}g can be written as linear 
combinations
of the integrals defined in Eqs.\ (\ref{K6}), (\ref{K8}) and (\ref{K7}), respectively. The Feynman rule for 4-point operator insertions
contains two terms. For some diagrams, such as those in Figures~\ref{samplediagrams}h and \ref{samplediagrams}i, both parts of the diagram 
associated 
with each term can be written as a linear combination of the $K_5$-type integrals defined in Eq. (\ref{K5}). In the case of the diagram in
Figure~\ref{samplediagrams}j, one piece of the diagram can be written in terms of the $K_5$-type integrals, and the other piece in terms
of the $K_9$-type integrals defined in Eq. (\ref{K9}).
 
Any given diagram has at most eight propagators, so at least one of the propagators in the lists $D_1, \ldots, D_9$ or $D'_1, \ldots, D'_9$
plays the role of an auxiliary propagator, whose presence allows us to uniquely express 
all possible scalar products of momenta $k_i \cdot k_j$ and $k_i \cdot p$ ($i,j=1,2,3$) as linear
combinations of all inverse propagators $D_1$ to $D_9$ (or $D'_1$ to $D'_9$). 
Which one(s) of the nine propagators turn out to be auxiliary depends on the specific diagram under consideration.
Scalar integrals will be identified by the indices $\nu_1, \ldots, \nu_9$, $a$, $b$ and $c$, 
where some of the indices $\nu_1$ to $\nu_9$ can be negative, which will represent a scalar integral with irreducible numerators. 
The factors $(\Delta.k_1)^a$, $(\Delta.k_2)^b$ and $(\Delta.k_3)^c$ arise from contractions of an internal momentum with
a $\Delta$ appearing in the operator insertion Feynman rule. In the case of integrals (\ref{K1}),
(\ref{K2}) and (\ref{K7}), the indices $a$, $b$ and $c$ are bounded by $0 \leq a+b+c \leq 1$, 
while in the case of integrals (\ref{K3}), (\ref{K4}), (\ref{K6}) and (\ref{K8}), we have that $0 \leq a+b+c \leq 2$, 
and in the case of Eqs. (\ref{K5}) and (\ref{K9}) we get $0 \leq a+b+c \leq 3$.
\subsection{Integration by parts identities}
\label{IBPsection}

\vspace{1mm}
\noindent
The number of scalar integrals required in order to calculate the diagrams is quite large. We use integration by parts identities \cite{IBP}
in order to express all scalar integrals in terms of a much smaller set of master integrals. For this purpose we use
{\tt Reduze2} \cite{vonManteuffel:2012np}\footnote{The package {\tt Reduze2} uses the codes {\tt Fermat}  
\cite{FERMAT} and {\tt GiNac} \cite{Bauer:2000cp}.}, which is a {\tt C++} program based on Laporta's algorithm 
\cite{Laporta:1996mq,Laporta:2001dd,Tkachov:1981wb,Chetyrkin:1981qh}.
It is somewhat difficult to adapt this algorithm to the case where we
have operator insertions since it requires the integrals to be identified by definite indices,
and in the numerator of the integrals we have dot products of internal momenta with $\Delta$ raised to arbitrary parameters such as $N$, $j$ or $N-j$. 
For this reason, we introduce a generating
function in a new variable $x$, rewriting all operator insertions in terms of a sum in $N$, cf. \cite{Ablinger:2012qm}. For example, a 
fermion line insertion with
momentum $k$ going through the line will be re-expressed as\footnote{We suppress here all factors independent of potential loop momenta.}
\begin{equation}
(\Delta \cdot k)^{N-1} \rightarrow \sum^{\infty}_{N=1} x^{N-1} (\Delta \cdot k)^{N-1} = \frac{1}{1-x \Delta \cdot k} \, .
\label{N2x1}
\end{equation}
This then can be treated as an additional propagator\footnote{{\tt Reduze2} has been adapted to deal with this kind of propagators.}, 
and Laporta's algorithm can be applied without further modifications. 
Similarly, the 3-point and 4-point vertex operator insertions can be replaced by  products of two or three such 
artificial 
propagators, respectively. In the 3-point case, we get
\begin{eqnarray}
\sum^{N-2}_{j=0} (\Delta \cdot q_1)^j (\Delta \cdot q_2)^{N-j-2} & \rightarrow & 
\sum^{\infty}_{N=2} x^{N-2} \sum^{N-2}_{j=0} (\Delta \cdot q_1)^j (\Delta \cdot q_2)^{N-j-2} \nonumber \\ 
& = & \frac{1}{(1-x \Delta \cdot q_1) (1-x \Delta \cdot q_2)}
\label{N2x2}
\end{eqnarray}
and in the 4-point case, the replacement
\begin{equation}
\sum^{N-3}_{j=0} \sum^{N-2}_{l=j+1} (\Delta \cdot q_1)^j (\Delta \cdot q_2)^{N-l-2} (\Delta \cdot q_3)^{l-j-1}  \rightarrow 
\frac{1}{(1-x \Delta \cdot q_1) (1-x \Delta \cdot q_2) (1-x \Delta \cdot q_3)}
\label{N2x3}
\end{equation}
holds.
In this way, the five integrals given in Eqs. (\ref{K1}--\ref{K5}) can all be represented in terms of the general integral
\begin{equation}
J^{\rm B1a}_{\nu_1, \ldots, \nu_{12}}(x) = \int dk  \,\, \frac{1}{{D}_1^{\nu_1} \cdots {D}_{12}^{\nu_{12}}},
\label{IntJ}
\end{equation}
where
\begin{equation}
D_{10} = 1-x (\Delta.k_3-\Delta.k_1), \quad D_{11} = 1-x \Delta.k_3, \quad D_{12} = 1-x (\Delta.k_3-\Delta.k_2),
\end{equation}
and the propagators $D_1$ to $D_9$ are the ones defined in Eq. (\ref{propsB1a}).

Notice that the set of (inverse) propagators $D_1$ to $D_{12}$ is complete and minimal, which means that
any scalar product of a loop momentum with $\Delta$, $p$ or loop momenta can be uniquely expressed as a linear
combination of these propagators. A set of propagators satisfying this condition is called an {\it integral family}.
The superscript {\sf B1a} in Eq. (\ref{IntJ}) labels the particular integral family defined by the propagators  $D_1$ to $D_{12}$.
A given scalar integral will be completely identified by specifying the integral family and the 
set of indices $\nu_1$ to $\nu_{12}$.
There is a total of 24 integral families needed for the calculation of {\it all} operator matrix elements, although,
as we will see, only three of them are needed for the calculation of $A_{Qq}^{(3), \rm PS}$.

\begin{figure}[ht]
\begin{minipage}[c]{1\linewidth}
\centering
\includegraphics[width=0.7\textwidth]{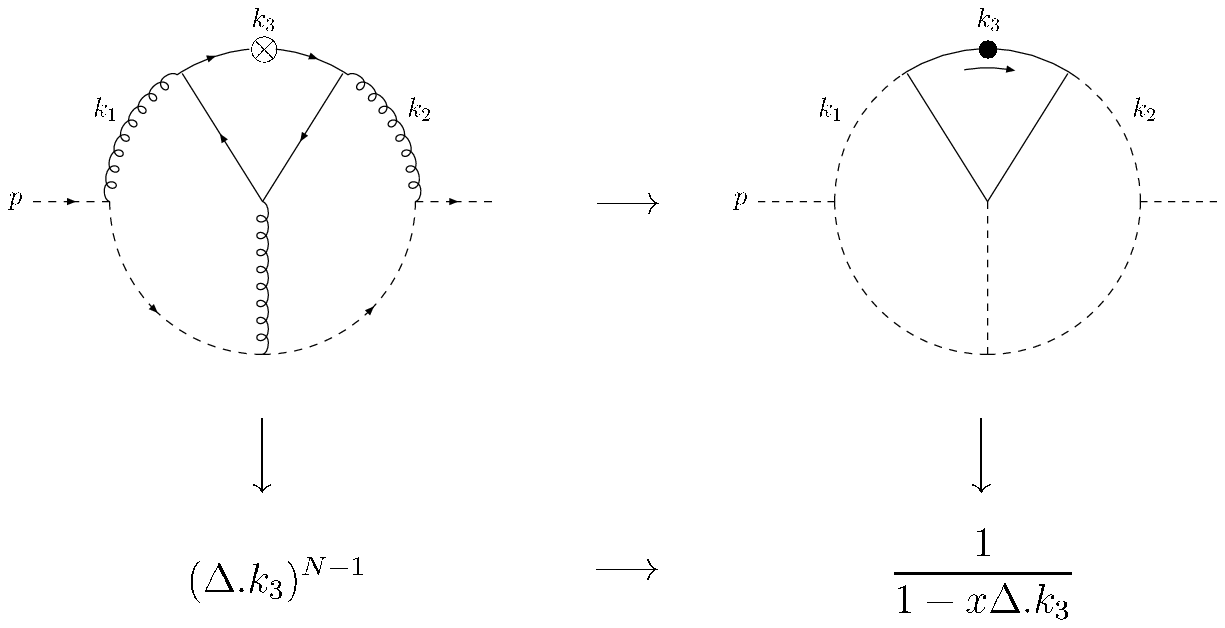}
\end{minipage}

\begin{center}
(a)
\end{center}

\begin{minipage}[c]{1\linewidth}
\centering
\includegraphics[width=0.7\textwidth]{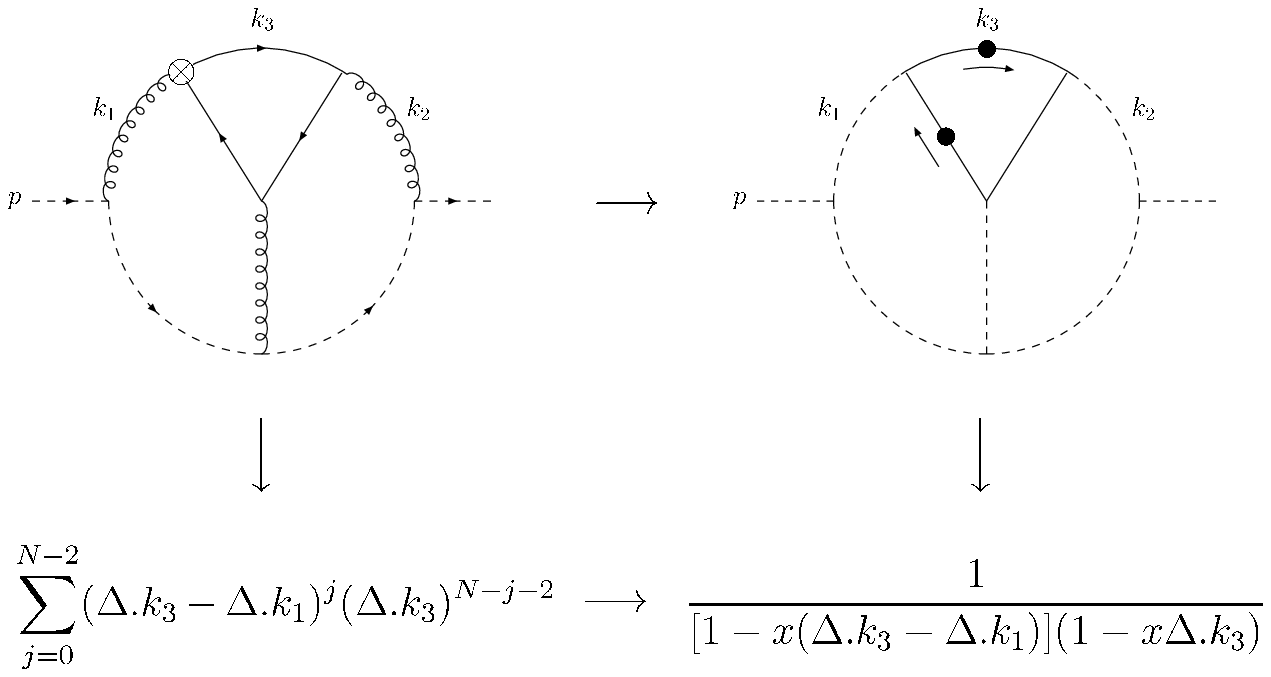}
\end{minipage}

\begin{center}
(b)
\end{center}
\caption{\sf \small Graphical illustration of the transformation to the $x$-representation of the integrals. In (a) we show a line 
insertion,
while in (b) a vertex insertion is depicted. The full circles denote the $x$-dependent generating functions. 
}
\label{Corresp}
\end{figure}

We can see from the replacements given in Eqs. (\ref{N2x1}--\ref{N2x3}) that
\begin{eqnarray}
K_1(\{\nu_i\};0,0,0;N) \rightarrow J^{\rm B1a}_{\nu_1, \ldots, \nu_9,0,1,0}(x), \label{A2J} \\
K_2(\{\nu_i\};0,0,0;N) \rightarrow J^{\rm B1a}_{\nu_1, \ldots, \nu_9,1,0,0}(x), \label{B2J} \\
K_3(\{\nu_i\};0,0,0;N) \rightarrow J^{\rm B1a}_{\nu_1, \ldots, \nu_9,1,1,0}(x), \label{F2J} \\
K_4(\{\nu_i\};0,0,0;N) \rightarrow J^{\rm B1a}_{\nu_1, \ldots, \nu_9,1,0,1}(x), \label{G2J} \\
K_5(\{\nu_i\};0,0,0;N) \rightarrow J^{\rm B1a}_{\nu_1, \ldots, \nu_9,1,1,1}(x), \label{H2J}.
\end{eqnarray}
This is represented diagrammatically in Figures~\ref{Corresp}a and \ref{Corresp}b, where we illustrate as examples the transformations 
corresponding
to Eqs. (\ref{A2J}) and (\ref{F2J}), respectively. The diagrams on the left-hand side of Figure~\ref{Corresp} must be interpreted as the corresponding
scalar integral with no numerator other than the term coming from the operator insertion shown below the diagram. The diagrams 
on the right-hand side
of Figure~\ref{Corresp} represent the scalar integrals after the transformations in Eqs. (\ref{N2x1}--\ref{N2x3}) are done. Solid and 
dashed lines
represent massive and massless propagators, respectively. A large dot on a line 
in these diagrams represents an artificial propagator of the form $(1-x \Delta.q)^{-1}$, where $q$ is the momentum going through the line in the depicted direction.

If any of the indices $a$, $b$, $c$ in Eqs. (\ref{K1}--\ref{K5}) is different from zero,
one can always represent the corresponding numerators via the  propagators $D_{10}$, $D_{11}$ and
$D_{12}$, leading to linear combinations of the integrals defined in Eq. (\ref{IntJ}). For example,
\begin{equation}
K_1(\{\nu_i\};1,0,0;N) \rightarrow \frac{1}{x} \left(J^{\rm B1a}_{\nu_1, \ldots, \nu_9,0,0,0}(x) - J^{\rm B1a}_{\nu_1, \ldots, \nu_9,1,-1,0}(x)\right)
\label{K1100}
\end{equation}
or
\begin{eqnarray}
K_3(\{\nu_i\};1,0,1;N) &\rightarrow& 
\frac{1}{x^2} \left(J^{\rm B1a}_{\nu_1, \ldots, \nu_9,0,1,0}(x) - J^{\rm B1a}_{\nu_1, \ldots, \nu_9,1,0,0}(x) - J^{\rm B1a}_{\nu_1, \ldots, \nu_9,0,0,0}(x)
\right. \nonumber \\ && \phantom{\frac{1}{x^2} \left( \right.} \left. 
 + J^{\rm B1a}_{\nu_1, \ldots, \nu_9,1,-1,0}(x)\right).
\label{K3101}
\end{eqnarray}

Similarly, the integrals defined in Eqs. (\ref{K6}--\ref{K7}) can be re-expressed as
\begin{equation}
J^{\rm B5a}_{\nu_1, \ldots, \nu_{12}}(x) = \int dk  \,\, \frac{1}{{D'}_1^{\nu_1} \cdots {D'}_{12}^{\nu_{12}}},
\label{IntJ5a}
\end{equation}
where
\begin{equation}
D'_{10} = 1-x \Delta.k_1, \quad D'_{11} = 1-x \Delta.k_3, \quad D'_{12} = 1-x \Delta.k_2,
\end{equation}
and the integrals defined in Eq. (\ref{K8}) and (\ref{K9}) are associated with
\begin{equation}
J^{\rm B5c}_{\nu_1, \ldots, \nu_{12}}(x) = \int dk  \,\, \frac{1}{{D'}_1^{\nu_1} \cdots {D'}_{9}^{\nu_{9}} {D''}_{10}^{\nu_{10}} {D''}_{11}^{\nu_{11}} {D''}_{12}^{\nu_{12}}},
\label{IntJ5c}
\end{equation}
where
\begin{equation}
{D''}_{10} = 1-x \Delta.k_1, \quad {D''}_{11} = 1-x (\Delta.k_1-\Delta.k_3), \quad {D''}_{12} = 1-x (\Delta.k_2-\Delta.k_3).
\end{equation}
We can see that
\begin{eqnarray}
K_6(\{\nu_i\};0,0,0;N) \rightarrow J^{\rm B5a}_{\nu_1, \ldots, \nu_9,1,0,1}(x), \label{L2J} \\
K_7(\{\nu_i\};0,0,0;N) \rightarrow J^{\rm B5a}_{\nu_1, \ldots, \nu_9,1,0,0}(x), \label{M2J} \\
K_8(\{\nu_i\};0,0,0;N) \rightarrow J^{\rm B5c}_{\nu_1, \ldots, \nu_9,1,1,0}(x), \label{R2J} \\
K_9(\{\nu_i\};0,0,0;N) \rightarrow J^{\rm B5c}_{\nu_1, \ldots, \nu_9,1,1,1}(x), \label{V2J} 
\end{eqnarray}
and in cases where any of the indices $a$, $b$, $c$ is different from zero, we again get linear combinations similar to those 
of Eqs.~(\ref{K1100}--\ref{K3101}).
The integral families {\sf B1a}, {\sf B5a} and {\sf B5c} are shown again in Appendix~\ref{K3amAppendix}, where we depict the different 
topologies that they cover.

In total, 66 master integrals were required for the reduction of all integrals appearing in the calculation of $A_{Qq}^{(3), \rm PS}$. 
Of those, 55 belong to family {\sf B1a} and 11 to family {\sf B5a}. In Table~\ref{masterslistB5a}, we list the master integrals in family 
{\sf B5a}. 
The list of integrals in family {\sf B1a} is a bit long and will be omitted here. No master integrals in family {\sf B5c} were required, 
since all
integrals in this family were reduced to master integrals in family {\sf B5a}. This is a peculiarity of $A_{Qq}^{(3), \rm PS}$. For other
operator matrix elements where family {\sf B5c} appears, a few master integrals belonging to this family will be required.

\begin{table}
\begin{center}
\begin{tabular}{|cccccccccccc|c|}
\hline
$\nu_1$ & $\nu_2$ & $\nu_3$ & $\nu_4$ & $\nu_5$ & $\nu_6$ &  $\nu_7$ & $\nu_8$ & $\nu_9$ & $\nu_{10}$ & $\nu_{11}$ & $\nu_{12}$     & order in $\varepsilon$  \\
\hline
    1    &    0    &    1    &    0    &    0    &    1    &     1    &    0    &    0    &     0      &     0      &     0          &         2            \\ 
    0    &    0    &    1    &    0    &    0    &    1    &     1    &    0    &    1    &     1      &     0      &     0          &         2            \\ 
    1    &    0    &    1    &    0    &    0    &    1    &     1    &    0    &    1    &     1      &     0      &     0          &         1            \\ 
    1    &    0    &    1    &    0    &    0    &    1    &     1    &    0    &    2    &     1      &     0      &     0          &         1            \\ 
    1    &    0    &    2    &    0    &    0    &    1    &     1    &    0    &    1    &     1      &     0      &     0          &         1            \\ 
    2    &    0    &    1    &    0    &    0    &    1    &     1    &    0    &    1    &     1      &     0      &     0          &         1            \\ 
    0    &    0    &    1    &    0    &    0    &    1    &     1    &    1    &    1    &     1      &     0      &     0          &         3            \\ 
    0    &    0    &    1    &    0    &    0    &    2    &     1    &    1    &    1    &     1      &     0      &     0          &         3            \\
    0    &    0    &    2    &    0    &    0    &    1    &     1    &    1    &    1    &     1      &     0      &     0          &         3            \\
    1    &    0    &    1    &    0    &    0    &    1    &     1    &    0    &    1    &     1      &     0      &     1          &         0            \\ 
    2    &    0    &    1    &    0    &    0    &    1    &     1    &    0    &    1    &     1      &     0      &     1          &         1            \\ 
\hline
\end{tabular}
\end{center}
\caption{\sf \small List of master integrals in family {\sf B5a} identified by the indices $\nu_1$ to $\nu_{12}$ according to Eq. 
(\ref{IntJ5a}).
In the last column, we indicate the order in the dimensional parameter 
$\varepsilon = D-4$ to which each integral needs to be expanded.} 
\label{masterslistB5a}
\end{table}

Any given integral $I(x)$ appearing in the diagrams can be expressed as a linear combination of the master integrals:
\begin{equation}
I(x) = \sum_{i=1}^{66} c_i(x) J_i(x),
\label{J2masters}
\end{equation}
where the $J_i(x)$'s denote the master integrals. The coefficients $c_i(x)$ are rational functions of $x$, $\Delta.p$, the mass $m$ and the dimension $D$.
Since the coefficients $c_i(x)$ may contain poles in $\varepsilon$, the corresponding master integrals may be required to higher orders beyond $\varepsilon^0$.

The diagrams themselves end up being expressed as a linear combination of master integrals as in Eq. (\ref{J2masters}). Therefore, once we calculate
the master integrals as functions of $x$, we can obtain an expression for the diagrams also as functions of this variable. At the end, we obtain
the diagrams as functions of the Mellin variable $N$ by extracting the $N$th term in the corresponding Taylor expansion in $x$ of the diagrams, and then
shifting $N$ depending on the type of operator insertion present in the diagram, according to Eqs.~(\ref{N2x1}--\ref{N2x3}).

It must be pointed out that the basis of master integrals we have chosen is arbitrary, and we can in principle choose any other basis. 
For convenience, we have chosen a basis where no master integral with negative indices appear. This choice was motivated by the fact that this type of 
integrals are easier to handle for many of the methods we have used to solve them. These methods are discussed in the next section. 
\subsection{Calculation of the master integrals}

\vspace{1mm}
\noindent
For the calculation of the master integrals we used a variety of methods. For the simplest cases, we combined the propagators using Feynman parameters,
leading to expressions that can be solved in terms of generalized hypergeometric functions \cite{GHYP,Slater,Appell}
or by introducing a Mellin-Barnes \cite{MELB} representation.
For more complicated integrals, we used the differential equations method. Below we will describe these methods using a few illustrative 
examples.

As we have seen, the introduction of the variable $x$ has allowed us
to turn all operator insertions into artificial propagators, making the application of Laporta's algorithm straightforward.
The calculation of the integrals in this representation using Feynman parameters, although possible in principle, can be somewhat difficult, since
integrals become more complicated as more propagators are present (more Feynman parameters need to be introduced). For this reason, in these cases we 
calculate the master integrals in the original $N$-dependent representation. Once the integrals are calculated as functions of $N$, one can always 
go to the $x$-representation when needed by performing the transformations given in Eqs. (\ref{N2x1}--\ref{N2x3}). 
On the other hand, in the case of the differential
equations method, we will see that the introduction of the variable $x$ turns out to be actually quite advantageous, although this method ultimately
leads to difference equations in the variable $N$, and we end up obtaining the integrals as functions of $N$, 
just like in the other methods. In calculating the master integrals and for their assembly to the individual Feynman diagrams
we made also use of the package {\tt Matad} \cite{Steinhauser:2000ry} and have performed checks for fixed moments.
\subsubsection{Hypergeometric functions and summation methods}

\vspace*{1mm}
\noindent
The majority of the master integrals were calculated in terms of hypergeometric functions evaluated at 1, or multiple sums of such
functions where the summation indices, the dimensional parameter $\varepsilon=D-4$ and $N$ may appear in the parameters of the function. 
If the corresponding series representation is convergent, the resulting multiple sums can then be evaluated with the {\tt Mathematica}
packages {\tt Sigma}, {\tt HarmonicSums}, {\tt EvaluateMultiSums} and {\tt SumProduction}. These packages implement summation algorithms 
based on difference fields \cite{Karr:81,Schneider:01,Schneider:05a,Schneider:07d,Schneider:08c,Schneider:10a,Schneider:10b,Schneider:10c,
Schneider:13b} and can deal with finite and infinite sums, simplifying the expressions in terms of definite nested sums and products.

Let us consider, for example, the following master integral
\begin{equation}
M_1(N) = K_1(0,1,0,1,1,1,1,0,0;0,0,0;N).
\end{equation}
After we introduce Feynman parameters and perform the loop momentum integrals, we obtain the following expression
\begin{eqnarray}
M_1(N) &=& \int_0^1 dx \int_0^1 dy \int_0^1 dz \int_0^1 dw \,\, \Gamma\left(5-\frac{3}{2} D\right) \sum_{k=0}^{N-1} \binom{N-1}{k} z^{N-k-1} (y (1-z) (1-w))^k 
\nonumber \\ && \phantom{\int_0^1 dx \int_0^1 dy \int_0^1 dz \int_0^1 dw}
\times x^{-2+D/2} (1-x)^{3-D} y^{1-D/2} (1-y)^{3-D}  
\nonumber \\ && \phantom{\int_0^1 dx \int_0^1 dy \int_0^1 dz \int_0^1 dw}
\times z^{1-D/2} w^{-2+D/2} (1-x (1-z (1-y))^{-5+\frac{3}{2} D}.
\end{eqnarray}
The integral in $w$ gives just a Beta-function, while the integral in $x$ can be done in terms of a 
hypergeometric ${}_2F_1$ function. We get,
\begin{eqnarray}
M_1(N) &=& \int_0^1 dy \int_0^1 dz \,\, \Gamma\left(5-\frac{3}{2} D\right) \sum_{k=0}^{N-1} \binom{N-1}{k} \frac{\Gamma(D/2-1)^2 \Gamma(k+1) \Gamma(4-D)}{\Gamma(k+D/2) \Gamma(3-D/2)}
\nonumber \\ && \phantom{\int_0^1 dy \int_0^1 dz}
\times y^{k+1-D/2} (1-y)^{3-D} z^{N-k-D/2} (1-z)^k
\nonumber \\ && \phantom{\int_0^1 dy \int_0^1 dz}
\times {}_2F_1\left[5-\frac{3}{2} D, \frac{D}{2}-1; 3-\frac{D}{2}; 1-z (1-y)\right].
\end{eqnarray}
We can now use the following analytic continuation \cite{RG},
\begin{eqnarray}
{}_2F_1[\alpha,\beta;\gamma;z] &=& \frac{\Gamma(\gamma) \Gamma(\gamma-\alpha-\beta)}
{\Gamma(\gamma-\alpha)\Gamma(\gamma-\beta)}
{}_2F_1[\alpha, \beta; \alpha+\beta-\gamma+1; 1-z]
\nonumber \\ &&
+(1-z)^{\gamma-\alpha-\beta} \frac{\Gamma(\gamma) \Gamma(\alpha+\beta-\gamma)}{\Gamma(\alpha) \Gamma(\beta)}
{}_2F_1[\gamma-\alpha, \gamma-\beta, \gamma-\alpha-\beta+1; 1-z],
\nonumber\\ 
\end{eqnarray}
which leads in our case to
\begin{eqnarray}
M_1(N) &=& \int_0^1 dy \int_0^1 dz \,\, \Gamma\left(5-\frac{3}{2} D\right) \sum_{k=0}^{N-1} \binom{N-1}{k} \frac{\Gamma(D/2-1) \Gamma(k+1) \Gamma(4-D)}{\Gamma(k+D/2)}
\nonumber \\ && 
\times y^{k+1-D/2} (1-y)^{3-D} z^{N-k-D/2} (1-z)^k
\nonumber \\ && 
\times \left\{\frac{\Gamma(D/2-1)^2}{\Gamma(D-2) \Gamma(4-D)} \, {}_2F_1\left[5-\frac{3}{2} D, \frac{D}{2}-1; 
2-\frac{D}{2}; z (1-y)\right] \right.
\nonumber \\ && 
+(z (1-y))^{D/2-1}  \frac{\Gamma(1-D/2)}{\Gamma(5-3D/2)} 
\left. {}_2F_1\left[D-2, 4-D; \frac{D}{2}; z (1-y)\right]\right\}.
\end{eqnarray}
Now we can do the shift $y \rightarrow 1-y$, and the remaining integrals in $y$ and $z$ can be evaluated using \cite{Slater}
\begin{eqnarray}
\int_0^1 dt \,\, t^{\alpha-1} (1-t)^{\beta-1} {}_AF_B[(a);(b);tz] = \frac{\Gamma(\alpha) \Gamma(\beta)}{\Gamma(\alpha+\beta)} \, {}_{A+1}F_{B+1}[\alpha,(a); \alpha+\beta,(b); z].
\end{eqnarray}
We obtain an expression in terms of ${}_4F_3$ hypergeomteric functions evaluated at 1,
\begin{eqnarray}
M_1(N) &=& \Gamma\left(5-\frac{3}{2} D\right) \sum_{k=0}^{N-1} \binom{N-1}{k} \frac{\Gamma(D/2-1) \Gamma(k+1) \Gamma(4-D)}{\Gamma(k+D/2)} 
\nonumber \\ &&
\times \Biggl\{\frac{\Gamma(D/2-1)^2}{\Gamma(D-2) \Gamma(4-D)} B\left(N-k+1-\frac{D}{2},k+1\right) B\left(k+2-\frac{D}{2},4-D\right) 
\nonumber \\ && \phantom{\times \biggl\{ }
\times {}_4F_3\left[\begin{matrix} 5-\frac{3}{2} D,4-D,N-k+1-\frac{D}{2},\frac{D}{2}-1 \\ \phantom{\rule{1mm}{5mm}} k+6-\frac{3}{2} D,N+2-\frac{D}{2},2-\frac{D}{2} \end{matrix} \, ; 1\right]
\nonumber \\ && \phantom{\times \biggl\{ }
+\frac{\Gamma(1-D/2)}{\Gamma(5-3D/2)} B\left(N-k,k+1\right) B\left(3-\frac{D}{2},k+2-\frac{D}{2}\right)
\nonumber \\ && \phantom{\times \biggl\{ }
\times {}_4F_3\left[\begin{matrix} 3-\frac{D}{2},N-k,D-2,4-D \\ \phantom{\rule{1mm}{5mm}} k+5-D,N+1,\frac{D}{2} 
\end{matrix} \, ; 1\right]\Biggr\}.
\end{eqnarray}
The parameters of the hypergeometric functions above satisfy the criteria for convergence, so we can use the corresponding series representations. We find
\begin{eqnarray}
M_1(N) &=& \Gamma^2\left(1+\frac{\varepsilon}{2}\right) \Gamma\left(-\frac{\varepsilon}{2}\right) \sum_{k=0}^{N-1} 
\binom{N-1}{k} \frac{\Gamma(k+1)^2 \Gamma(k-\frac{\varepsilon}{2})}{\Gamma(2 + 
\varepsilon) \Gamma(k+2+\frac{\varepsilon}{2}) }
\nonumber \\ && 
\times 
\biggl\{
\sum_{j=0}^{\infty} 
\frac{\Gamma(j-1-\frac{3}{2} \varepsilon) \Gamma(j-\varepsilon) \Gamma(j+N-k-1-\frac{\varepsilon}{2}) 
\Gamma(j+1+\frac{\varepsilon}{2})}{j! \Gamma(j+k-\frac{3}{2} \varepsilon) \Gamma(j+N-\frac{\varepsilon}{2}) 
\Gamma(j-\frac{\varepsilon}{2})}
\nonumber \\ && 
-
\sum_{j=0}^{\infty} 
\frac{\Gamma(j+1-\frac{\varepsilon}{2}) \Gamma(j+N-k) \Gamma(j+2+\varepsilon) 
\Gamma(j - \varepsilon)}{j! \Gamma(j+k+1-\varepsilon) \Gamma(j+N+1) \Gamma(j+2+\frac{\varepsilon}{2})}
\biggr\}.
\end{eqnarray}
So we get an expression in terms of a double sum (one of them finite and the other infinite). 
This double sum can be done using the packages we mentioned at the beginning of this section. 
These packages can perform the $\varepsilon$ expansion of the expression given above to the required order, 
and then calculate the sums.
The final result is 
\begin{eqnarray}
M_1(N) &=& \frac{1}{N} \biggl\{
\frac{8}{3 \varepsilon^3}
-\frac{4 \left(5 N^2-2 N-1\right)}{3 \varepsilon^2 (N-1) N}
+\frac{1}{\varepsilon} \biggl[
\frac{2 \left(19 N^4-25 N^3-6 N^2+13 N-5\right)}{3 (N-1)^2 N^2}
\nonumber \\ &&
+4 S_2+\zeta_2\biggr]
-\frac{2 \left(3 N^2-N-1\right)}{(N-1) N} S_2
-2 (N-3) S_3
-\frac{\left(5 N^2-2 N-1\right)}{2 (N-1) N} \zeta_2
\nonumber \\ &&
-\frac{65 N^6-152 N^5+63 N^4+86 N^3-95 N^2+54 N-13}{3 (N-1)^3 N^3}
+\frac{1}{3} (6 N+5) \zeta_3
\nonumber \\ &&
+\varepsilon \biggl[
\frac{\left(6 N^3-13 N^2-N+5\right) S_3}{(N-1) N}
+\frac{6 N^4-4 N^3-10 N^2+9 N-3}{(N-1)^2 N^2} S_2
\nonumber \\ &&
+\frac{3}{2} \zeta_2 S_2
+(N-2) S_1 (\zeta_3-S_3)
+\frac{1}{2} (4-N) S_2^2
+\frac{1}{2} (22-9 N) S_4
\nonumber \\ &&
-\frac{36 N^3-23 N^2-22 N+7}{6 (N-1) N} \zeta_3
+\frac{19 N^4-25 N^3-6 N^2+13 N-5}{4 (N-1)^2 N^2} \zeta_2
\nonumber \\ &&
+\frac{-715 N^7+737 N^6+27 N^5-586 N^4+557 N^3-373 N^2+155 N-29}{6 (N-1)^4 N^4}
\nonumber \\ &&
+\frac{211 N^4}{6 (N-1)^4}
+(N-2) S_{3,1}
+\frac{1}{80} (144 N+115) \zeta_2^2\biggr]
+\varepsilon^2 \biggl[
-\frac{3}{2} (5 N-11) S_5
\nonumber \\ &&
+(N-2) S_1 \biggl(
\frac{1}{2} S_{3,1}
-\frac{1}{4} S_2^2
-\frac{9}{4} S_4
+\frac{9}{10} \zeta_2^2
\biggr)
+\frac{27 N^3-51 N^2-7 N+20}{2 (N-1) N} S_4
\nonumber \\ &&
+\frac{(3 N-2) \left(N^2-N-1\right)}{(N-1) N} \big((S_3-\zeta_3) S_1-S_{3,1}\big)
+\frac{3 N^3-8 N^2+3}{2 (N-1) N} S_2^2
\nonumber \\ &&
+\biggl(
\frac{1}{4} (22-9 N) S_3
+\frac{1}{2} (N-2) S_{2,1}
-\frac{3 \left(3 N^2-N-1\right)}{4 (N-1) N} \zeta_2
+\frac{1}{4} (5 N-8) \zeta_3
\nonumber \\ &&
-\frac{8 N^6-52 N^4+82 N^3-57 N^2+30 N-7}{2 (N-1)^3 N^3}
\biggr) S_2
-S_3 \biggl(\frac{3}{4} (N-3) \zeta_2
\nonumber \\ &&
+\frac{24 N^5-54 N^4+4 N^3+58 N^2-37 N+11}{2 (N-1)^2 N^2}
\biggr)
+\frac{1}{4} (N-2) \big(\zeta_3-S_3\big) S_1^2
\nonumber \\ &&
+\frac{1}{2} (N-2) \big(5 S_{4,1}-3 S_{2,3}-S_{2,2,1}-S_{3,1,1}\big)
-\frac{665 N^5-2954 N^4}{12 (N-1)^5}
\nonumber \\ &&
+\frac{1}{20} (180 N+121) \zeta_5
-\frac{864 N^3-577 N^2-518 N+173}{160 (N-1) N} \zeta_2^2
-\frac{397 N^3}{(N-1)^5}
\nonumber \\ &&
+\frac{1}{8} (6 N+5) \zeta_2 \zeta_3
+\frac{144 N^5-277 N^4+79 N^3+90 N^2-43 N+11}{12 (N-1)^2 N^2} \zeta_3
\nonumber \\ &&
-\frac{65 N^6-152 N^5+63 N^4+86 N^3-95 N^2+54 N-13}{8 (N-1)^3 N^3} \zeta_2
+\frac{1327 N^2}{6 (N-1)^5}
\nonumber \\ &&
-\frac{-1425 N^6+3246 N^5-2919 N^4+2130 N^3-1152 N^2+392 N-61}{12 (N-1)^5 N^5}
\biggr]
\biggr\}.
\label{EQ:M1}
\end{eqnarray}
Here $S_{\vec{a}}(N)$  denote the nested harmonic sums \cite{Blumlein:1998if,Vermaseren:1998uu}. They are defined by
\begin{eqnarray}
S_{b,\vec{a}}(N) = \sum_{k=1}^N \frac{({\rm sign}(b))^k}{k^{|b|}} S_{\vec{a}}(k),~~~S_\emptyset = 1,~b, a_i~\in \mathbb{Z} \backslash 
\{0\}~.
\end{eqnarray}
We use the shorthand notation $S_{\vec{a}}(N) \equiv S_{\vec{a}}$.
Note that we have omitted an overall factor of $i S_{\varepsilon}^3$, where $S_{\varepsilon}$ is the
spherical factor given by 
\begin{eqnarray}
S_{\varepsilon} = \exp\left(\frac{\varepsilon}{2}(\gamma_E-\ln(4 \pi))\right).
\label{Sep}
\end{eqnarray}
Here $\gamma_E$ denotes the Euler-Mascheroni constant.

The expression given in Eq.~(\ref{EQ:M1}) is divergent for $N=1$, so we have to calculate this value separately. We get,
\begin{eqnarray}
M_1(1) &=& \frac{16}{3 \varepsilon^3}
-\frac{44}{3 \varepsilon^2}
+\frac{2 \zeta_2+28}{\varepsilon}
-\frac{11}{2} \zeta_2
+\frac{10}{3} \zeta_3
-\frac{139}{3}
\\ && \nonumber
+\varepsilon \left(
\frac{23}{8} \zeta_2^2
+\frac{21}{2} \zeta_2
-\frac{43}{6} \zeta_3
+\frac{215}{3}
\right).
\end{eqnarray}
\subsubsection{Mellin-Barnes integral representations}

\vspace*{1mm}
\noindent
A few master integrals were calculated using a Mellin-Barnes integral representation. In particular, we used this method for seven $K_7$-type master integrals, corresponding
in the $x$-representation to the integrals in Table~\ref{masterslistB5a} starting from the third row until the ninth row, together with 
the first integral appearing 
in this Table (which is independent of $x$).
Let us consider the case
\begin{equation}
M_2(N) = K_7(2,0,1,0,0,1,1,0,1;0,0,0;N).
\end{equation}
After Feynman parameterization we obtain
\begin{eqnarray}
M_2(N) &=& -\int_0^1 dx \int_0^1 dy \int_0^1 dz \int_0^1 dw \,\,
\Gamma\left(-\frac{3}{2} \varepsilon\right) x^{N-2+\varepsilon/2} (1-x)^{\varepsilon/2} y^{\varepsilon/2} (1-y)^{\varepsilon/2}
\nonumber \\ && \phantom{aaaaaaaaaaaaaaaaaaaaa}
\times
\frac{w^{\varepsilon/2} (1-w)^{N-1}}{z^{\varepsilon/2} (1-z)^{1+\varepsilon/2}}
\left[\frac{z}{x (1-x)}+\frac{1-z}{y (1-y)}\right]^{\frac{3}{2} \varepsilon}.
\end{eqnarray}
Now we split the last term using,
\begin{equation}
\frac{1}{(A+B)^{\nu}} = \frac{1}{2 \pi i} \int_{\gamma-i \infty}^{\gamma+i \infty} d\sigma \,\, 
\frac{\Gamma(-\sigma) \Gamma(\sigma+\nu)}{\Gamma(\nu)} A^{\sigma} B^{-\sigma-\nu},
\end{equation}
cf.~\cite{MELB}.
This makes it possible to integrate the Feynman parameters at the expense of introducing the contour integral in $\sigma$. We obtain,
\begin{eqnarray}
M_2(N) &=& -\frac{1}{2 \pi i} \int_{\gamma-i \infty}^{\gamma+i \infty} d\sigma \,\, 
\Gamma(-\sigma) \Gamma\left(\sigma-\frac{3}{2} \varepsilon\right)
\frac{\Gamma(-\sigma+N-1+\varepsilon/2) \Gamma(-\sigma+1+\varepsilon/2)}{\Gamma(-2 \sigma+N+\varepsilon)}
\nonumber \\ && \phantom{-\frac{1}{2 \pi i} \int_{-i \infty}^{+i \infty}}
\times
\Gamma(\sigma+1-\varepsilon/2) \Gamma(-\sigma+\varepsilon)
\frac{\Gamma(\sigma+1-\varepsilon)^2}{\Gamma(2 \sigma+2-2 \varepsilon)}
\frac{\Gamma(N)}{\Gamma(N+1+\varepsilon/2)}.
\label{MB1}
\end{eqnarray}
At this point we use the Mathematica package {\tt MB} \cite{Czakon:2005rk} to find a value for $\gamma$ and $\varepsilon$ such that the 
integral in Eq. (\ref{MB1}) is well defined, 
and then analytically continue to $\varepsilon \rightarrow 0$ and later expand this expression in $\varepsilon$. We get,
\begin{equation}
M_2(N) = a_0(N)+b_0(N)+\varepsilon b_1(N), 
\end{equation}
where $a_0(N)$ is a term produced by {\tt MB} after taking a residue at $\sigma=\frac{3}{2} \varepsilon$ in order to perform 
the analytic continuation in $\varepsilon$. It is given by
\begin{eqnarray}
a_0(N) &=& \frac{1}{N (N-1)} \biggl\{
-\frac{4}{3 \varepsilon^2}
+\frac{1}{\varepsilon}\biggl[\frac{4 \big(N^2-N-1\big)}{3 (N-1) N}-\frac{2}{3} S_1\biggr]
+\frac{2 \big(N^2-N-1\big)}{3 (N-1) N} S_1
-\frac{1}{6} S_1^2
\nonumber \\ &&
-\frac{13}{6} S_2
-\frac{4 \big(N^4-2 N^3-N^2+4 N-1\big)}{3 (N-1)^2 N^2}
-\frac{\zeta_2}{2}
+\varepsilon \biggl[
-\frac{1}{36} S_1^3
-\frac{55}{18} S_3
+\frac{7}{6} \zeta_3
\nonumber \\ &&
+\frac{\big(N^2-N-1\big)}{6 (N-1) N} \big(S_1^2+13 S_2+3 \zeta_2\big)
-\frac{2 \big(N^4-2 N^3-N^2+4 N-1\big)}{3 (N-1)^2 N^2} S_1
\nonumber \\ &&
-\left(\frac{13}{12} S_2+\frac{\zeta_2}{4}\right) S_1
+\frac{4 \big(N^6-3 N^5+N^4+7 N^3-11 N^2+5 N-1\big)}{3 (N-1)^3 N^3}
\biggr]\biggr\}
\end{eqnarray}
The functions $b_0(N)$ and $b_1(N)$ are the following contour integrals,
\begin{equation}
b_0(N)=\frac{1}{2\pi i} \int_{-i \infty}^{+i \infty} d\sigma \,\, \frac{\Gamma(-\sigma)^3 \Gamma(\sigma+1)^4 \Gamma(-\sigma+N-1)}{N \, \Gamma(2 \sigma+2) \Gamma(N-2 \sigma)},
\end{equation}
and
\begin{eqnarray}
b_1(N) &=&
-\frac{1}{2 \pi i} \int_{-i \infty}^{+i \infty} d\sigma \,\,
\frac{\Gamma(-\sigma)^3 \Gamma(\sigma+1)^4 \Gamma(-\sigma+N-1)}{2 N \, \Gamma(2 \sigma+2) \Gamma(N-2 \sigma)} \bigl(3 \gamma_E-2 \psi(-\sigma)+3 \psi(\sigma)
\nonumber \\ && 
\phantom{d\sigma} \;\;\;\;\;\;\;\;\;\;\;\;\;\;\;\;\;\;\;\;
+2 \psi(N-2 \sigma)-\psi(N-1-\sigma)+\psi(N+1)-\psi(1-\sigma)
\nonumber \\ && 
\phantom{d\sigma} \;\;\;\;\;\;\;\;\;\;\;\;\;\;\;\;\;\;\;\;
+5 \psi(\sigma+1)-4 \psi(2 \sigma+2)\bigr).
\end{eqnarray}
Here $\psi(z)$ denotes the Digamma function.
We can now close the contours to the left (or to the right), express the integrals in terms of a sum of residues
and obtain	
\begin{eqnarray}
b_0(N) &=& \sum_{k=1}^{\infty}
\frac{\Gamma (2 k-1) \Gamma (k+N-1)}{N \, \Gamma (k) \Gamma (2 k+N)}
\bigl\{
S_1(k+N-2)^2
+4 S_1(2 k+N-1)^2
\nonumber \\ &&  \phantom{aa}
-S_1(k+N-2) \big[4 S_1(2 k+N-1)+2 S_1(k-1)-4 S_1(2 k-2)\big]
\nonumber \\ &&  \phantom{aa}
+4 \big[S_1(k-1)-2 S_1(2 k-2)\big] S_1(2 k+N-1)
-S_2(k+N-2)
\nonumber \\ &&  \phantom{aa}
+4 S_2(2 k+N-1)
+S_1(k-1)^2
-4 S_1(2 k-2) S_1(k-1)
\nonumber \\ &&  \phantom{aa}
+4 S_1(2 k-2)^2
+S_2(k-1)
-4 S_2(2 k-2)
\bigr\},
\end{eqnarray}
and
\begin{eqnarray}
b_1(N) &=& \sum_{k=1}^{\infty}
\frac{\Gamma (2 k-1) \Gamma (k+N-1)}{N^2 \Gamma (k) \Gamma (2 k+N)}
\biggl\{
-\frac{1}{2} S_1(k+N-2)^2
-2 S_2(2 k+N-1)
\nonumber \\ &&  \phantom{aa}
-\frac{1}{2} S_1(k-1)^2
-2 S_1(2 k-2)^2
-2 S_1(2 k+N-1)^2
+2 S_2(2 k-2)
\nonumber \\ &&  \phantom{aa}
+\big[4 S_1(2 k-2)-2 S_1(k-1)\big] S_1(2 k+N-1)
+\frac{1}{2} S_2(k+N-2)
\nonumber \\ &&  \phantom{aa}
+S_1(k+N-2) \big[S_1(k-1)-2 S_1(2 k-2)+2 S_1(2 k+N-1)\big]
\nonumber \\ &&  \phantom{aa}
+N S_1(N-1) \biggl[-\frac{1}{2} S_1(k-1)^2
                  +2 S_1(2 k-2) S_1(k-1)
                  -2 S_1(2 k-2)^2
\nonumber \\ &&  \phantom{aa}
                  -\frac{1}{2} S_1(k+N-2)^2
                  +(4 S_1(2 k-2)-2 S_1(k-1)) S_1(2 k+N-1)
\nonumber \\ &&  \phantom{aa}
                  -2 S_1(2 k+N-1)^2
                  -\frac{1}{2} S_2(k-1) 
                  +\frac{1}{2} S_2(k+N-2)
                  +2 S_2(2 k-2)
\nonumber \\ &&  \phantom{aa}
                  +S_1(k+N-2) (S_1(k-1)-2 S_1(2 k-2)+2 S_1(2 k+N-1))
\nonumber \\ &&  \phantom{aa}
                  -2 S_2(2 k+N-1)
             \biggr]
-\frac{1}{2} S_2(k-1)
+2 S_1(2 k-2) S_1(k-1)
\nonumber \\ &&  \phantom{aa}
+N \biggl[-\frac{1}{2} S_1(k-1)^3
         -S_1(k) S_1(k-1)^2 
          -\frac{1}{2} S_1(k+N-2)^3
\nonumber \\ &&  \phantom{aa}
          +\biggl(-\frac{3}{2} S_2(k-1)-2 S_2(k)\biggr) S_1(k-1)
          +4 S_1(2 k+N-1)^3
\nonumber \\ &&  \phantom{aa}
          -\big(2 S_1(k-1)+4 S_1(k)\big) S_1(2 k-2)^2
          +\big(2 S_1(k-1)+4 S_1(k)\big) S_2(2 k-2)
\nonumber \\ &&  \phantom{aa}
          +\big(2 S_1(k-1)-4 S_1(k)-8 S_1(2 k-2)\big) S_1(2 k+N-1)^2
          -S_1(k) S_2(k-1)
\nonumber \\ &&  \phantom{aa}
          +S_1(k+N-2)^2 \biggl(\frac{1}{2} S_1(k-1)-S_1(k)-2 S_1(2 k-2)+3 S_1(2 k+N-1)\biggr)
\nonumber \\ &&  \phantom{aa}
          +S_1(2 k-2) \big(2 S_1(k-1)^2+4 S_1(k) S_1(k-1)+2 S_2(k-1)+4 S_2(k)\big)
\nonumber \\ &&  \phantom{aa}
          +\biggl(-\frac{1}{2} S_1(k-1)+S_1(k)+2 S_1(2 k-2)\biggr) S_2(k+N-2)
\nonumber \\ &&  \phantom{aa}
          +S_1(k+N-2) \biggl(\frac{1}{2} S_1(k-1)^2+2 S_1(k) S_1(k-1)-2 S_1(2 k-2)^2
\nonumber \\ &&  \phantom{aa}
                             -6 S_1(2 k+N-1)^2-4 S_1(k) S_1(2 k-2)+\frac{1}{2} S_2(k-1)+2 S_2(2 k-2)
\nonumber \\ &&  \phantom{aa}
                             +(-2 S_1(k-1)+4 S_1(k)+8 S_1(2 k-2)) S_1(2 k+N-1)+2 S_2(k)
\nonumber \\ &&  \phantom{aa}
                             +\frac{3}{2} S_2(k+N-2)-6 S_2(2 k+N-1)\biggr)
          +8 S_3(2 k+N-1)
          -2 S_3(k)
\nonumber \\ &&  \phantom{aa}
          +(2 S_1(k-1)-4 S_1(k)-8 S_1(2 k-2)) S_2(2 k+N-1)
          -S_3(k+N-2)
\nonumber \\ &&  \phantom{aa}
          +S_1(2 k+N-1) \biggl(-S_1(k-1)^2-4 S_1(k) S_1(k-1)+4 S_1(2 k-2)^2
\nonumber \\ &&  \phantom{aa}
          +8 S_1(k) S_1(2 k-2)-S_2(k-1)-4 S_2(k)+12 S_2(2 k+N-1)
\nonumber \\ &&  \phantom{aa}
          -3 S_2(k+N-2)-4 S_2(2 k-2)\biggr)
          -S_3(k-1)
          -4 \zeta_3
   \biggr]
\biggr\}.
\end{eqnarray}

The above sums can now be performed using the Mathematica package {\tt Sigma}. The final result is
\begin{eqnarray}
M_2(N) &=& \frac{1}{(N-1) N} \biggl\{
-\frac{4}{3 \varepsilon^2}
+\frac{2}{3 \varepsilon} \biggl[\frac{2 (N^2-N-1)}{N (N-1)}-S_1\biggr]
-\frac{13 N-18}{6 N} S_2
\nonumber \\ &&
+\frac{2 \big(N^3-N^2-4 N+3\big)}{3 (N-1) N^2} S_1
-\frac{2 \big(2 N^5-4 N^4-2 N^3+11 N^2-8 N+3\big)}{3 (N-1)^2 N^3}
\nonumber \\ &&
-\frac{N-6}{6 N} S_1^2
-\frac{\zeta_2}{2}
+\varepsilon \biggl[
\frac{4}{N^2} S_{1,1}
+\frac{2 (N-3)}{N} S_{2,1}
+\frac{\big(N^2-N-1\big)}{2 (N-1) N} \zeta_2
\nonumber \\ &&
+\frac{2^{-N+2}}{N} \big(S_3(2)-S_{1,2}(2,1)+S_{2,1}(2,1)-S_{1,1,1}(2,1,1)-7 \zeta_3\big)
\nonumber \\ &&
+\frac{N^3-7 N^2+14 N-3}{6 (N-1) N^2} S_1^2
+\frac{13 N^3-31 N^2+38 N-15}{6 (N-1) N^2} S_2
\nonumber \\ &&
-S_1 \biggl(\frac{2 \big(N^5-2 N^4-N^3+10 N^2-10 N+3\big)}{3 (N-1)^2 N^3}
+\frac{(13 N-36)}{12 N} S_2
+\frac{\zeta_2}{4}\biggr)
\nonumber \\ &&
+\frac{4 N^7-12 N^6+4 N^5+28 N^4-59 N^3+59 N^2-37 N+9}{3 (N-1)^3 N^4}
-\frac{1}{36} S_1^3
\nonumber \\ &&
-\frac{(19 N-18)}{18 N} S_3
-\frac{(29 N-84)}{6 N} \zeta_3
\biggr]
\biggr\}.
\label{M2N}
\end{eqnarray}
The generalized harmonic sums \cite{Moch:2001zr,Ablinger:2013cf}
are defined by
\begin{eqnarray}
S_{b,\vec{a}}\big(c, \vec{d} \hspace*{0.7mm}\big)(N) = \sum_{k=1}^N \frac{c^k}{k^b} S_{\vec{a}}\big(\vec{d} 
\hspace*{0.7mm}\big)(k),~~
b, a_i \in \mathbb{N} \backslash \{0\},~~c, d_i \in \mathbb{Z}  \backslash \{0\},~S_\emptyset = 1~,
\end{eqnarray}
and we use the shorthand notation $S_{\vec{a}}(\vec{b},N) \equiv S_{\vec{a}}(\vec{b})$.
Here we have again omitted an overall factor of $i S_{\varepsilon}^3$ defined in Eq.~(\ref{Sep}).

The expression in Eq.~(\ref{M2N}) is divergent for $N=1$, so this value has to be computed separately. We obtain 
\begin{eqnarray} M_2(1) &=& \frac{4}{3 \varepsilon^3}-\frac{2}{\varepsilon^2}+\frac{3 \zeta_2+14}{6 \varepsilon} -\frac{3}{4} 
\zeta_2+\frac{29}{6} \zeta_3-\frac{5}{2} \nonumber \\ && +\varepsilon \left(-B_4+\frac{691}{160} \zeta_2^2+\frac{7}{8} 
\zeta_2-\frac{29}{4} \zeta_3+\frac{31}{12}\right). \label{M20} \end{eqnarray} 
The constant $B_4$ is given by 
\begin{eqnarray} B_4 = - 4 \zeta_2 \ln^2(2) + \frac{2}{3} \ln^4(2) - \frac{13}{2} \zeta_4 + 16 \Li_4\left(\frac{1}{2}\right) 
\equiv 8\left[ \sigma_{-1,-3} - \sigma_{-1,3}\right] - 12 \zeta_4, 
\end{eqnarray} 
with 
$\sigma_{\vec{a}} = \lim_{N \rightarrow \infty} S_{\vec{a}}(N)$, and belongs to the multiple zeta values \cite{Blumlein:2009cf}. 
Here $\Li_n(x)$ denotes the polylogarithm \cite{LEWIN}. 
\subsubsection{Differential equations}

\vspace*{1mm}
\noindent
In the $x$-representation of the integrals, we have the possibility to take derivatives of the integrals with respect to $x$.
If we do this to a master integral $J(x,\varepsilon)$, the result can then be rewritten using integration by parts (IBP) reductions in 
terms 
of the master integrals $J_i(x,\varepsilon)$ themselves.

\begin{equation}
\frac{d}{dx} J(x, \ep) = \sum_i \frac{p_i(x, \ep)}{q_i(x, \ep)} J_i(x, \ep),
\end{equation}
where $p_i(x,\varepsilon)$ and $q_i(x,\varepsilon)$ are polynomials in $x$ and $\ep$. Here and in the following we set $m^2 = \Delta.p = 
1$. The $\ep$-dependence 
will only 
be made explicit when needed. Integrals in a given sector (i.e., integrals for which the set of indices $\nu_i$ that are positive is the 
same) will produce a system of coupled differential equations, 
which we can solve after an expansion in $\varepsilon$. In Table~\ref{masterslistDE}, we show the list of integrals solved using this 
method. They all belong to the 
integral family {\sf B1a}. We have included a few horizontal lines separating the different sectors.

\begin{table}
\begin{center}
\begin{tabular}{|cccccccccccc|c|}
\hline
$\nu_1$ & $\nu_2$ & $\nu_3$ & $\nu_4$ & $\nu_5$ & $\nu_6$ &  $\nu_7$ & $\nu_8$ & $\nu_9$ & $\nu_{10}$ & $\nu_{11}$ & $\nu_{12}$     & order in $\varepsilon$  \\
\hline
    0    &    1    &    0    &    1    &    1    &    1    &     1    &    0    &    0    &     1      &     1      &     0          &         1            \\ 
    0    &    2    &    0    &    1    &    1    &    1    &     1    &    0    &    0    &     1      &     1      &     0          &         2            \\    
\hline
    0    &    0    &    0    &    2    &    1    &    1    &     1    &    1    &    0    &     1      &     1      &     0          &         1            \\
\hline 
    0    &    1    &    0    &    1    &    1    &    1    &     1    &    0    &    0    &     1      &     0      &     1          &         1            \\ 
    0    &    1    &    0    &    1    &    2    &    1    &     1    &    0    &    0    &     1      &     0      &     1          &         1            \\ 
    0    &    2    &    0    &    1    &    1    &    1    &     1    &    0    &    0    &     1      &     0      &     1          &         1            \\ 
\hline
    1    &    1    &    0    &    1    &    1    &    0    &     1    &    1    &    0    &     1      &     1      &     0          &         0            \\ 
\hline
    1    &    1    &    0    &    1    &    1    &    1    &     1    &    0    &    0    &     1      &     0      &     1          &         1            \\
    2    &    1    &    0    &    1    &    1    &    1    &     1    &    0    &    0    &     1      &     0      &     1          &         2            \\
\hline
    0    &    1    &    1    &    1    &    1    &    1    &     0    &    1    &    0    &     1      &     0      &     1          &         1            \\ 
    0    &    2    &    1    &    1    &    1    &    1    &     0    &    1    &    0    &     1      &     0      &     1          &         2            \\
\hline
\end{tabular}
\end{center}
\caption{\sf \small List of master integrals in family {\sf B1a} solved using the differential equations method.
In the last column, we indicate the order in $\varepsilon$ to which each integral needs to be expanded.}
\label{masterslistDE}
\end{table}

Let us consider the first two integrals in Table~\ref{masterslistDE}, and use the following shorthand notation,
\begin{eqnarray}
J_1(x) &=& J^{\sf B1a}_{0,1,0,1,1,1,1,0,0,1,1,0}(x)\, , \\
J_2(x) &=& J^{\sf B1a}_{0,2,0,1,1,1,1,0,0,1,1,0}(x)\, .
\end{eqnarray}
Taking derivatives with respect to $x$ we obtain
\begin{eqnarray}
\frac{d}{dx} J_1(x) &=& \frac{1}{1-x} \left(2+\varepsilon -\frac{1}{x} \right) J_1(x)+\frac{2 x}{1-x} J_2(x) + \frac{T_1(x)}{1-x}, \label{diffeq1} \\
\frac{d}{dx} J_2(x) &=& -\frac{1}{1-x} \left(\frac{1-2\varepsilon}{x}+\frac{3}{2}\varepsilon-2\right) J_2(x) \nonumber \\ &&
                         +\frac{\varepsilon}{4}(2+3\varepsilon) \frac{1}{1-x} \left( \frac{1}{x^2} - \frac{1}{x} \right) J_1(x) + \frac{T_2(x)}{1-x}, \label{diffeq2}
\end{eqnarray}
where $T_1(x)$ and $T_2(x)$ are linear combinations of sub-sector master integrals that have been solved previously. 

$T_1(x)$ and $T_2(x)$ can be turned into the $N$-representation using {\tt Sigma}. Then using the fact that
\begin{eqnarray}
J_1(x) &=& \sum_{N=0}^{\infty} x^N F_1(N) \, , \\
J_2(x) &=& \sum_{N=0}^{\infty} x^N F_2(N) \, ,
\end{eqnarray}
where
\begin{eqnarray}
F_1(N) &=& K_3(0,1,0,1,1,1,1,0,0,1,1,0;0,0,0;N+1) , \\
F_2(N) &=& K_3(0,2,0,1,1,1,1,0,0,1,1,0;0,0,0;N+1),
\end{eqnarray}
we get the following system of coupled difference equations:
\begin{eqnarray}
(N+2) F_1(N+1) - (N+2+\varepsilon) F_1(N) -2 F_2(N-1) &=& \hat{T}_1(N) \, , \\
(N+2-2\varepsilon) F_2(N+1) -\left( N+2-\frac{3}{2}\varepsilon \right) F_2(N) && \nonumber \\ 
   -\frac{\varepsilon}{4}(2+3\varepsilon) \left( F_1(N+2)-F_1(N+1) \right) &=& \hat{T}_2(N) \, ,
\end{eqnarray}
where $\hat{T}_1(N)$ and $\hat{T}_2(N)$ are the $N$th terms of the Taylor expansions of $T_1(x)$ and $T_2(x)$, respectively.
The system can be solved using the Mathematica packages {\tt Sigma} \cite{SIG1,SIG2} and 
{\tt OreSys} \cite{ORESYS}. In order to do so, we need to provide a few initial values. This can be done along the lines of 
Ref.~\cite{Ablinger:2014uka}. We obtain
\begin{eqnarray}
F_1(1) &=& -\frac{1}{3 \varepsilon^3}
-\frac{7}{12 \varepsilon^2}
+\frac{1}{\varepsilon} \left(\frac{139}{48}-\frac{\zeta_2}{8}\right)
-\frac{7}{32} \zeta_2+\frac{79}{24} \zeta_3-\frac{1243}{192}
\nonumber \\ &&
+\varepsilon \left(\frac{1901}{640} \zeta_2^2+\frac{139}{128} \zeta_2-\frac{791}{96} \zeta_3+\frac{8651}{768}\right) \\
F_1(2) &=& \frac{49}{27 \varepsilon^3}
-\frac{445}{108 \varepsilon^2}
+\frac{1}{\varepsilon} \left(\frac{49}{72} \zeta_2+\frac{38353}{3888}\right)
-\frac{445}{288} \zeta_2+\frac{689}{216} \zeta_3-\frac{85435}{5184} 
\nonumber \\ && 
+\varepsilon \left(\frac{16291}{5760} \zeta_2^2+\frac{38353}{10368} \zeta_2-\frac{6461}{864} \zeta_3+\frac{15227977}{559872}\right).
\end{eqnarray}

The results are given in terms of standard harmonic sums. We give the results up to $\varepsilon^0$, although $F_1(N)$
and $F_2(N)$ are needed and were calculated up to orders $\varepsilon$ and $\varepsilon^2$, respectively. We obtain
\begin{eqnarray}
F_1(N) &=& \frac{4}{3 \varepsilon^3} \biggl[-S_2-2 S_{-2}+\frac{N}{(N+1)^2}+\frac{2 (-1)^N (N+2)}{(N+1)^2}\biggr]
+\frac{2}{3 \varepsilon^2} \biggl[
2 S_{2,1}+8 S_{-2,1}
\nonumber \\ &&
-\frac{5 N^2+10 N+4}{(N+1)^3}
-\frac{2 (-1)^N \big(5 N^2+12 N+9\big)}{(N+1)^3}
+S_{-2} \biggl(\frac{2 (3 N+1)}{N+1}-4 S_1\biggr)
\nonumber \\ &&
+S_1 \biggl(\frac{2 (-1)^N (N-1)}{(N+1)^2}-2 S_2\biggr)
+\frac{(3 N+1) S_2}{N+1}+S_3
\biggr]
+\frac{1}{\varepsilon} \biggl[
\frac{10}{3} S_{3,1}
-2 S_{-3,1}
\nonumber \\ &&
+S_1 \biggl(\frac{4}{3} S_{2,1}+\frac{4}{3} S_{-2,1}+\frac{2 (3 N+1)}{3 (N+1)} S_2-\frac{10}{3} S_3\biggr)
-\frac{2 (3 N+1)}{3 (N+1)} S_{2,1}
-\frac{4}{3} S_{-2,2}
\nonumber \\ &&
-\frac{4 (3 N+2)}{3 (N+1)} S_{-2,1}
-\frac{4}{3} S_{2,1,1}
+\frac{4}{3} S_{-2,1,1}
-\frac{9 N^2+10 N+7}{3 (N+1)^2} S_2
-\frac{2}{3} S_2 S_1^2
\nonumber \\ &&
+(-1)^N \biggl(
\frac{2 \big(19 N^3+61 N^2+68 N+30\big)}{3 (N+1)^4}
+\frac{N-1}{3 (N+1)^2} S_1^2
-\frac{2 N (5 N+9)}{3 (N+1)^3} S_1
\nonumber \\ &&
+\frac{7 N+5}{3 (N+1)^2} S_2
\biggr)
+\frac{19 N^3+63 N^2+69 N+24}{3 (N+1)^4}
+\zeta_2 \biggl(\frac{(-1)^N (N+2)}{(N+1)^2}-\frac{S_2}{2}
\nonumber \\ &&
-S_{-2}+\frac{N}{2 (N+1)^2}\biggr)
-\frac{1}{3} S_2^2
+S_{-3} \biggl(\frac{2 N}{N+1}-2 S_1\biggr)
+S_{-2} \biggl(-\frac{4}{3} S_1^2+\frac{4 S_2}{3}
\nonumber \\ &&
+\frac{4 (3 N+1)}{3 (N+1)} S_1-\frac{2 (9 N+7)}{3 (N+1)}\biggr)
+\frac{(9 N-1)}{3 (N+1)} S_3
+\frac{2}{3} S_4
\biggr]
+\biggl(\frac{3 N+1}{3 (N+1)} S_2
\nonumber \\ &&
-\frac{8}{3} S_3
+\frac{2}{3} S_{2,1}
-\frac{1}{3} S_{-2,1}
\biggr) S_1^2
+\biggl(-\frac{1}{3} S_2^2
-\frac{9 N+7}{3 (N+1)} S_2
+\frac{21 N+5}{3 (N+1)} S_3
\nonumber \\ &&
-\frac{22}{3} S_4
-\frac{2 (3 N+1)}{3 (N+1)} S_{2,1}
+\frac{16}{3} S_{3,1}
-\frac{2}{3 (N+1)} S_{-2,1}
-\frac{4}{3} S_{-2,2}
-S_{-3,1}
\nonumber \\ &&
-\frac{4}{3} S_{2,1,1}
+\frac{10}{3} S_{-2,1,1}
\biggr) S_1
-\frac{2}{9} S_2 S_1^3
+\frac{3 N+1}{6 (N+1)} S_2^2
+S_{-4} \biggl(\frac{3 N}{N+1}-3 S_1\biggr)
\nonumber \\ &&
-\frac{65 N^4+290 N^3+486 N^2+362 N+100}{6 (N+1)^5}
+S_{-3} \biggl(-\frac{3}{2} S_1^2
+\frac{4 N+1}{N+1} S_1
\nonumber \\ &&
-\frac{5 N^2+9 N+3}{(N+1)^2}
-\frac{9}{2} S_2
\biggr)
-\frac{51 N^2+74 N+41}{6 (N+1)^2} S_3
+\frac{21 N-1}{3 (N+1)} S_4
+\frac{19}{6} S_5
\nonumber \\ &&
-7 S_{-5}
+S_{-2} \biggl[-\frac{4}{9} S_1^3
+\frac{2 (3 N+1)}{3 (N+1)} S_1^2
-\biggl(\frac{2 (9 N+7)}{3 (N+1)}+\frac{8}{3} S_2\biggr) S_1
-\frac{8}{9} S_3
\nonumber \\ &&
+\frac{27 N+29}{3 (N+1)}
+\frac{2 (3 N-1)}{3 (N+1)} S_2
-\frac{2}{3} S_{2,1}
\biggr]
+\frac{9 N+7}{3 (N+1)} S_{2,1}
-\frac{21 N+5}{3 (N+1)} S_{3,1}
\nonumber \\ &&
+(-1)^N \biggl[\frac{N-1}{18 (N+1)^2} S_1^3
-\frac{N (5 N+9)}{6 (N+1)^3} S_1^2
+\biggl(\frac{19 N^3+64 N^2+68 N+15}{3 (N+1)^4}
\nonumber \\ &&
+\frac{7 (N-1)}{6 (N+1)^2} S_2
\biggr) S_1
-\frac{35 N^2+75 N+36}{6 (N+1)^3} S_2
+\frac{19 N-1}{9 (N+1)^2} S_3
+\frac{N-1}{(N+1)^2} S_{2,1}
\nonumber \\ &&
-\frac{65 N^4+268 N^3+417 N^2+295 N+89}{3 (N+1)^5}
\biggr]
+\frac{2 (3 N+1)}{3 (N+1)} S_{-2,2}
+\frac{25}{3} S_{4,1}
\nonumber \\ &&
-4 S_{2,3}
+\frac{7}{3} S_{2,-3}
+\frac{2 \big(3 N^2+5 N+5\big)}{3 (N+1)^2} S_{-2,1}
+S_2 \biggl(\frac{29}{9} S_3
+\frac{4}{3} S_{2,1}
+7 S_{-2,1}
\nonumber \\ &&
+\frac{27 N^3+53 N^2+25 N+5}{6 (N+1)^3}
\biggr)
+\frac{14}{3} S_{-2,3}
+3 S_{-4,1}
+\frac{14}{3} S_{2,1,-2}
+\frac{4}{3} S_{2,1,1,1}
\nonumber \\ &&
-2 S_{2,2,1}
-\frac{16}{3} S_{3,1,1}
+\frac{2 (3 N+1)}{3 (N+1)} S_{2,1,1}
+\frac{2 N+1}{N+1} S_{-3,1}
-\frac{2 (6 N+1)}{3 (N+1)} S_{-2,1,1}
\nonumber \\ &&
+\frac{8}{3} S_{-2,2,1}
+3 S_{-3,1,1}
-\frac{14}{3} S_{-2,1,1,1}
+\biggl[-\frac{5 N^2+10 N+4}{4 (N+1)^3}
+\frac{3 N+1}{4 (N+1)} S_2
\nonumber \\ &&
+S_{-2} \biggl(\frac{3 N+1}{2 (N+1)}-S_1\biggr)
+(-1)^N \biggl(-\frac{5 N^2+12 N+9}{2 (N+1)^3}
+\frac{N-1}{2 (N+1)^2} S_1
\biggr)
\nonumber \\ &&
-\frac{1}{2} S_1 S_2
+\frac{S_3}{4}
+\frac{1}{2} S_{2,1}
+2 S_{-2,1}
\biggr] \zeta_2
+\biggl(\frac{7}{6} S_2
-\frac{5}{3} S_{-2}
+\frac{(-1)^N (4-N)}{3 (N+1)^2}
\nonumber \\ &&
+\frac{5 N+12}{6 (N+1)^2}
\biggr) \zeta_3
\end{eqnarray}
\begin{eqnarray}
F_2(N) &=& \frac{1}{N+2} \biggl\{
-\frac{8}{3 \varepsilon^3} \bigl[1+(-1)^N\bigr]
+\frac{4}{3 \varepsilon^2} \biggl[\frac{(-1)^N \big(2 N^2+2 N-1\big)}{(N+1) (N+2)}-(-1)^N S_1
\nonumber \\ &&
+\frac{2 N+1}{N+2}\biggr]
+\frac{1}{\varepsilon} \biggl[
(-1)^N \biggl(-\frac{4 \big(2 N^4+8 N^3+15 N^2+19 N+13\big)}{3 (N+1)^2 (N+2)^2}-\zeta_2\biggr)
\nonumber \\ &&
-\frac{2 \big(4 N^4+18 N^3+33 N^2+26 N+6\big)}{3 (N+1)^2 (N+2)^2}
-\frac{(-1)^N}{3} S_1^2
+\frac{2 (-1)^N (2 N+1)}{3 (N+1)} S_1
\nonumber \\ &&
+\frac{1}{3} \big(2-7 (-1)^N\big) S_2
+\frac{4}{3} S_{-2}
-\zeta_2
\biggr]
-\biggl((-1)^N+\frac{2}{3}\biggr) S_{2,1}
-\frac{2 (4 N+7)}{3 (N+1)} S_{-2,1}
\nonumber \\ &&
+S_1 \biggl[
-2 (N+2) \big(S_{-2,1}+\zeta_3\big)
-(-1)^N \biggl(\frac{4 N^3+10 N^2+20 N+17}{3 (N+1)^3}
+\frac{\zeta_2}{2}
\biggr)
\nonumber \\ &&
+\biggl(\frac{2}{3}-\frac{7}{6} (-1)^N\biggr) S_2
\biggr]
+2 (N+2) \big(S_{-3,1}+2 S_{-2,1,1}\big)
+S_{-2} \biggl(
-2 (N+2) S_2
\nonumber \\ &&
-\frac{2 \big(2 N^2+5 N+6\big)}{3 (N+1)^2}
+\frac{4}{3} S_1
\biggr)
-S_{-3} \biggl((N+2) S_1+\frac{1}{N+1}\biggr)
-\frac{(-1)^N}{18} S_1^3
\nonumber \\ &&
+\biggl(\frac{(-1)^N \big(14 N^2+23 N+2\big)}{6 (N+1) (N+2)}
-\frac{2 N^3+9 N^2+4 N-6}{3 (N+1)^2 (N+2)}
\biggr) S_2
+\frac{2 N+1}{2 (N+2)} \zeta_2
\nonumber \\ &&
+\frac{(-1)^N (2 N+1)}{6 (N+1)} S_1^2
-\frac{5 N+11}{3 (N+1)} \zeta_3
+(-1)^N \biggl(\frac{2 N^2+2 N-1}{2 (N+1) (N+2)} \zeta_2
+\frac{\zeta_3}{3}
\nonumber \\ &&
+\frac{8 N^7+64 N^6+228 N^5+453 N^4+486 N^3+162 N^2-164 N-124}{3 (N+1)^4 (N+2)^3}
\biggr)
\nonumber \\ &&
+\frac{8 N^7+68 N^6+250 N^5+508 N^4+639 N^3+560 N^2+360 N+124}{3 (N+1)^4 (N+2)^3}
\nonumber \\ &&
+\biggl(\frac{11}{3}-\frac{19}{9} (-1)^N\biggr) S_3
+\frac{1}{2} (N+2) \big(S_2^2+S_4\big)
\biggr\}~.
\end{eqnarray}
Here again we have omitted an overall factor of $i S_{\varepsilon}^3$ in both expressions.
Further details on our differential equation method are outlined in~\cite{BFS:14}.
\section{The Pure Singlet Anomalous Dimension}
\label{sec:4}

\vspace*{1mm}
\noindent
The pure singlet anomalous dimension at 3-loop order can be calculated in complete form from the term $\propto \ln(m^2/\mu^2)$ of the 
renormalized OME
Eq.~(\ref{AQq3PSMSren}) since $\gamma_{qq}^{(2), \rm PS} \propto T_F N_F$. Likewise, the two-loop anomalous dimension is 
obtained 
from the term $\propto \ln^2(m^2/\mu^2)$. We define
\begin{eqnarray}
F = \frac{(2+N+N^2)^2}{(N-1)N^2(N+1)^2(N+2)}
\end{eqnarray}
as shorthand notation. In the present calculation we obtained for the 2- and 3-loop anomalous dimensions
\begin{eqnarray}
\label{eq:gPS2L}
\lefteqn{\gamma_{qq}^{(1), \rm PS}(N) = - \tfrac{1}{2}[1 + (-1)^N]
\textcolor{blue}{C_F T_F N_F} \frac{16 
\big(N^2+5 N+2\big) \big(5 N^3+7 N^2+4 N+4\big)}{(N-1) N^3 (N+1)^3 (N+2)^2}
}
\\
\label{eq:gPS3L}
\lefteqn{\gamma_{qq}^{(2), \rm PS}(N) = \tfrac{1}{2}[1 + (-1)^N]} \nonumber\\ &&
\times \Biggl\{\textcolor{blue}{C_F^2 T_F N_F} 
\Biggl\{
-\frac{8 \big(N^2+N+2\big) Q_1}
      {(N-1) N^3 (N+1)^3 (N+2)} S_1^2
+\frac{32 Q_8}
      {(N-1) N^4 (N+1)^4 (N+2)^3} S_1
\nonumber\\ &&
-\frac{8 Q_{12}}{(N-1) N^5 (N+1)^5 (N+2)^3}
-\frac{8 Q_6}{(N-1) N^3 (N+1)^3 (N+2)^2} S_2
\nonumber\\ &&
+ 32 F
\Biggl[
\frac{1}{3} S_1^3
-S_2 S_1
-\frac{7}{3} S_3
+2 S_{2,1}
+ 6\zeta_3
\Biggr]
\Biggr\}
\nonumber\\ &&
+
\textcolor{blue}{C_F T_F^2 N_F^2}
\Biggl\{
-\frac{64 Q_9}{27 (N-1) N^4 (N+1)^4 (N+2)^3}
+\frac{64 Q_7}{9 (N-1) N^3 (N+1)^3 (N+2)^2} S_1
\nonumber\\ &&
- F \frac{32}{3} [S_1^2 + S_2]
\Biggr\}
+
\textcolor{blue}{C_F C_A T_F N_F}
\Biggl\{
 \frac{8 \big(N^2+N+2\big) Q_3}{3 (N-1)^2 N^3 (N+1)^3 (N+2)^2} S_1^2
\nonumber\\ &&
-\frac{16 Q_{11}}{9 (N-1)^2 N^4 (N+1)^4 (N+2)^2} S_1
+\frac{16 Q_{13}}{27 (N-1)^2 N^5 (N+1)^5 (N+2)^4}
\nonumber\\ &&
+(-1)^N 
\Biggl[
\frac{128 Q_2}{3 (N-1) N^2 (N+1)^3 (N+2)^3} S_1
-\frac{128 Q_{10}}{9 (N-1) N^3 (N+1)^5 (N+2)^4}
\Biggr]
\nonumber\\ &&
+\frac{8 \big(N^2+N+2\big) Q_4}{3 (N-1)^2 N^3 (N+1)^3 (N+2)^2} S_2
+\frac{16 \big(N^2+N+2\big) \big(23 N^2+23 N+58\big)}{3 (N-1) N^2 (N+1)^2 (N+2)} S_3
\nonumber\\ &&
+\frac{32 Q_5}{(N-1) N^3 (N+1)^3 (N+2)^2} S_{-2}
+\frac{32 \big(N^2+N+2\big) \big(7 N^2+7 N+10\big)}{(N-1) N^2 (N+1)^2 (N+2)} S_{-3}
\nonumber\\ &&
-\frac{64 \big(N^2+N+2\big) \big(3 N^2+3 N+2\big)}{(N-1) N^2 (N+1)^2 (N+2)} S_{-2,1}
\nonumber\\ &&
+ 32 F
\Biggl[
-\frac{1}{3} S_1^3
+ S_2 S_1
+2 S_{-2} S_1
-2 S_{2,1}
-6\zeta_3
\Biggr]
\Biggr\} 
\Biggr\},
\end{eqnarray}
with the polynomials
\begin{eqnarray}
Q_1 &=& 5 N^4+10 N^3+25 N^2+20 N+4
\\
Q_2 &=& 5 N^6+29 N^5+78 N^4+118 N^3+114 N^2+72 N+16
\\
Q_3 &=& 17 N^6+51 N^5+99 N^4+113 N^3-32 N^2-80 N-24
\\
Q_4 &=& 29 N^6+99 N^5+39 N^4+65 N^3+64 N^2-128 N-24
\\
Q_5 &=& 2 N^7+14 N^6+37 N^5+102 N^4+155 N^3+158 N^2+132 N+40
\\
Q_6 &=& 5 N^7+25 N^6+11 N^5-213 N^4-420 N^3-416 N^2-352 N-112
\\
Q_7 &=& 8 N^7+37 N^6+68 N^5-11 N^4-86 N^3-56 N^2-104 N-48
\\
Q_8 &=& 9 N^{10}+69 N^9+219 N^8+345 N^7+410 N^6+724 N^5+1124 N^4+1116 N^3+824 N^2
\NN\\ &&
+400 N+96
\\
Q_9 &=& 52 N^{10}+392 N^9+1200 N^8+1353 N^7-317 N^6-1689 N^5-2103 N^4-2672 N^3
\NN\\ &&
-1496 N^2 -48 N+144
\\
Q_{10} &=& 77 N^{10}+646 N^9+2553 N^8+6903 N^7+14498 N^6+22898 N^5+24861 N^4
\NN\\ &&
+17068 N^3 +7040 N^2+1760 N+192
\\
Q_{11} &=& 127 N^{10}+713 N^9+1458 N^8+78 N^7-2360 N^6-2352 N^5-3663 N^4-3359 N^3
\nonumber\\ &&
+298 N^2 +924 N+72
\\
Q_{12} &=& 49 N^{12}+417 N^{11}+1619 N^{10}+3868 N^9+6831 N^8+10189 N^7+13445 N^6
+14934 N^5
\NN\\ &&
+12760 N^4+8160 N^3+4176 N^2+1504 N+256
\\
Q_{13} &=& 731 N^{14}+8804 N^{13}+40614 N^{12}+90274 N^{11}+102402 N^{10}+67882 N^9
+23170 N^8
\NN\\ &&
-120782 N^7-357069 N^6-421954 N^5-293880 N^4-183088 N^3-109968 N^2
\NN\\ &&
-42912 N-6912~.
\end{eqnarray}
Here the color factors are given by $C_F = (N_c^2 - 1)/(2 N_c), C_A = N_c, T_F = 1/2$ for 
$SU(N_c)$ and $N_c = 3$ in case of QCD.

The three-loop pure singlet anomalous dimension depends on the following harmonic 
sums only
\begin{eqnarray}
S_{1},
S_{2},
S_{-2},
S_{3},
S_{-3},
S_{2,1},
S_{-2,1},
\label{eq:HS1}
\end{eqnarray}
if one reduces the final result algebraically \cite{Blumlein:2003gb}, cf. also 
\cite{Blumlein:2009tj}. These sums, furthermore, obey structural relations \cite{Blumlein:2009ta,Blumlein:2009fz},
leading to a further reduction to 
\begin{eqnarray}
S_{1}, S_{2,1}, S_{-2,1}~.
\end{eqnarray}
In $x$-space the pure singlet anomalous dimensions read~:
\begin{eqnarray}
\label{eq:PPS2L}
\lefteqn{\gamma_{qq}^{(1), \rm PS}(x) =} \nonumber\\ &&
\textcolor{blue}{C_F T_F N_F} 
\Biggl\{
16 (x+1) H_{0}^2
-\frac{16}{3} \big(
8 x^2+15 x+3\big) H_0 +\frac{32 (x-1) \big(28 x^2+x+10\big)}
{9 x}
\Biggr\}
\\
\label{eq:PPS3L}
\lefteqn{\gamma_{qq}^{(2), \rm PS}(x) =} \nonumber\\ &&
\textcolor{blue}{C_F^2 T_F N_F}
\Biggl\{
\frac{32}{3} (x+1) H_0^4
-\frac{16}{9} \big(44 x^2+15 x+9\big) H_0^3
+\Biggl[
\frac{8}{3} \big(88 x^2-53 x+128\big)
-192 (x+1) \zeta_2
\Biggr] 
\nonumber\\ && 
\times H_0^2
+\Biggl[
\frac{8}{27} \big(52 x^2-8229 x-2265\big)
+\frac{64}{3} \big(16 x^2+9 x-6\big) \zeta_2
\Biggr]
H_0
-\frac{32 (x-1) \big(4 x^2+7 x+4\big)}
      {9 x} H_1^3
\nonumber\\ &&
+\Biggl[
 \frac{8  (x-1) \big(16 x^2+23 x+16\big)}{3 x}
-\frac{32 (x-1) \big(4 x^2+7 x+4\big)}{3 x} H_0
\Biggr]
H_1^2
-64 (x+1) H_{0,1}^2
\nonumber\\ &&
+\frac{8 (x-1) \big(3924 x^2+2255 x+990\big)}{27 x}
+592 (x+1) \zeta_4
+\Biggl[
 \frac{16 (x-1) \big(4 x^2+7 x+4\big)}{3 x} H_0^2
\nonumber\\ &&
+\frac{64 (x-1) \big(50 x^2-x+23\big)}{9 x} H_0
+\frac{16 (x-1) \big(242 x^2-2017 x-523\big)}{27 x}
\Biggr]
H_1
+\Biggl[
-32 (x+1) H_0^2
\nonumber\\ &&
-\frac{32 \big(28 x^3+45 x^2+9 x-4\big)}{3 x} H_0
-\frac{16 \big(88 x^3-249 x^2-516 x-92\big)}{9 x}
\nonumber\\ &&
+\frac{64 (x-1) \big(4 x^2+7 x+4\big)}{3 x} H_1
\Biggr] H_{0,1}
+\Biggl[
\frac{32 \big(20 x^3+69 x^2+33 x-4\big)}{3 x}
+384 (x+1) H_0
\Biggr]
H_{0,0,1}
\nonumber\\ &&
+\Biggl[
128 (x+1) H_0
-\frac{64 \big(8 x^3+15 x^2+6 x-4\big)}{3 x}
\Biggr]
H_{0,1,1}
-576 (x+1) H_{0,0,0,1}
\nonumber\\ &&
+64 (x+1) H_{0,0,1,1}
+128 (x+1) H_{0,1,1,1}
-\frac{16}{9} \big(112 x^2+45 x+612\big) \zeta_2
+\Biggl[
448 (x+1) H_0
\nonumber\\ &&
-\frac{128 \big(8 x^3+15 x^2-6\big)}{3 x}
\Biggr]
\zeta_3
\Biggr\}
\nonumber\\ &&
+ \textcolor{blue}{C_F T_F^2 N_F^2}
\Biggl\{
-\frac{64}{9} (x+1) H_0^3
-\frac{928}{9} (x+1) H_0^2
+\Biggl[
\frac{128}{27} \big(57 x^2+35 x-37\big)
-\frac{256}{3} (x+1) \zeta_2
\Biggr] 
H_0
\nonumber\\ &&
+\frac{32 (x-1) \big(4 x^2+7 x+4\big)}{9 x} H_1^2
-\frac{64 (x-1) \big(100 x^2-85 x-8\big)}{27 x}
+\frac{64 (x-1) \big(74 x^2-43 x+20\big)}{27 x} H_1
\nonumber\\ &&
-\frac{128}{9} \big(6 x^2+4 x-5\big) H_{0,1}
+\frac{256}{3} (x+1) H_{0,0,1}
-\frac{128}{3} (x+1) H_{0,1,1}
+\frac{128}{9} \big(6 x^2+4 x-5\big) 
\zeta_2
\nonumber\\ &&
-\frac{128}{3} (x+1) 
\zeta_3
\Biggr\}
\nonumber\\ &&
+\textcolor{blue}{C_F C_A T_F N_F}
\Biggl\{
 \frac{8}{3} (3 x-4) H_0^4
-\frac{16}{9} (31 x-29) H_0^3
+\Biggl[
-\frac{8}{9} \big(498 x^2-397 x+269\big)
\nonumber\\ &&
+\frac{16 (x+1) \big(16 x^2-19 x+16\big)}{3 x} H_{-1}
+16 (x-1) \zeta_2
\Biggr]
H_0^2
+\Biggl[
\frac{32 (x+1) \big(4 x^2-7 x+4\big)}{3 x} H_{-1}^2
\nonumber\\ &&
-\frac{128 (x+1) \big(53 x^2-2 x+26\big)}{9 x} H_{-1}
+\frac{16 \big(4598 x^3+3304 x^2+3565 x+224\big)}{27 x}
\nonumber\\ && 
+\frac{16}{3} \big(8 x^2-7 x+59\big) 
\zeta_2
\Biggr]
H_0
+\frac{32 (x-1) \big(4 x^2+7 x+4\big)}{9 x} H_1^3
+\Biggl[
\frac{32 (x-1) \big(4 x^2+7 x+4\big)}{3 x} H_0
\nonumber\\ && 
-\frac{8 (x-1) \big(4 x^2+31 x+4\big)}{9 x}\Biggr] H_1^2
-64 (x-1) H_{0,-1}^2
+64 (x+1) H_{0,1}^2
-8 (117 x+107) 
\zeta_4
\nonumber\\ && 
-\frac{16 (x-1) \big(20558 x^2+3494 x+6761\big)}{81 x}
-\frac{16\big(112 x^3+226 x^2-479 x+128\big)}{9x} \zeta_2
\nonumber\\ && 
-\frac{32 (x+1) \big(4 x^2+11 x+4\big)}{3x} \zeta_2 H_{-1}
+\Biggl[
-\frac{8 (x-1) \big(8 x^2+17 x+8\big)}{x} H_0^2
\nonumber\\ && 
+\frac{32 (x-1) \big(27 x^2+22 x+9\big)}{3 x} H_0
-\frac{16 (x-1) \big(302 x^2-1882 x-571\big)}{27 x}
\nonumber\\ && 
+\frac{32 (x-1) \big(4 x^2+7 x+4\big)}{3 x} \zeta_2
\Biggr]
H_1
+\Biggl[
\frac{64 (x+1) \big(4 x^2-7 x+4\big)}{3 x}
+128 (x-1) H_0
\Biggr] 
H_{0,-1,-1}
\nonumber\\ && 
-\frac{128 (x+1) \big(2 x^2+x+2\big)}{3 x} H_{0,-1,1}
+\Biggl[
-\frac{32 \big(32 x^3-75 x^2+21 x-16\big)}{3 x}
-64 (5 x-1) H_0
\Biggr]
\nonumber\\ && 
\times H_{0,0,-1}
+\Biggl[
-\frac{32 \big(8 x^3+19 x^2+4 x-12\big)}{3 x}
-448 (x+1) H_0
\Biggr]
H_{0,0,1}
\nonumber\\ && 
-\frac{128 (x+1) \big(2 x^2+x+2\big)}{3 x} H_{0,1,-1}
+\Biggl[
\frac{32 \big(12 x^3+23 x^2+5 x-12\big)}{3 x}
-128 (x+1) H_0
\Biggr]
H_{0,1,1}
\nonumber\\ && 
+384 (x+1) H_{0,0,0,-1}
+704 (x+1) H_{0,0,0,1}
-64 (x+1) H_{0,0,1,1}
-128 (x+1) H_{0,1,1,1}
\nonumber\\ && 
+H_{0,-1}
\Biggl[
96 (x-1) H_0^2
+\frac{128 \big(2 x^3-9 x^2+3 x-4\big)}{3 x} H_0
+\frac{128 (x+1) \big(53 x^2-2 x+26\big)}{9 x}
\nonumber\\ && 
-\frac{64 (x+1) \big(4 x^2-7 x+4\big)}{3 x} H_{-1}
+64 (x-1) \zeta_2
\Biggr]
+H_{0,1} 
\Biggl[
112 (x+1) H_0^2
\nonumber\\ && 
+\frac{16 \big(16 x^3-3 x^2-39 x-24\big)}{3 x} H_0
-\frac{16 \big(50 x^3+152 x^2+401 x+26\big)}{9 x}
\nonumber\\ && 
+\frac{128 (x+1) \big(2 x^2+x+2\big)}{3 x} H_{-1}
-\frac{64 (x-1) \big(4 x^2+7 x+4\big)}{3 x} H_1
-64 (x+1) \zeta_2
\Biggr]
\nonumber\\ && 
+
\Biggl[
\frac{32 \big(52 x^3-31 x^2+2 x-36\big)}{3 x}
-64 (9 x+13) H_0
\Biggr]
\zeta_3
\Biggr\}.
\end{eqnarray}
Here we used the shorthand notation $H_{\vec{a}}(x) \equiv H_{\vec{a}}$ for the 
harmonic polylogarithms \cite{Remiddi:1999ew}. They are defined by
\begin{eqnarray}
H_{b,\vec{a}}(x) &=& \int_0^x dy f_b(y) H_{\vec{a}}(y),~~~H_\emptyset = 1,~a_i, b~\in~\{0,1,-1\}~. 
\end{eqnarray}
The letters $f_b(x)$ are given by
\begin{eqnarray}
f_0(x) = \frac{1}{x},~~~f_1(x) = \frac{1}{1-x},~~~f_{-1}(x) = \frac{1}{1+x}~.
\end{eqnarray}
Furthermore, the $k$th iteration of the letter $0$ leads to $\ln^k(x)/k!$. Again, we have used the algebraic 
relations \cite{Blumlein:2003gb}. The pure singlet anomalous dimension up to 3-loop order depends on the following 
harmonic polylogarithms
\begin{eqnarray}
&&
H_{0},
H_{1},
H_{-1},
H_{0,1},
H_{0,-1},
H_{0,0,1},
H_{0,0,-1},
H_{0,1,1},
H_{0,-1,-1},
H_{0,-1,1},
H_{0,1,-1},
H_{0,0,0,1},
H_{0,0,0,-1},
\nonumber\\ &&
H_{0,0,1,1},
H_{0,1,1,1}
\label{eq:HPL1}
\end{eqnarray}
only. Since the functions $H_{0,-1,1}$ and $H_{0,1,-1}$ emerge as a sum, all these
harmonic polylogarithms can be represented as Nielsen integrals 
\cite{Nielsen1909,Kolbig:1969zza,Kolbig:1983qt}, with argument $x, -x$ 
and $x^2$, respectively, cf.~\cite{Blumlein:2009ta,Ablinger:2014lka}. These are the functions
\begin{eqnarray}
&& \ln(x), \ln(1-x), \ln(1+x), \Li_2(x), \Li_3(x), \Li_3(-x), S_{1,2}(x), S_{1,2}(-x),
S_{1,2}(x^2), \Li_4(x), \Li_4(-x), 
\nonumber\\ &&
S_{2,2}(x), S_{1,3}(x)~,
\end{eqnarray}
where
\begin{eqnarray}
S_{n,p}(x) &=& \frac{(-1)^{n+p-1}}{(n-1)! p!} \int_0^1 \frac{dz}{z} \ln^{n-1}(z) 
\ln^p(1-zx)~, \\
\Li_n(x)    &=& S_{n-1,1}(x)~.
\end{eqnarray}
This behaviour is observed for all 3--loop splitting functions, see Refs.~\cite{Blumlein:2004bb,Blumlein:2009ta}.
Despite of the algebraic reduction, the representations in $x$-space request more basic special functions 
than the case in Mellin-$N$ space. Eqs.~(\ref{eq:gPS2L}, \ref{eq:PPS2L}) agree with the results given in 
Refs.~\cite{NLO,Moch:1999eb} and Eqs.~(\ref{eq:gPS3L}, \ref{eq:PPS3L}) with the corresponding results given in 
\cite{Vogt:2004mw}. For the latter case we present the first independent recalculation here. 
\section{The Pure Singlet Massive Operator Matrix Element}
\label{sec:5}
In the following we derive the result for the massive OME $A_{Qq}^{\rm PS}$ to 3-loop order, Eqs.~(\ref{AQq2PSMSren},
\ref{AQq3PSMSren}). As outlined in Ref.~\cite{Bierenbaum:2009mv} all contributions apart from the constant part 
$a_{Qq}^{(3),\rm PS}$ of the unrenormalized OME \cite{Bierenbaum:2009mv} Eq.~(4.24), are given by renormalization 
and factorization \cite{Behring:2014eya}. We first consider this quantity. In $N$-space it is given by
\begin{eqnarray}
\label{eq:aQq3PS}
\lefteqn{a_{Qq}^{(3), \rm PS}(N) =}
\nonumber \\ &&
\textcolor{blue}{C_F T_F^2} \Biggl[
        \frac{32}{27 (N-1) (N+3) (N+4) (N+5)} \biggl(
                \frac{P_{15}}{N^3 (N+1)^2 (N+2)^2} S_2
\nonumber \\ &&
                -\frac{P_{19}}{N^3 (N+1)^3 (N+2)^2} S_1^2
                +\frac{2 P_{28}}{3 N^4 (N+1)^4 (N+2)^3} S_1
                -\frac{2 P_{32}}{9 N^5 (N+1)^4 (N+2)^4}
        \biggr)
\nonumber \\ &&
        -\frac{32 P_3}{9 (N-1) N^3 (N+1)^2 (N+2)^2} \zeta_2
        +\biggl(
                \frac{32}{27} S_1^3
                -\frac{160}{9} S_1 S_2
                -\frac{512}{27} S_3
                +\frac{128}{3} S_{2,1}
\nonumber \\ &&
                +\frac{32}{3} S_1 \zeta_2
                -\frac{1024}{9} \zeta_3
        \biggr) F
\Biggr]
+\textcolor{blue}{C_F N_F T_F^2} \Biggl[
        \frac{16 P_7}{27 (N-1) N^3 (N+1)^3 (N+2)^2} S_1^2
\nonumber \\ &&
        +\frac{208 P_7}{27 (N-1) N^3 (N+1)^3 (N+2)^2} S_2
        -\frac{32 P_{21}}{81 (N-1) N^4 (N+1)^4 (N+2)^3} S_1
\nonumber \\ &&
        +\frac{32 P_{29}}{243 (N-1) N^5 (N+1)^5 (N+2)^4}
        +\biggl(
                -\frac{16}{27} S_1^3
                -\frac{208}{9} S_1 S_2
                -\frac{1760}{27} S_3
                -\frac{16}{3} S_1 \zeta_2
\nonumber \\ &&
                +\frac{224}{9} \zeta_3
        \biggr) F
        +\frac{1}{(N-1) N^3 (N+1)^3 (N+2)^2} \frac{16 P_7}{9} \zeta_2
\Biggr]
\nonumber \\ &&
+\textcolor{blue}{C_F^2 T_F} \Biggl[
        \frac{32 P_9}{3 (N-1) N^3 (N+1)^3 (N+2)^2} S_{2,1}
        -\frac{16 P_{14}}{9 (N-1) N^3 (N+1)^3 (N+2)^2} S_3
\nonumber \\ &&
        -\frac{4 P_{17}}{3 (N-1) N^4 (N+1)^4 (N+2)^3} S_1^2
        +\frac{4 P_{23}}{3 (N-1) N^4 (N+1)^4 (N+2)^3} S_2
\nonumber \\ &&
        +\frac{4 P_{31}}{3 (N-1) N^6 (N+1)^6 (N+2)^4}
        +\biggl(
                \biggl(
                        \frac{2 P_5}{N^2 (N+1)^2}
                        -\frac{4 P_1}{N (N+1)} S_1
                \biggr) \zeta_2
\nonumber \\ &&
                -\frac{4 P_1}{9 N (N+1)} S_1^3
        \biggr) G
        +\biggl(
                \biggl(
                        \frac{80}{9} S_3
                        -64 S_{2,1}
                \biggr) S_1
                -\frac{2}{9} S_1^4
                -\frac{20}{3} S_1^2 S_2
                +\frac{46}{3} S_2^2
                +\frac{124}{3} S_4
\nonumber \\ &&
                +\frac{416}{3} S_{2,1,1}
                +64 \biggl(
                        \biggl(
                                S_3({{2}})
                                -S_{1,2}({{2,1}})
                                +S_{2,1}({{2,1}})
                                -S_{1,1,1}({{2,1,1}})
                        \biggr) S_1\left({{\frac{1}{2}}}\right)
\nonumber \\ &&
                        -S_{1,3}\left({{2,\frac{1}{2}}}\right)
                        +S_{2,2}\left({{2,\frac{1}{2}}}\right)
                        -S_{3,1}\left({{2,\frac{1}{2}}}\right)
                        +S_{1,1,2}\left({{2,\frac{1}{2},1}}\right)
                        -S_{1,1,2}\left({{2,1,\frac{1}{2}}}\right)
\nonumber \\ &&
                        -S_{1,2,1}\left({{2,\frac{1}{2},1}}\right)
                        +S_{1,2,1}\left({{2,1,\frac{1}{2}}}\right)
                        -S_{2,1,1}\left({{2,\frac{1}{2},1}}\right)
                        -S_{2,1,1}\left({{2,1,\frac{1}{2}}}\right)
\nonumber \\ &&
                        +S_{1,1,1,1}\left({{2,\frac{1}{2},1,1}}\right)
                        +S_{1,1,1,1}\left({{2,1,\frac{1}{2},1}}\right)
                        +S_{1,1,1,1}\left({{2,1,1,\frac{1}{2}}}\right)
                \biggr)
                +\biggl(
                        12 S_2
                        -4 S_1^2
                \biggr) \zeta_2
\nonumber \\ &&
                +\biggl(
                        \frac{112}{3} S_1
                        -448 S_1\left({{\frac{1}{2}}}\right)
                \biggr) \zeta_3
                +144 \zeta_4
                -32 B_4
        \biggr) F
        +\frac{32 P_2 2^{-N}}{(N-1) N^3 (N+1)^2} \biggl(
                -S_3({{2}})
\nonumber \\ &&
                +S_{1,2}({{2,1}})
                -S_{2,1}({{2,1}})
                +S_{1,1,1}({{2,1,1}})
                +7 \zeta_3
        \biggr)
        +\biggl(
                -\frac{4 P_8}{3 (N-1) N^3 (N+1)^3 (N+2)^2} S_2
\nonumber \\ &&
                +\frac{8 P_{27}}{3 (N-1) N^5 (N+1)^5 (N+2)^4}
        \biggr) S_1
        +\frac{1}{(N-1) N^3 (N+1)^3 (N+2)^2} \frac{4 P_{16}}{3} \zeta_3
\Biggr]
\nonumber \\ &&
+\textcolor{blue}{C_A C_F T_F} \Biggl[
        -\frac{8 P_{10}}{3 (N-1) N^3 (N+1)^3 (N+2)^2} S_{2,1}
        +\frac{8 P_{12}}{3 (N-1) N^3 (N+1)^3 (N+2)^2} S_{-3}
\nonumber \\ &&
        +\frac{16 P_{13}}{3 (N-1) N^3 (N+1)^3 (N+2)^2} S_{-2,1}
        +\frac{8 P_{22}}{27 (N-1)^2 N^3 (N+1)^3 (N+2)^2} S_3
\nonumber \\ &&
        -\frac{4 P_{24}}{27 (N-1)^2 N^4 (N+1)^4 (N+2)^3} S_1^2
        -\frac{4 P_{26}}{27 (N-1)^2 N^4 (N+1)^4 (N+2)^3} S_2
\nonumber \\ &&
        -\frac{8 P_{33}}{243 (N-1)^2 N^6 (N+1)^6 (N+2)^5}
        +\biggl(
                \frac{4 P_4}{27 (N-1) N (N+1) (N+2)} S_1^3
\nonumber \\ &&
                +\biggl(
                        \frac{8}{9} \big(137 N^2+137 N+334\big) S_3
                        -\frac{16}{3} \big(35 N^2+35 N+18\big) S_{-2,1}
                \biggr) S_1
\nonumber \\ &&
                +\frac{8}{3} \big(69 N^2+69 N+94\big) S_{-3} S_1
                +\frac{64}{3} \big(7 N^2+7 N+13\big) S_{-2} S_2
                +\frac{2}{3} \big(29 N^2+29 N+74\big) S_2^2
\nonumber \\ &&
                +\frac{4}{3} \big(143 N^2+143 N+310\big) S_4
                -\frac{16}{3} \big(3 N^2+3 N-2\big) S_{-2}^2
                +\frac{16}{3} \big(31 N^2+31 N+50\big) S_{-4}
\nonumber \\ &&
                -8 \big(7 N^2+7 N+26\big) S_{3,1}
                -64 \big(3 N^2+3 N+2\big) S_{-2,2}
                -\frac{32}{3} \big(23 N^2+23 N+22\big) S_{-3,1}
\nonumber \\ &&
                +\frac{64}{3} \big(13 N^2+13 N+2\big) S_{-2,1,1}
                +\frac{4 P_4}{3 (N-1) N (N+1) (N+2)} S_1 \zeta_2
\nonumber \\ &&
                -\frac{8}{3} \big(11 N^2+11 N+10\big) S_1 \zeta_3
        \biggr) G
        +\biggl(
                \frac{112}{3} S_{-2} S_1^2
                +\frac{2}{9} S_1^4
                +\frac{68}{3} S_1^2 S_2
                -\frac{80}{3} S_{2,1,1}
\nonumber \\ &&
                +32 \biggl(
                        \biggl(
                                -S_3({{2}})
                                +S_{1,2}({{2,1}})
                                -S_{2,1}({{2,1}})
                                +S_{1,1,1}({{2,1,1}})
                        \biggr) S_1\left({{\frac{1}{2}}}\right)
                        +S_{1,3}\left({{2,\frac{1}{2}}}\right)
\nonumber \\ &&
                        -S_{2,2}\left({{2,\frac{1}{2}}}\right)
                        +S_{3,1}\left({{2,\frac{1}{2}}}\right)
                        -S_{1,1,2}\left({{2,\frac{1}{2},1}}\right)
                        +S_{1,1,2}\left({{2,1,\frac{1}{2}}}\right)
                        +S_{1,2,1}\left({{2,\frac{1}{2},1}}\right)
\nonumber \\ &&
                        -S_{1,2,1}\left({{2,1,\frac{1}{2}}}\right)
                        +S_{2,1,1}\left({{2,\frac{1}{2},1}}\right)
                        +S_{2,1,1}\left({{2,1,\frac{1}{2}}}\right)
                        -S_{1,1,1,1}\left({{2,\frac{1}{2},1,1}}\right)
\nonumber \\ &&
                        -S_{1,1,1,1}\left({{2,1,\frac{1}{2},1}}\right)
                        -S_{1,1,1,1}\left({{2,1,1,\frac{1}{2}}}\right)
                \biggr)
                +\biggl(
                        4 S_1^2
                        +12 S_2
                        +24 S_{-2}
                \biggr) \zeta_2
\nonumber \\ &&
                +224 S_1\left({{\frac{1}{2}}}\right) \zeta_3
                -144 \zeta_4
                +16 B_4
        \biggl) F
        +\frac{16 P_2 2^{-N}}{(N-1) N^3 (N+1)^2} \biggl(
                 S_3({{2}})
                -S_{1,2}({{2,1}})
\nonumber \\ &&
                +S_{2,1}({{2,1}})
                -S_{1,1,1}({{2,1,1}})
                -7 \zeta_3
        \biggr)
        +\biggl(
                \frac{4 P_{11}}{9 (N-1)^2 N^3 (N+1)^3 (N+2)^2} S_2
\nonumber \\ &&
                +\frac{4 P_{30}}{81 (N-1)^2 N^5 (N+1)^5 (N+2)^4}
        \biggr) S_1
        +\biggl(
                \frac{32 P_6}{3 (N-1) N^3 (N+1)^3 (N+2)^2} S_1
\nonumber \\ &&
                -\frac{8 P_{18}}{3 (N-1) N^4 (N+1)^4 (N+2)^3}
        \biggr) S_{-2}
        -\frac{4 P_{25}}{9 (N-1)^2 N^4 (N+1)^4 (N+2)^3} \zeta_2
\nonumber \\ &&
        -\frac{8 P_{20}}{9 (N-1)^2 N^3 (N+1)^3 (N+2)^2} \zeta_3
\Biggr].
\end{eqnarray}

Here we defined
\begin{eqnarray}
        G &=& \frac{(N^2+N+2)}{(N-1) N^2 (N+1)^2 (N+2)} 
\end{eqnarray}
and the polynomials $P_i$ are given by
\begin{eqnarray}
P_1 &=& 5 N^4+4 N^3+N^2-10 N-8 \\
P_2 &=& N^5-N^3+10 N^2-2 N+4 \\
P_3 &=& 8 N^6+29 N^5+84 N^4+193 N^3+162 N^2+124 N+24 \\
P_4 &=& 17 N^6+51 N^5+27 N^4+77 N^3+76 N^2-80 N-24 \\
P_5 &=& 38 N^6+108 N^5+151 N^4+106 N^3+21 N^2-28 N-12 \\
P_6 &=& 3 N^7+24 N^6+49 N^5+122 N^4+154 N^3+104 N^2+120 N+32 \\
P_7 &=& 8 N^7+37 N^6+68 N^5-11 N^4-86 N^3-56 N^2-104 N-48 \\
P_8 &=& 81 N^7+271 N^6+229 N^5-159 N^4-530 N^3-844 N^2-904 N-288 \\
P_9 &=& 6 N^8+40 N^7+84 N^6+59 N^5+114 N^4+283 N^3+250 N^2+180 N+88 \\
P_{10} &=& 6 N^8+48 N^7+100 N^6-5 N^5+194 N^4+763 N^3+626 N^2+356 N+152 \\
P_{11} &=& 269 N^8+1064 N^7+1342 N^6+2552 N^5+3273 N^4+1896 N^3+516 N^2
        \nonumber \\ &&
        -2560 N-864 \\
P_{12} &=& 6 N^9+39 N^8+89 N^7+148 N^6+85 N^5+147 N^4+286 N^3+248 N^2
        \nonumber \\ &&
        +440 N+112 \\
P_{13} &=& 6 N^9+39 N^8+105 N^7+76 N^6-91 N^5-293 N^4-338 N^3-248 N^2
        \nonumber \\ &&
        -264 N-80 \\
P_{14} &=& 36 N^9+216 N^8+478 N^7+293 N^6-663 N^5-2063 N^4-2859 N^3-1074 N^2
        \nonumber \\ &&
        +444 N+56 \\
P_{15} &=& 40 N^9+625 N^8+3284 N^7+5392 N^6-7014 N^5-33693 N^4-47454 N^3
        \nonumber \\ &&
        -46100 N^2-26280 N+7200 \\
P_{16} &=& 48 N^9+192 N^8-45 N^7-1089 N^6-1487 N^5-3299 N^4-7320 N^3-4120 N^2
        \nonumber \\ &&
        -1008 N-1072 \\
P_{17} &=& 3 N^{10}+75 N^9+363 N^8+735 N^7+662 N^6+490 N^5+944 N^4+840 N^3
        \nonumber \\ &&
        +176 N^2+256 N+192 \\
P_{18} &=& 5 N^{10}+44 N^9+82 N^8+214 N^7+259 N^6+14 N^5-346 N^4-2096 N^3
        \nonumber \\ &&
        -3680 N^2-1952 N-416 \\
P_{19} &=& 8 N^{10}+133 N^9+1095 N^8+5724 N^7+18410 N^6+34749 N^5+40683 N^4
        \nonumber \\ &&
        +37370 N^3+22748 N^2-3960 N-7200 \\
P_{20} &=& 9 N^{10}-229 N^8-367 N^7+1135 N^6-472 N^5-5661 N^4-837 N^3+1098 N^2
        \nonumber \\ &&
        +260 N+1032 \\
P_{21} &=& 25 N^{10}+176 N^9+417 N^8+30 N^7-20 N^6+1848 N^5+2244 N^4+1648 N^3
        \nonumber \\ &&
        +3040 N^2+2112 N+576 \\
P_{22} &=& 135 N^{10}+702 N^9+1745 N^8+2039 N^7+1345 N^6+2618 N^5-4923 N^4
        \nonumber \\ &&
        -9939 N^3-11598 N^2-10516 N-2136 \\
P_{23} &=& 153 N^{10}+1049 N^9+2811 N^8+3411 N^7+1084 N^6-3976 N^5-11660 N^4
        \nonumber \\ &&
        -16088 N^3-12272 N^2-6240 N-1664 \\
P_{24} &=& 46 N^{11}+145 N^{10}+406 N^9+1566 N^8+1411 N^7-4318 N^6-12231 N^5
        \nonumber \\ &&
        -14165 N^4-6636 N^3+3200 N^2+4512 N+1872 \\
P_{25} &=& 127 N^{11}+856 N^{10}+2323 N^9+2484 N^8-317 N^7-106 N^6+4779 N^5
        \nonumber \\ &&
        +8470 N^4+11112 N^3+9680 N^2+4656 N+864 \\
P_{26} &=& 1696 N^{11}+10993 N^{10}+27688 N^9+26208 N^8-773 N^7+17000 N^6
        \nonumber \\ &&
        +62901 N^5+81499 N^4+114180 N^3+106112 N^2+55200 N+12240 \\
P_{27} &=& 12 N^{13}+151 N^{12}+819 N^{11}+2549 N^{10}+4893 N^9+7260 N^8+11172 N^7
        \nonumber \\ &&
        +15420 N^6+16388 N^5+16824 N^4+16352 N^3+10880 N^2+4672 N+896 \\
P_{28} &=& 52 N^{13}+746 N^{12}+4658 N^{11}+20431 N^{10}+79990 N^9+251778 N^8
        \nonumber \\ &&
        +553796 N^7+837697 N^6+886552 N^5+599060 N^4+155864 N^3-82368 N^2
        \nonumber \\ &&
        -76896 N-17280 \\
P_{29} &=& 158 N^{13}+1663 N^{12}+7714 N^{11}+23003 N^{10}+56186 N^9+89880 N^8+59452 N^7
        \nonumber \\ &&
        -8896 N^6-12856 N^5-24944 N^4-84608 N^3-77952 N^2-35712 N-6912 \\
P_{30} &=& 247 N^{14}+2518 N^{13}+12147 N^{12}+29936 N^{11}+47061 N^{10}+66314 N^9
        \nonumber \\ &&
        +15119 N^8-144034 N^7+1854 N^6+528058 N^5+571260 N^4+113008 N^3
        \nonumber \\ &&
        -61248 N^2-22752 N+1728 \\
P_{31} &=& 88 N^{15}+978 N^{14}+4569 N^{13}+11443 N^{12}+18236 N^{11}+25694 N^{10}+41400 N^9
        \nonumber \\ &&
        +57974 N^8+50675 N^7+9415 N^6-48500 N^5-88676 N^4-83504 N^3
        \nonumber \\ &&
        -45232 N^2-13504 N-1728 \\
P_{32} &=& 293 N^{15}+4670 N^{14}+32280 N^{13}+145948 N^{12}+559575 N^{11}+1871440 N^{10}
        \nonumber \\ &&
        +4877344 N^9+9333994 N^8+12958212 N^7+12693884 N^6+8472792 N^5
        \nonumber \\ &&
        +4514336 N^4+3109248 N^3+2192832 N^2+1026432 N+207360 \\
P_{33} &=& 3244 N^{17}+40465 N^{16}+218915 N^{15}+671488 N^{14}+1331937 N^{13}+1654143 N^{12}
        \nonumber \\ &&
        +374900 N^{11}-2526162 N^{10}-3045065 N^9+1320584 N^8+6186057 N^7
        \nonumber \\ &&
        +9141018 N^6+12149124 N^5+13312808 N^4+10121520 N^3+4812768 N^2
        \nonumber \\ &&
        +1308096 N+155520.
\end{eqnarray}

A new result if compared to the calculation of other massive operator matrix elements given in
Refs.~\cite{Ablinger:2014lka,Ablinger:2014vwa} is that generalized harmonic sums \cite{Moch:2001zr,Ablinger:2013cf}
contribute to the final result. 
In addition to the sums contributing to (\ref{eq:HS1}) the following sums occur
\begin{eqnarray}
&&
S_{-4},
S_{4},
S_{-3,1},
S_{3,1},
S_{-2,2},
S_{-2,1,1}, 
S_{2,1,1},
S_1\left(\frac{1}{2}\right),
S_3\left(2\right),
S_{1,3}\left(2,\frac{1}{2}\right),
S_{1,2}\left(2,1\right),
S_{2,1}\left(2,1\right),
\nonumber\\ &&
S_{2,2}\left(2,\frac{1}{2}\right),
S_{3,1}\left(2,\frac{1}{2}\right),
S_{1,1,1}\left(2,1,1\right),
S_{1,1,2}\left(2,\frac{1}{2},1\right),
S_{1,1,2}\left(2,1,\frac{1}{2}\right),
S_{1,2,1}\left(2,\frac{1}{2},1\right),
\nonumber\\ &&
S_{1,2,1}\left(2,1,\frac{1}{2}\right),
S_{2,1,1}\left(2,\frac{1}{2},1\right),
S_{2,1,1}\left(2,1,\frac{1}{2}\right),
S_{1,1,1,1}\left(2,\frac{1}{2},1,1\right),
S_{1,1,1,1}\left(2,1,\frac{1}{2},1\right),
\nonumber\\ &&
S_{1,1,1,1}\left(2,1,1,\frac{1}{2}\right)~.
\end{eqnarray}
As some of them also contribute to the scalar ladder diagrams, a series of Mellin 
inversions has been given in \cite{Ablinger:2012qm} already. Those of the remaining 
sums will be presented in Appendix~\ref{app:B}. Individual sums, as e.g. $S_3(2)$, diverge exponentially in the limit 
$N \rightarrow \infty$. However, a regular asymptotic behaviour of the combination of the corresponding sums 
for large values of $N$ is obtained for $a_{Qq}^{(3),\rm PS}$. It behaves like
\begin{eqnarray}
a_{Qq}^{(3),\rm PS} &\propto& \frac{2}{9} \textcolor{blue}{C_F T_F (C_A-C_F)} \frac{\ln^4(\bar{N})}{N^2}
+ \frac{4}{27} \textcolor{blue}{C_F T_F} \left[17 \textcolor{blue}{C_A} - 15 \textcolor{blue}{C_F} + (8 - 4 \textcolor{blue}{N_F})
\textcolor{blue}{T_F}\right] \frac{\ln^3(\bar{N})}{N^2} 
\nonumber\\ &&
+ O\left(\frac{\ln^2(\bar{N})}{N^2}\right)
\end{eqnarray}
outside the singularities. Here $\bar{N}$ is defined as $\bar{N} = N \exp(\gamma_E)$.
The Mellin inversion of Eq.~(\ref{eq:aQq3PS}) leads to generalized harmonic polylogarithms of argument $x$ \cite{Ablinger:2013cf}.
However, one may trade the index set in terms of different arguments in this special case and end up with the usual harmonic 
polylogarithms over the alphabet $\{0,-1, 1\}$. A corresponding method has been implemented in the package {\tt HarmonicSums}.
In the given physical combination we obtain the usual harmonic polylogarithms at argument $x$
and a series of harmonic polylogarithms at argument $1 - 2x$, which we denote by
\begin{eqnarray}
\tilde{H}_{\vec{a}} = H_{\vec{a}}(1 - 2 x)
\end{eqnarray}
in the following. This representation is of advantage for later numerical representations\footnote{Note that using the 
harmonic polylogarithms at a different continuous argument implies in general a new class of functions with only exceptional 
relations.}. The above mapping needs not to occur always and it is even possible in individual cases in more extended
calculations where iterated integrals contribute, which have support on a subset of the interval $[0,1]$ only. This property 
will carry over to the OME and asymptotic Wilson coefficient and require a modification of the Mellin convolution. In 
intermediary steps we observed the supports $[0,1/2]$ and $[1/2,1]$. The corresponding Mellin convolutions with parton 
distribution functions of support $[0,1]$ read~:
\begin{eqnarray}
\left[A_1(x) \theta\left(\tfrac{1}{2} - x\right)\right] \otimes f(x) &=& \theta\left(\tfrac{1}{2} - x\right) 
\int_{2x}^1 \frac{dy}{y} A_1\left(\frac{x}{y}\right) f(y)
\\ 
\left[A_2(x) \theta\left(x - \tfrac{1}{2}\right)\right] \otimes f(x) &=& \int_x^1 \frac{dy}{y} A_2\left(\frac{x}{y}\right)
f(y) - \theta\left(\tfrac{1}{2} - x\right) 
\int_{2x}^1 \frac{dy}{y} A_2\left(\frac{x}{y}\right) f(y)~.
\end{eqnarray}

We will split $a_{Qq}^{(3),\rm PS}(x)$ into a part being represented by the harmonic 
polylogarithms of only argument $x$ and a part containing also harmonic polylogarithms of argument $1 - 2 x$. One obtains

\begin{eqnarray}
&&
                        +\frac{32}{3} (x+1) H_0
                \biggr) H_{0,1}
                -128 (x-1) H_{0,-1,-1}
                +\frac{544}{3} (x-1) H_{0,0,-1}
                +\frac{16}{3} (7 x+3) H_{0,0,1}
\nonumber\\ &&
                +\frac{80}{3} (x+1) H_{0,1,1}
        \biggr) \zeta_2
        -\frac{512}{3} (x+1) \zeta_2 \ln^3(2)
        +\biggl(
                \frac{8}{9} \frac{x-1}{x} \big(88 x^2+235 x+88\big) H_1
\nonumber\\ &&
                -\frac{16}{9} \frac{1}{x} \big(52 x^3+819 x^2-144 x+28\big) H_0
                -\frac{4}{27} \frac{1}{x} \big(4612 x^3+15262 x^2+8524 x+2559\big)
\nonumber\\ &&
                +\frac{16 (x-4) (x+1) (4 x-1) H_{-1}}{x}
                +\frac{64}{3} (5 x-6) H_0^2
                +\frac{608}{3} (x-1) H_{0,-1}
                -\frac{496}{3} (x+1) 
\nonumber\\ && \times
H_{0,1}
        \biggr) \zeta_3
        -896 (x+1) \zeta_3 \ln^2(2)
        +\biggl(
                \frac{448}{3} \frac{1}{x} \big(4 x^3-9 x^2-6 x-2\big)
                -896 (x+1) H_0
        \biggr) 
\nonumber\\ && \times
\zeta_3 \ln(2)
        +\biggl(
                -\frac{2}{9} \frac{1}{x} \big(1752 x^3+11325 x^2+1401 x+1828\big)
                -\frac{76}{3} (17 x-15) H_0
        \biggr) \zeta_4
\nonumber\\ && 
       -832 (x+1) \zeta_4 \ln(2)
        +\biggl(
                \frac{16}{3} \frac{1}{x} \big(8 x^3-30 x^2-15 x-2\big)
                -64 (x+1) H_0
        \biggr) B_4
\nonumber\\ &&
        -128 (x+1) B_4 \ln(2)
        +\frac{8}{3} (31 x-127) \zeta_2 \zeta_3
        -12 (47 x-145) \zeta_5
        -2048 \Li_5\left(\frac{1}{2}\right) (x+1)
\nonumber\\ &&
        +\frac{256}{15} (x+1) \ln^5(2)
\Biggr] + \tilde{a}_{Qq}^{(3),\rm PS}
\\
\text{and}
\nonumber\\ && 
\lefteqn{\tilde{a}_{Qq}^{(3),\rm PS} =} \nonumber\\ &&
\textcolor{blue}{C_F T_F \left(\frac{C_A}{2}-C_F\right)} \Biggl[
        -64 (x+1) H_0^2 \tilde{H}_{0,-1,-1} 
        +\frac{64}{3} \frac{x-1}{x} \big(4 x^2+7 x+4\big) H_1 \tilde{H}_{0,-1,-1}
\nonumber\\ &&
        +\frac{16}{3} \frac{1}{x} \big(32 x^3+200 x^2
-104 x+1\big) \tilde{H}_{0,-1,-1}
        +\frac{64}{3} \big(4 x^2-21 x-9\big) H_0 \tilde{H}_{0,-1,-1}
\nonumber\\ &&
        -128 (x+1) H_{0,1} \tilde{H}_{0,-1,-1}
        +\biggl(
                -\frac{64}{3} \frac{x-1}{x}  \big(4 x^2+7 x+4\big) H_1
                -\frac{16}{3} \frac{1}{x} \big(32 x^3+200 x^2
\nonumber\\ &&
-104 x+1\big)
                -\frac{64}{3} \big(4 x^2-21 x-9\big) H_0
                +64 (x+1) H_0^2
                +128 (x+1) H_{0,1}
        \biggr) \tilde{H}_{0,-1,1}
\nonumber\\ &&
        +\biggl(
                \frac{64}{3} \frac{x-1}{x} \big(4 x^2+7 x+4\big) H_1
                +\frac{16}{3} \frac{1}{x} \big(32 x^3+200 x^2-104 x+1\big)
                +\frac{64}{3} \big(4 x^2-21 x
\nonumber\\ &&
-9\big) H_0
                -64 (x+1) H_0^2
                -128 (x+1) H_{0,1}
        \biggr) \tilde{H}_{0,1,-1}
        +\biggl(
                -\frac{64}{3} \frac{x-1}{x} \big(4 x^2+7 x+4\big) H_1
\nonumber\\ &&               
 -\frac{16}{3} \frac{1}{x} \big(32 x^3+200 x^2-104 x+1\big)
                -\frac{64}{3} \big(4 x^2-21 x-9\big) H_0
                +64 (x+1) H_0^2
                +128 (x+1) 
\nonumber\\ && \times
H_{0,1}
        \biggr) \tilde{H}_{0,1,1}
        +\biggl(
                \frac{64 (x-1) \big(4 x^2+7 x+4\big)}{x}
                -384 (x+1) H_0
        \biggr) \tilde{H}_{0,-1,-1,-1}
        +\biggl(
                -\frac{64}{3} \frac{1}{x} 
\nonumber\\ && \times
\big(4 x^3+27 x^2+3 x-8\big)
                +128 (x+1) H_0
        \biggr) \tilde{H}_{0,-1,-1,1}
        +\biggl(
                \frac{64}{3} \big(4 x^2-21 x-9\big)
\nonumber\\ &&                
-128 (x+1) H_0
        \biggr) \tilde{H}_{0,-1,1,-1}
        +\biggl(
                -\frac{64}{3} \frac{1}{x} \big(12 x^3-39 x^2-21 x-4\big)
                +384 (x+1) H_0
        \biggr) 
\nonumber\\ && \times
\tilde{H}_{0,-1,1,1}
        +\biggl(
                \frac{64}{3} \frac{1}{x} \big(12 x^3-15 x^2-15 x-8\big)
                -384 (x+1) H_0
        \biggr) \tilde{H}_{0,1,-1,-1}
        +\biggl(
                -\frac{64}{3} \frac{1}{x} 
\nonumber\\ && \times 
(x-1) \big(4 x^2+7 x+4\big)
                +128 (x+1) H_0
        \biggr) \tilde{H}_{0,1,-1,1}
        +\biggl(
                \frac{64}{3} \frac{1}{x} \big(4 x^3-45 x^2-15 x+4\big)
\nonumber\\ &&
                -128 (x+1) H_0
        \biggr) \tilde{H}_{0,1,1,-1}
        +\biggl(
                -64 \big(4 x^2-21 x-9\big)
  +384 (x+1) H_0
        \biggr) \tilde{H}_{0,1,1,1}
\nonumber
\end{eqnarray}
\begin{eqnarray}
&&               
        -384 (x+1) \tilde{H}_{0,-1,-1,-1,1}
        -256 (x+1) \tilde{H}_{0,-1,-1,1,-1}
        +384 (x+1) \tilde{H}_{0,-1,-1,1,1}
\nonumber\\ && 
        -128 (x+1) \tilde{H}_{0,-1,1,-1,-1}
        -384 (x+1) \tilde{H}_{0,-1,1,1,-1}
        +768 (x+1) \tilde{H}_{0,-1,1,1,1}
\nonumber\\ && 
        -384 (x+1) \tilde{H}_{0,1,-1,-1,1}
        -256 (x+1) \tilde{H}_{0,1,-1,1,-1}
        +384 (x+1) \tilde{H}_{0,1,-1,1,1}
\nonumber\\ && 
        -128 (x+1) \tilde{H}_{0,1,1,-1,-1}
        -384 (x+1) \tilde{H}_{0,1,1,1,-1}
        +768 (x+1) \tilde{H}_{0,1,1,1,1}
        +
                64 (x+1) 
\nonumber\\ && 
\times \biggl(\tilde{H}_{0,-1,-1}
                -\tilde{H}_{0,-1,1}
                +\tilde{H}_{0,1,-1}
                -\tilde{H}_{0,1,1}
        \biggr) \zeta_2
                -128 (x+1) \biggl(\tilde{H}_{0,-1}
                +\tilde{H}_{0,1}
        \biggr) \zeta_2 \ln(2)
\nonumber\\ && 
        +\biggl(
                -\frac{128}{3} \frac{x-1}{x} \big(4 x^2+7 x+4\big) H_1 \tilde{H}_{0,-1}
                -\frac{32}{3} \frac{1}{x} \big(32 x^3+200 x^2-104 x+1\big) \tilde{H}_{0,-1}
\nonumber\\ && 
                -\frac{128}{3} \big(4 x^2-21 x-9\big) H_0 \tilde{H}_{0,-1}
                +128 (x+1) H_0^2 \tilde{H}_{0,-1}
                +\biggl(
                        -\frac{128}{3} \frac{x-1}{x} \big(4 x^2+7 x
\nonumber\\ && 
+4\big) H_1
                        -\frac{32}{3} \frac{1}{x} \big(32 x^3+200 x^2-104 x+1\big)
                        -\frac{128}{3} \big(4 x^2-21 x-9\big) H_0
                        +128 (x+1) H_0^2
                \biggr) \tilde{H}_{0,1}
\nonumber\\ && 
                +\biggl(
                        256 (x+1) \tilde{H}_{0,-1}
                        +256 (x+1) \tilde{H}_{0,1}
                \biggr) H_{0,1}
                +\biggl(
                        -\frac{128}{3} \frac{1}{x} \big(8 x^3+18 x^2-3 x-10\big)
\nonumber\\ && 
                        +512 (x+1) H_0
                \biggr) \tilde{H}_{0,-1,-1}
                +\biggl(
                        -\frac{128}{3} \frac{1}{x} \big(8 x^3-30 x^2-15 x-2\big)
                        +512 (x+1) H_0
                \biggr) \tilde{H}_{0,-1,1}
\nonumber\\ && 
 +\biggl(
                        -\frac{128}{3} \frac{1}{x} \big(8 x^3-6 x^2-9 x-6\big)
                        +512 (x+1) H_0
                \biggr) \tilde{H}_{0,1,-1}
                +\biggl(
                        -\frac{128}{3} \frac{1}{x} \big(8 x^3-54 x^2
\nonumber\\ && 
-21 x+2\big)
                        +512 (x+1) H_0
                \biggr) \tilde{H}_{0,1,1}
                -384 (x+1) \tilde{H}_{0,-1,-1,-1}
                +640 (x+1) \tilde{H}_{0,-1,-1,1}
\nonumber\\ && 
                +128 (x+1) \tilde{H}_{0,-1,1,-1}
                +1152 (x+1) \tilde{H}_{0,-1,1,1}
                -384 (x+1) \tilde{H}_{0,1,-1,-1}
                +640 (x+1) \tilde{H}_{0,1,-1,1}
\nonumber\\ && 
                +128 (x+1) \tilde{H}_{0,1,1,-1}
                +1152 (x+1) \tilde{H}_{0,1,1,1}
        \biggr) \ln(2)
        +\biggl(
                \frac{256 \big(12 x^2+3 x-2\big)}{3 x} [\tilde{H}_{0,-1}+\tilde{H}_{0,1}]
\nonumber\\ &&               
 +512 (x+1) [\tilde{H}_{0,-1,-1}
                + \tilde{H}_{0,-1,1}
                + \tilde{H}_{0,1,-1}
                + \tilde{H}_{0,1,1}]
        \biggr) \ln^2(2)
\Biggr]~.
\end{eqnarray}
Here the following harmonic polylogarithms contribute
\begin{eqnarray}
&&
H_{0,-1,-1,-1},
H_{0,-1,-1,1},
H_{0,-1,0,1},
H_{0,-1,1,-1},
H_{0,-1,1,1},
H_{0,0,-1,-1},
H_{0,0,-1,1},
H_{0,0,1,-1},
\nonumber\\ &&
H_{0,1,-1,-1},
H_{0,1,-1,1},
H_{0,1,1,-1},
H_{0,-1,-1,0,1},
H_{0,-1,0,-1,-1},
H_{0,0,-1,-1,-1},
H_{0,0,-1,0,-1},
\nonumber\\ &&
H_{0,0,-1,0,1},
H_{0,0,0,-1,-1},
H_{0,0,0,-1,1},
H_{0,0,0,0,-1},
H_{0,0,0,0,1},
H_{0,0,0,1,-1},
H_{0,0,0,1,1},
H_{0,0,1,0,-1},
\nonumber\\ &&
H_{0,0,1,0,1},
H_{0,0,1,1,1},
H_{0,1,0,1,1},
H_{0,1,1,1,1}
\end{eqnarray}
and
\begin{eqnarray}
&&
\tilde{H}_{0,-1}, \tilde{H}_{0,1}, \tilde{H}_{0,-1,-1}, \tilde{H}_{0,-1,1}, \tilde{H}_{0,1,-1}, \tilde{H}_{0,1,1},
\tilde{H}_{0,-1,-1,-1}, \tilde{H}_{0,-1,-1,1}, \tilde{H}_{0,-1,1,-1}, \tilde{H}_{0,-1,1,1}, 
\nonumber\\ &&
\tilde{H}_{0,1,-1,-1},
\tilde{H}_{0,1,-1,1}, \tilde{H}_{0,1,1,-1}, \tilde{H}_{0,1,1,1}, \tilde{H}_{0,-1,-1,-1,1}, \tilde{H}_{0,-1,-1,1,-1},
\tilde{H}_{0,-1,-1,1,1}, \tilde{H}_{0,-1,1,-1,-1}, 
\nonumber\\ &&
\tilde{H}_{0,-1,1,1,-1}, \tilde{H}_{0,-1,1,1,1},
\tilde{H}_{0,1,-1,-1,1}, \tilde{H}_{0,1,-1,1,-1}, \tilde{H}_{0,1,-1,1,1}, \tilde{H}_{0,1,1,-1,-1},
\tilde{H}_{0,1,1,1,-1},
\tilde{H}_{0,1,1,1,1}
\nonumber\\
\end{eqnarray} 
beyond those in Eq.~(\ref{eq:HPL1}). Note that $\tilde{a}_{Qq}^{(3),\rm PS}(x)$ vanishes for $x = 1/2$. Both functions
${a}_{Qq}^{(3),\rm PS}(x)$ and $\tilde{a}_{Qq}^{(3),\rm PS}(x)$ move to a constant of opposite value for $x \rightarrow 1$.

It is useful to have a precise and compact numeric representation of ${a}_{Qq}^{(3),\rm PS}$. The usual numerical 
implementations in {\tt Fortran} \cite{FORTRAN} are given at double precision accuracy (16 digits) and can be obtained 
in a systematic way based on elementary functions. For very large values of $x \lsim 1$ we apply the analytic series expansion of 
$a_{Qq}^{(3),\rm PS}$ and $\tilde{a}_{Qq}^{(3),\rm PS}(x)$. In the remaining $x$-region the 
so-called series-improvement based on Bernoulli- and Euler-numbers \cite{BERN}, see \cite{Gehrmann:2001pz}, is applied. 
It is related to the Euler-Mac Laurin \cite{EML} representation of linear combinations of harmonic polylogarithms and can be extended to 
arbitrary precision. It improves earlier representations based on Chebyshev polynomials \cite{CHEB} as applied to Nielsen integrals
in Ref.~\cite{Kolbig:1969zza}. The method has been applied to polylogarithms in Ref.~\cite{'tHooft:1978xw}. In the physical literature it 
dates back to Debye's work on the specific heat \cite{Debye:1912} in 1912. It is important to separate the cuts of the polylogarithms, 
which 
are either located on the positive or negative real axis. In deriving representation also mixed terms appear. Moreover, 
in $\tilde{a}_{Qq}^{(3),\rm PS}(x)$ the harmonic polylogarithms ${\rm H}_{\vec{a}}(x)$ appear together with $\tilde{\rm H}_{\vec{b}}(x)$. 

We decompose ${a}_{Qq}^{(3),\rm PS}$ as
\begin{eqnarray}
{a}_{Qq}^{(3),\rm PS}(x) = \hat{a}_{Qq}^{(3),\rm PS} + \tilde{a}_{Qq}^{(3),\rm PS}~.
\end{eqnarray}
Here $\hat{a}_{Qq}^{(3),\rm PS}$ denotes the part of ${a}_{Qq}^{(3),\rm PS}$ consisting only of harmonic polylogarithms of argument $x$.
The following representation is obtained
\begin{eqnarray}
\hat{a}_{Qq}^{(3),\rm PS}(x) &=& 
\sum_{k,l=0}^{15} a_{k,l}(x) u^k v^l,~~~x \in [0, \sqrt{2} - 1], \nonumber\\ &&
u = -\ln(1-x),~~v = \ln(1+x),~~a_{k,l}(x) = \sum_{m=0}^5 r^{(1)}_{k,l,m}(x) \ln^m(x)~
\label{EQ:REP1}
\\
\hat{a}_{Qq}^{(3),\rm PS}(x) &=& \sum_{k,l=0}^{16} \bar{a}_{k,l}(x,t) \bar{u}^k \bar{v}^l,~~~t = \frac{1-x}{1+x},~~~x \in [\sqrt{2} - 1, 
0.9],
\nonumber\\ && \bar{u} = \ln(1+x) - \ln(x) - \ln(2),~~\bar{v} = \ln(2) - \ln(1+x),
\nonumber\\ &&
\bar{a}_{k,l}(x,t) 
= \sum_{m=0}^4 r^{(2)}_{k,l,m}(x) 
\ln^m(t) 
\\
\hat{a}_{Qq}^{(3),\rm PS}(x) &=& c + \sum_{k=1}^{18} (1-x)^k 
\bigl[a_k + b_k \ln(1-x) + d_k \ln^2(1-x) 
+ e_k \ln^3(1-x) 
\nonumber\\ && 
+ f_k \ln^4(1-x)
\bigr],~~~x \in [0.9, 1]~,
\\
\tilde{a}_{Qq}^{(3),\rm PS}(x) &=& \sum_{k,l=0}^{15} \tilde{b}_{k,l}(x) \tilde{u}^k \tilde{v}^l,~~~x \in \left[0,1 - 
\frac{1}{\sqrt{2}}\right]
\nonumber\\ &&
\tilde{u} = \ln(1-x) - \ln(1 - 2x),~~~\tilde{v} = - \ln(1-x)
\\
\tilde{a}_{Qq}^{(3),\rm PS}(x) &=& 
\sum_{k,l=0}^{25} b_{k,l}(x) \hat{u}^k \hat{v}^l   ,~~~x \in \left[1 - \frac{1}{\sqrt{2}}, 0.9\right],
\nonumber\\ &&
\hat{u} = - \ln(2) - \ln(x),~~~\hat{v} = \ln(2) + \ln(1-x)
\\
\label{EQ:REP2}
\tilde{a}_{Qq}^{(3),\rm PS}(x) &=& -c 
+ \sum_{k=1}^{18} (1-x)^k \left[\tilde{a}_k + \tilde{b}_k \ln(1-x) + \tilde{d}_k \ln^2(1-x)\right],~~~x \in [0.9, 
1]~.
\end{eqnarray}
Here $b_{k,l}(x)$ and $\tilde{b}_{k,l}(x)$ are low order polynomials in $\ln(x), \ln(1-x)$ and $\Li_2(x)$  with rational coefficients 
in $x$. For the function $\Li_2(x)$ one uses the well-known Bernoulli-representation \cite{'tHooft:1978xw,Gehrmann:2001pz}. 
The functions $r^{(i)}_{k,l,m}(x)$ are rational in $x$ and $c, a_k, b_k, d_k, e_k, f_k, \tilde{a}_k, \tilde{b}_k$ and 
$\tilde{d}_k$ are constants. In part of the region the above representation yields an even  higher accuracy than double precision. 
The polynomial representations (\ref{EQ:REP1}--\ref{EQ:REP2}) may be further compactified using Horner's method \cite{HORNER} and are well 
suited to efficiently 
generate grids for further numerical use. This also applies to the corresponding OME and Wilson coefficient. 
All expressions have been derived using the package {\tt HarmonicSums}. Numerical checks were performed using the code of 
Ref.~\cite{Vollinga:2004sn} and the code {\tt HPL 2.0} \cite{Maitre:2007kp}.

\begin{figure}[H]
\centering
\includegraphics[width=0.8\textwidth]{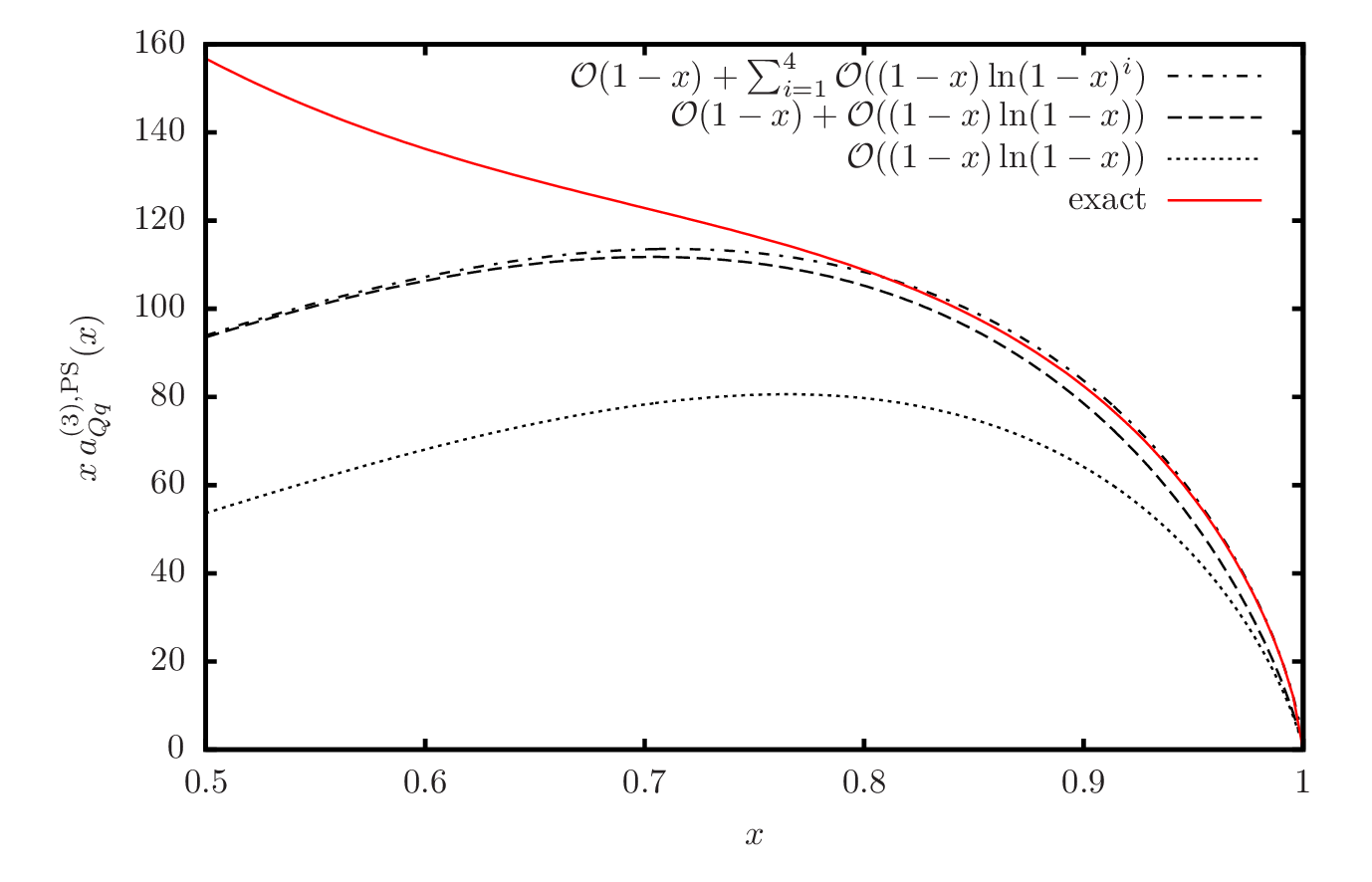}
\caption{\sf \small $xa_{Qq}^{(3),\rm PS}(x)$ in the large $x$ region (solid red line). Dotted line: leading large $x$ behaviour
$O((1-x) \ln(1-x)$, dashed line: adding the term $\propto (1-x)$, dash-dotted line: adding all higher logarithms $O((1-x) \ln^k(1-x)$.}
\label{fig:2}
\end{figure}

We now consider the limiting behaviour of $a_{Qq}^{(3),\rm PS}(x)$ for large and small values of the momentum fraction $x$.
In the limit $x \rightarrow 1$ one obtains
\begin{eqnarray}
a_{Qq}^{(3),\rm PS}(x) &\propto& \textcolor{blue}{C_F T_F} (1-x) \Biggl\{
\frac{2}{9}(\textcolor{blue}{C_A - C_F}) \ln^4(1-x) 
\nonumber\\
&&
- \frac{4}{27} (23 \textcolor{blue}{C_A} - 21 \textcolor{blue}{C_F}
+ 4(2 - \textcolor{blue}{N_F}) \textcolor{blue}{T_F}
) \ln^3(1-x)
\Biggr\} + O((1-x)\ln^2(1-x))
\nonumber\\ 
&\propto& (1-x) \Biggl[
   0.24691358 \ln^4(1-x)
+(-4.44444444
  +0.19753086~N_F) \ln^3(1-x)
\nonumber\\ &&
+(-2.28230742
  +0.98765432~N_F) \ln^2(1-x)
\nonumber\\ &&
+(-357.426943
  +15.9385086~N_F) \ln(1-x)
+ 116.478169 
  +14.3167889~N_F \Biggr]~.
\nonumber\\
&& + O((1-x)^2 \ln^3(1-x))~.
\label{eq:aQqLAX}
\end{eqnarray}
The last expressions correspond to the numerical values for $SU(3)_c$. $a_{Qq}^{(3),\rm PS}(x)$ vanishes at $x = 1$. 
The behaviour of $a_{Qq}^{(3),\rm PS}$ in the large-$x$ region is
illustrated in Figure~\ref{fig:2}. 
One should note that the factors of the various expansion coefficients are partly rather different, which gives 
preference to less singular terms.

\noindent
For small values $x$, the function $a_{Qq}^{(3),\rm PS}$ behaves like
\begin{eqnarray}
a_{Qq}^{(3),\rm PS}(x) &\propto& \frac{64}{243} \textcolor{blue}{C_F T_F C_A} \left[1312 + 135 \zeta_2 - 189 \zeta_3\right] 
\frac{\ln(x)}{x}
\nonumber\\ 
&& + \Biggl[\textcolor{blue}{C_F T_F C_A} \Biggl(
        \frac{2198960}{729}
        +\frac{64}{3} B_4
        -\frac{5600}{9} \zeta_4
        -\frac{3496}{9} \zeta_3
        -\frac{5624}{81} \zeta_2
\Biggr)
\nonumber\\ &&
+\textcolor{blue}{C_F^2 T_F} \Biggl(
        \frac{12304}{27}
        -\frac{128}{3} B_4
        +192 \zeta_4
        -\frac{2416}{9} \zeta_3
        +80 \zeta_2
\Biggr)
\nonumber\\ &&
+\textcolor{blue}{C_F T_F^2} \Biggl(
        -\frac{77696}{729}
        -\frac{4096}{27} \zeta_3
        -\frac{1280}{27} \zeta_2
\Biggr)
\nonumber\\ &&
+\textcolor{blue}{C_F T_F^2 N_F} \Biggl(
        -\frac{111104}{729}
        +\frac{896}{27} \zeta_3
        -\frac{320}{27} \zeta_2
\Biggr)\Biggr] \frac{1}{x} + O(\ln^5(x))
\nonumber\\ 
&\propto& 688.39630 \frac{\ln(x)}{x} + (
          3812.8990
         -44.003690 N_F) \frac{1}{x}
         + 1.6 \ln^5(x) +
\nonumber\\ &&
        (-20.345679
         +0.7901235 N_F) \ln^4(x)
        +(165.11455 
         +2.6337449 N_F) \ln^3(x)
\nonumber\\ && +
        (-604.63554
         +30.502827 N_F) \ln^2(x)
        +(3524.9967 
        + 33.908944 N_F) \ln(x)
\nonumber\\ &&
+ O(x^0)~.
\label{eq:aQqLOX}
\end{eqnarray}

\begin{figure}[H]
\centering
\includegraphics[width=0.8\textwidth]{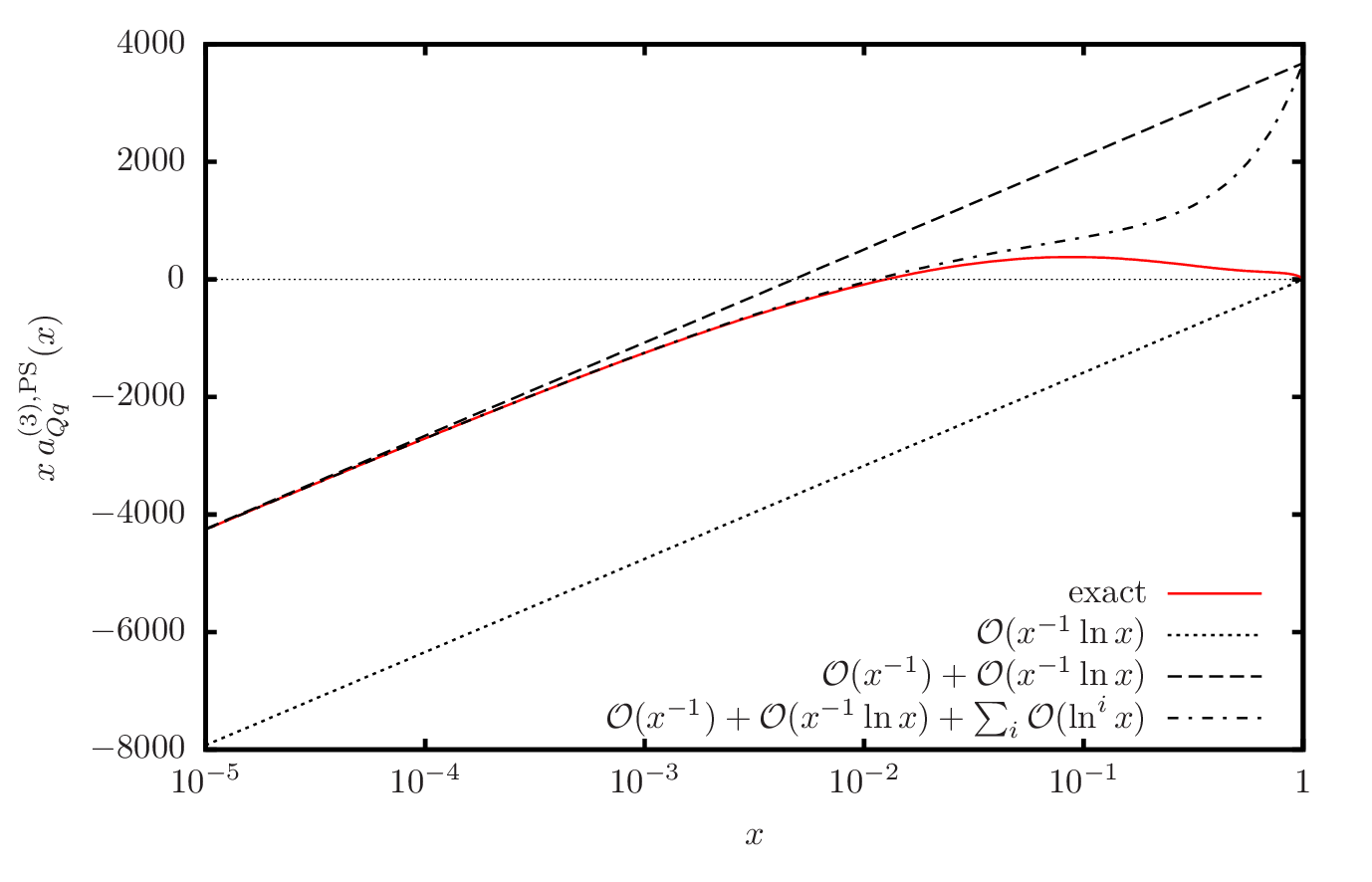}
\caption{\sf \small $xa_{Qq}^{(3),\rm PS}(x)$ in the low $x$ region (solid red line) and leading terms approximating this 
quantity; dotted line: `leading' small $x$ approximation $O(\ln(x)/x)$, dashed line: adding the $O(1/x)$-term, dash-dotted
line: adding all other logarithmic contributions.}
\label{fig:1}
\end{figure}

The asymptotic behaviour at small values of $x$ is depicted in Figure~\ref{fig:1}.
The leading term $\propto \ln(x)/x$ does nowhere describe $a_{Qq}^{(3),\rm PS}(x)$, which is a common experience 
in many small-$x$ studies, cf.~\cite{Blumlein:1997em,Blumlein:1999ev}. An important term in the small $x$ region is the `next dominant'
one $\propto 1/x$, which has firstly been calculated in this paper\footnote{The terms $\propto N_F$ were given in \cite{Ablinger:2010ty} 
before.}. The first two terms give a sufficient description up to $x \simeq 5 \cdot 10^{-4}$. 
At larger values, up to $x \simeq
2 \cdot 10^{-2}$ in the small-$x$ region all corrections $\propto \ln^k(x),~~k = 5 ... 1$ are required.

Let us reconsider the term $\propto \ln(x)/x$ in $a_{Qq}^{(3),\rm PS}$ in view of the leading order small-$x$ approximation. We 
mention
that a rigorous theory of QCD in the small $x$ limit has still not been worked out. This concerns, in particular, non-leading terms. 
As has already been known from the most singular terms $\propto 1/x$, contributing to the pole $1/(N-1)$ in Mellin-$N$ space, 
in the unpolarized leading order splitting functions $P_{gg}(x)$ \cite{Gross:1974cs,Georgi:1951sr} and $P_{gq}(x)$ 
\cite{Williams,vonWeizsacker:1934sx}, they are related by the ratio $C_A/C_F$. In the ladder-approach using a physical 
gauge, this is the effect of exchanging one gluon-cell by that of a massless quark-cell \cite{Cheng:1985bj}. In this way,
leading small-$x$ contributions to the constant part of the unrenormalized OMEs $A_{Qg}$ and $A_{Qq}^{\rm PS}$ might be 
related.
This is indeed the case, which is easily seen for the expressions at 2-loop order in Mellin space \cite{Bierenbaum:2007qe}, see also 
\cite{Buza:1995ie} for the expressions in $x$-space. The leading small-$x$ contribution to the massive 
Wilson coefficient $H_{2,g}$ has been calculated in \cite{Catani:1990eg}. Using the representation of the asymptotic 
heavy flavor Wilson 
coefficient given in Ref.~\cite{Bierenbaum:2009mv} and expanding the result of \cite{Catani:1990eg} in the limit $Q^2 \gg m^2$
using the corresponding perturbative representations of the BFKL resummed anomalous dimension \cite{Kuraev:1977fs,Jaroszewicz:1980mq},
following e.g. \cite{Blumlein:1997em}, one obtains a prediction for the term of $O(\ln(x)/x)$ in $a_{Qg}^{(3)}$. This 
has been performed in 
\cite{Kawamura:2012cr}. Multiplying this expression by $C_F/C_A$ agrees with the small-$x$ limit of the exact 
expression (\ref{eqaQq3}), which is a new result of the present calculation.
As has been shown above, the knowledge of this term is not enough for a quantitative
prediction even in the small-$x$ region and various more terms have to be computed. 
Following a method, developed by T.~van Ritbergen \cite{RITBERGEN}\footnote{
Other approximate methods based on orthogonal polynomials to reconstruct $x$-shapes from moments have been known and were widely used even 
earlier, see Refs.~\cite{APPROX}. We would like to mention that the determination of all these quantities, which are recurrent in $N$ can 
be determined exactly knowing a {\it finite} number of moments as has been shown in \cite{Blumlein:2009tj}. The corresponding number of 
moments is, however, for mathematical reasons far larger than $N=6$ as used in \cite{Kawamura:2012cr}.},
the authors of \cite{Kawamura:2012cr} used the fixed Mellin moments at 3-loop order calculated by Bierenbaum, Klein and 
one of the present authors in Ref.~\cite{Bierenbaum:2009mv} using some sets of functions to determine a band of possibilities for values of
$a_{Qq}^{(3), \rm PS}$. 
\begin{figure}[H]
\centering
\includegraphics[width=0.8\textwidth]{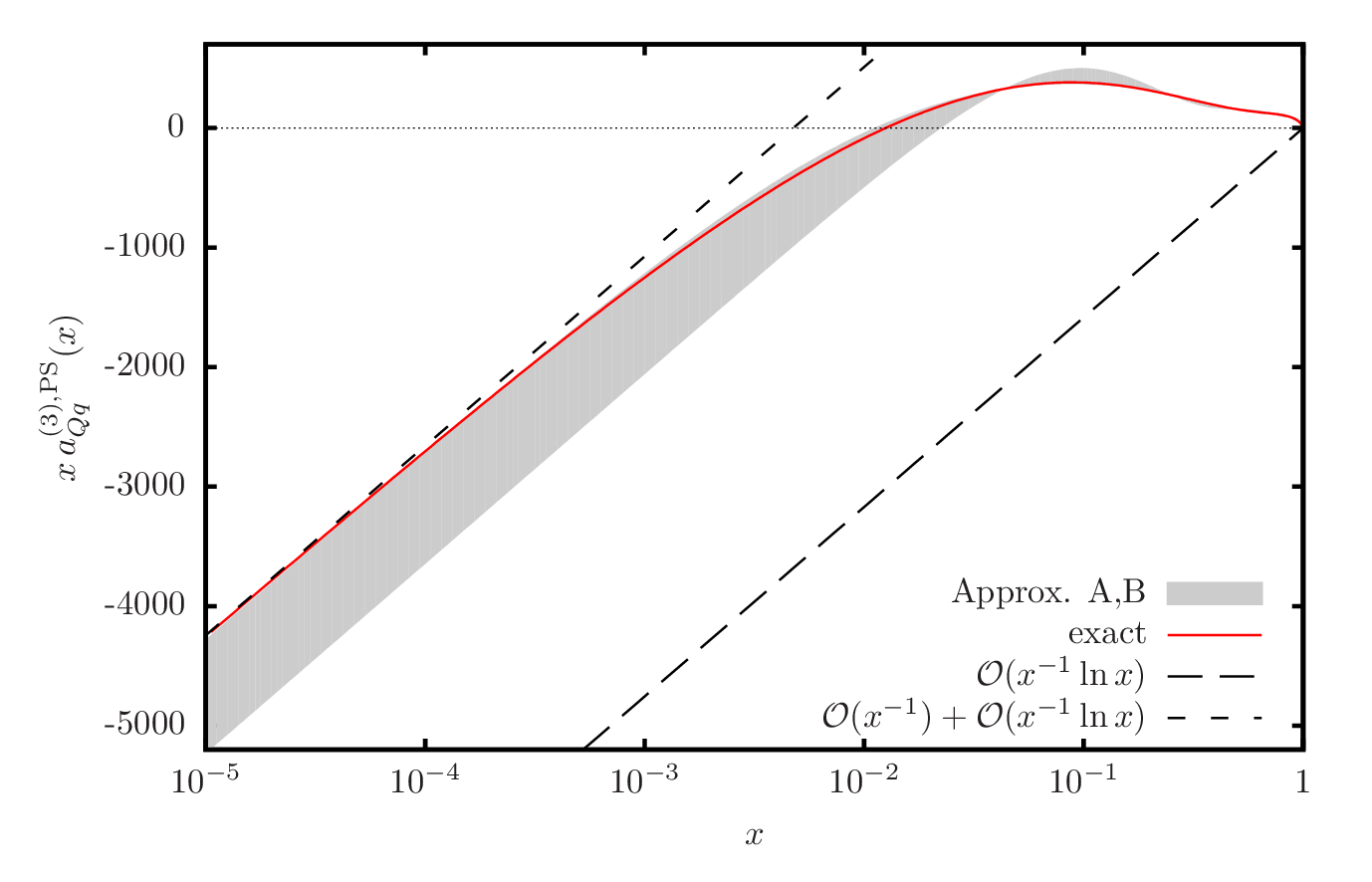}
\caption{\sf \small Comparison of the exact result for $xa_{Qq}^{3, \rm PS}(x)$ (solid red line) with an estimate in  
\cite{Kawamura:2012cr} (shaded area). Also shown are the first two leading small $x$ contributions.
\label{Fig:WILS3}}
\end{figure}

\noindent
To what extend these ans\"atze are compelling is hard to say, since the functional structure in a higher 
order 
calculation is difficult to predict and only understood after the calculation has been performed. In particular, the simplification
of the initially large amount of generalized harmonic polylogarithms to the functions $H$ and $\tilde{H}$ is far from 
obvious. 
This applies also to the general functional structure. Furthermore, the former estimate \cite{Kawamura:2012cr} has been based on the 
unproven hypotheses of $C_F/C_A$ scaling and the assumption that the leading small-$x$ term for the Wilson 
coefficient can be determined in the way being described 
above. In Figure~\ref{Fig:WILS3} we compare the range suggested for $a_{Qq}^{(3), \rm PS}$ in \cite{Kawamura:2012cr} with the 
exact result. 
In the important region of small values of $x$ the guess in \cite{Kawamura:2012cr} has an uncertainty of $\sim 20\%$.
At large values of $x$ the grey area in Figure~\ref{Fig:WILS3} winds around the exact result. This is enforced by known  
Mellin moments of Ref.~\cite{Bierenbaum:2009mv} used to solve a linear system for the particular function set selected. 

We now turn to the OME $A_{Qq}^{\rm PS}$ in the on-shell scheme for the quark mass. It receives contributions from $O(a_s^2)$ onward. We 
denote the logarithmic scale dependence by 
\begin{eqnarray}
L_M = \ln\left(\frac{m^2}{\mu^2}\right)~.
\end{eqnarray}
The matrix element reads in Mellin-$N$ space as follows
\begin{eqnarray}
A_{Qq}^{\rm PS}(N) &=&
a_s^2 \textcolor{blue}{C_F T_F} \Biggl\{
        -4 F L_M^2
        -L_M  \frac{8 \big(N^2+5 N+2\big) \big(5 N^3+7 N^2+4 N+4\big)}{(N-1) N^3 (N+1)^3 (N+2)^2}
        -8 F S_2
\nonumber \\ &&
        +\frac{4 P_{53}}{(N-1) N^4 (N+1)^4 (N+2)^3}
\Biggr\}
+a_s^3 \Biggl\{
        \textcolor{blue}{C_F T_F^2} \Biggl[
                -L_M^3 \frac{128}{9} F
                +L_M^2 \biggl[
                        \frac{32}{3} F S_1
\nonumber \\ &&
                        -\frac{32 P_3}{9 (N-1) N^3 (N+1)^2 (N+2)^2}
                \biggr]
                +L_M \biggl[
                        \frac{64 P_{44}}{9 (N-1) N^3 (N+1)^3 (N+2)^2} S_1
\nonumber \\ &&
                        -\frac{64 P_{61}}{27 (N-1) N^4 (N+1)^4 (N+2)^3}
                        +\biggl(
                                -\frac{32}{3} S_1^2
                                -32 S_2
                        \biggr) F
                \biggr]
\nonumber \\ &&
                +\frac{16}{27 (N-1) N^3 (N+1)^3 (N+2)^2 (N+3) (N+4) (N+5)} \biggl(
                        P_{56} S_1^2
                        +P_{62} S_2
\nonumber \\ &&
                        -\frac{2 P_{70}}{3 N (N+1) (N+2)} S_1
                        +\frac{2 P_{75}}{9 N^2 (N+1)^2 (N+2)^2}
                \biggr)
                +\biggl(
                        -\frac{16}{27} S_1^3
                        -\frac{208}{9} S_1 S_2
\nonumber \\ &&
                        -\frac{32}{27} S_3
                        +\frac{128}{3} S_{2,1}
                        -\frac{896}{9} \zeta_3
                \biggr) F
        \Biggr]
        +\textcolor{blue}{C_F N_F T_F^2} \Biggl[
                -L_M^3 \frac{32}{9} F
                +L_M^2 \biggl[
                        -\frac{32}{3} F S_1
\nonumber \\ &&
                        +\frac{32 P_{44}}{9 (N-1) N^3 (N+1)^3 (N+2)^2}
                \biggr]
                +L_M \biggl[
                        \frac{32 P_7}{9 (N-1) N^3 (N+1)^3 (N+2)^2} S_1
\nonumber \\ &&
                        -\frac{32 P_{60}}{27 (N-1) N^4 (N+1)^4 (N+2)^3}
                        +\biggl(
                                -\frac{16}{3} S_1^2
                                -\frac{80}{3} S_2
                        \biggr) F
                \biggr]
\nonumber \\ &&
                +\frac{16}{27 (N-1) N^3 (N+1)^3 (N+2)^2} \biggl(
                        P_7 S_1^2
                        +P_{46} S_2
                        -\frac{2 P_{21}}{3 N (N+1) (N+2)} S_1
\nonumber \\ &&
                        +\frac{2 P_{71}}{9 N^2 (N+1)^2 (N+2)^2}
                \biggr)
                +\biggl(
                        -\frac{16}{27} S_1^3
                        -\frac{208}{9} S_1 S_2
                        -\frac{1184}{27} S_3
                        +\frac{256}{9} \zeta_3
                \biggr) F
        \Biggr]
\nonumber \\ &&
        +\textcolor{blue}{C_F^2 T_F} \Biggl[
                L_M^3 F \biggl[
                        \frac{4 \big(3 N^2+3 N+2\big)}{3 N (N+1)}
                        -\frac{16}{3} S_1
                \biggr]
                +L_M^2 \biggl[
                        \frac{4 P_{37}}{N^2 (N+1)^2} G
                        +\biggl(
                                16 S_2
\nonumber \\ &&
                                -\frac{8 \big(5 N^2+N-2\big)}{N (N+1)} S_1
                        \biggr) F
                \biggr]
                +L_M \biggl[
                        -\frac{4 \big(5 N^3+4 N^2+9 N+6\big)}{N+1} S_1^2 G
\nonumber \\ &&
                        -\frac{4 P_{42}}{(N-1) N^3 (N+1)^3 (N+2)^2} S_2
                        +\frac{8 P_{58}}{(N-1) N^4 (N+1)^4 (N+2)^3} S_1
\nonumber \\ &&
                        -\frac{4 P_{68}}{(N-1) N^5 (N+1)^5 (N+2)^3} 
                        +\biggl(
                                \frac{8}{3} S_1^3
                                -24 S_1 S_2
                                -\frac{80}{3} S_3
                                +32 S_{2,1}
\nonumber \\ &&
                                +96 \zeta_3
                        \biggr) F
                \biggr]
                +\frac{4}{3 (N-1) N^3 (N+1)^3 (N+2)^2} \biggl(
                        8 P_{47} S_{2,1}
                        -\frac{2 P_{52}}{3} S_3
\nonumber \\ &&
                        -\frac{2 P_{54}}{N (N+1) (N+2)} S_1^2
                        +\frac{2 P_{59}}{N (N+1) (N+2)} S_2
                        +\frac{P_{74}}{N^3 (N+1)^3 (N+2)^2}
                \biggr)
\nonumber \\ &&
                +\biggl(
                        \biggl(
                                \frac{128}{9} S_3
                                -32 S_{2,1}
                        \biggr) S_1
                        +\frac{4}{9} S_1^4
                        -\frac{8}{3} S_1^2 S_2
                        +\frac{52}{3} S_2^2
                        +\frac{88}{3} S_4
                        +\frac{224}{3} S_{2,1,1}
\nonumber \\ &&
                        +32 S_{3,1}
                        +\biggl(
                                64 S_3({{2}})
                                -64 S_{1,2}({{2,1}})
                                +64 S_{2,1}({{2,1}})
                                -64 S_{1,1,1}({{2,1,1}})
                        \biggr) S_1\left({{\frac{1}{2}}}\right)
\nonumber \\ &&
                        +64 \biggl(
                                -S_{1,3}\left({{2,\frac{1}{2}}}\right)
                                +S_{2,2}\left({{2,\frac{1}{2}}}\right)
                                -S_{3,1}\left({{2,\frac{1}{2}}}\right)
                                +S_{1,1,2}\left({{2,\frac{1}{2},1}}\right)
\nonumber \\ &&
                                -S_{1,1,2}\left({{2,1,\frac{1}{2}}}\right)
                                -S_{1,2,1}\left({{2,\frac{1}{2},1}}\right)
                                +S_{1,2,1}\left({{2,1,\frac{1}{2}}}\right)
                                -S_{2,1,1}\left({{2,\frac{1}{2},1}}\right)
\nonumber \\ &&
                                -S_{2,1,1}\left({{2,1,\frac{1}{2}}}\right)
                                +S_{1,1,1,1}\left({{2,\frac{1}{2},1,1}}\right)
                                +S_{1,1,1,1}\left({{2,1,\frac{1}{2},1}}\right)
\nonumber \\ &&
                                +S_{1,1,1,1}\left({{2,1,1,\frac{1}{2}}}\right)
                        \biggr)
                        +\biggl(
                                \frac{128}{3} S_1
                                -448 S_1\left({{\frac{1}{2}}}\right)
                        \biggr) \zeta_3
                        -32 B_4
                        +144 \zeta_4
                \biggr) F
\nonumber \\ &&
                +2^{-N} \frac{32 P_2}{(N-1) N^3 (N+1)^2}  \biggl(
                        7 \zeta_3
                        -S_3({{2}})
                        + S_{1,2}({{2,1}})
                        - S_{2,1}({{2,1}})
                        + S_{1,1,1}({{2,1,1}})
                \biggr)
\nonumber \\ &&
                +\biggl(
                        -\frac{4 P_{45}}{3 (N-1) N^3 (N+1)^3 (N+2)^2} S_2
                        +\frac{8 P_{69}}{3 (N-1) N^5 (N+1)^5 (N+2)^4}
                \biggr) S_1
\nonumber \\ &&
                +\frac{32 P_{49}}{3 (N-1) N^3 (N+1)^3 (N+2)^2} \zeta_3
                -\frac{4 P_{34}}{9 N (N+1)} S_1^3 G 
        \Biggr]
\nonumber \\ &&
        +\textcolor{blue}{C_A C_F T_F} \Biggl[
                L_M^3 F \biggl[
                        \frac{8 P_{35}}{9 (N-1) N (N+1) (N+2)}
                        +\frac{16}{3} S_1
                \biggr]
\nonumber \\ &&
                +L_M^2 \biggl[
                        \frac{8 P_{38}}{3 (N-1) N (N+1) (N+2)} S_1 G
                        -\frac{8 P_{66}}{9 (N-1)^2 N^4 (N+1)^4 (N+2)^3}
\nonumber \\ &&
                        +\biggl(
                                16 S_2
                                +32 S_{-2}
                        \biggr) F
                \biggr]
                +L_M \biggl[
                        \frac{16 P_{41}}{(N-1) N^3 (N+1)^3 (N+2)^2} S_{-2}
\nonumber \\ &&
                        -\frac{8 P_{64}}{9 (N-1)^2 N^4 (N+1)^4 (N+2)^2} S_1
                        +\frac{8 P_{73}}{27 (N-1)^2 N^5 (N+1)^5 (N+2)^4}
\nonumber \\ &&
                        +\biggl(
                                \frac{4 P_{40}}{3 (N-1) N (N+1) (N+2)} S_2
                                +\frac{4 P_{36}}{3 (N-1) N} S_1^2
                                +\frac{8}{3} \big(31 N^2+31 N+74\big) S_3
\nonumber \\ &&
                                +64 \big(2 N^2+2 N+3\big) S_{-3}
                                -128 \big(N^2+N+1\big) S_{-2,1}
                        \biggr) G
                        +\biggl(
                                64 S_{-2} S_1
                                -\frac{8}{3} S_1^3
\nonumber \\ &&
                                +40 S_1 S_2
                                -32 S_{2,1}
                                -96 \zeta_3
                        \biggr) F
                \biggr]
                +\frac{4}{3 (N-1) N^3 (N+1)^3 (N+2)^2} \biggl(
                        -2 P_{10} S_{2,1}
\nonumber \\ &&
                        +2 P_{50} S_{-3}
                        +4 P_{51} S_{-2,1}
                        +\frac{2 P_{63}}{9 (N-1)} S_3
                        -\frac{P_{65}}{9 (N-1) N (N+1) (N+2)} S_1^2
\nonumber \\ &&
                        -\frac{P_{67}}{9 (N-1) N (N+1) (N+2)} S_2
                        -\frac{2 P_{76}}{81 (N-1) N^3 (N+1)^3 (N+2)^3}
                \biggr)
\nonumber \\ &&
                +\biggl(
                        \frac{4 P_{39}}{27 (N-1) N (N+1) (N+2)} S_1^3
                        +\biggl(
                                \frac{8}{9} \big(77 N^2+77 N+214\big) S_3
\nonumber \\ &&
                                -\frac{16}{3} \big(23 N^2+23 N-6\big) S_{-2,1}
                        \biggr) S_1
                        +\frac{8}{3} \big(57 N^2+57 N+70\big) S_{-3} S_1
\nonumber \\ &&
                        +\frac{32}{3} \big(11 N^2+11 N+20\big) S_{-2} S_2
                        +\frac{4}{3} \big(13 N^2+13 N+34\big) S_2^2
\nonumber \\ &&
                        +\frac{16}{3} \big(29 N^2+29 N+64\big) S_4
                        -\frac{16}{3} \big(3 N^2+3 N-2\big) S_{-2}^2
\nonumber \\ &&
                        +\frac{64}{3} \big(7 N^2+7 N+11\big) S_{-4}
                        -8 \big(5 N^2+5 N+22\big) S_{3,1}
\nonumber \\ &&
                        -32 \big(5 N^2+5 N+2\big) S_{-2,2}
                        -\frac{128}{3} \big(5 N^2+5 N+4\big) S_{-3,1}
\nonumber \\ &&
                        +\frac{128}{3} \big(5 N^2+5 N-2\big) S_{-2,1,1}
                        -\frac{8}{3} \big(13 N^2+13 N+14\big) S_1 \zeta_3
                \biggr) G
\nonumber \\ &&
                +\biggl(
                        \frac{16}{3} S_{-2} S_1^2
                        -\frac{4}{9} S_1^4
                        +\frac{8}{3} S_1^2 S_2
                        -\frac{32}{3} S_{2,1,1}
                        +\biggl(
                                -32 S_3({{2}})
                                +32 S_{1,2}({{2,1}})
\nonumber \\ &&
                                -32 S_{2,1}({{2,1}})
                                +32 S_{1,1,1}({{2,1,1}})
                        \biggr) S_1\left({{\frac{1}{2}}}\right)
                        +32 \biggl(
                                S_{1,3}\left({{2,\frac{1}{2}}}\right)
\nonumber \\ &&
                                - S_{2,2}\left({{2,\frac{1}{2}}}\right)
                                + S_{3,1}\left({{2,\frac{1}{2}}}\right)
                                - S_{1,1,2}\left({{2,\frac{1}{2},1}}\right)
                                + S_{1,1,2}\left({{2,1,\frac{1}{2}}}\right)
\nonumber \\ &&
                                + S_{1,2,1}\left({{2,\frac{1}{2},1}}\right)
                                - S_{1,2,1}\left({{2,1,\frac{1}{2}}}\right)
                                + S_{2,1,1}\left({{2,\frac{1}{2},1}}\right)
                                + S_{2,1,1}\left({{2,1,\frac{1}{2}}}\right)
\nonumber \\ &&
                                - S_{1,1,1,1}\left({{2,\frac{1}{2},1,1}}\right)
                                - S_{1,1,1,1}\left({{2,1,\frac{1}{2},1}}\right)
                                - S_{1,1,1,1}\left({{2,1,1,\frac{1}{2}}}\right)
\nonumber \\ &&
                                +7 S_1\left({{\frac{1}{2}}}\right) \zeta_3
                        \biggr)
                        +16 B_4
                        -144 \zeta_4
                \biggr) F
                +2^{-N} \frac{16 P_2}{(N-1) N^3 (N+1)^2} \biggl(
                        S_3({{2}})
\nonumber \\ &&
                        - S_{1,2}({{2,1}})
                        + S_{2,1}({{2,1}})
                        - S_{1,1,1}({{2,1,1}})
                        -7 \zeta_3
                \biggr)
\nonumber \\ &&
                +\biggl(
                        \frac{4 P_{48}}{9 (N-1)^2 N^3 (N+1)^3 (N+2)^2} S_2
                        -\frac{4 P_{72}}{81 (N-1)^2 N^5 (N+1)^5 (N+2)^4}
                \biggr) S_1
\nonumber \\ &&
                +\biggl(
                        -\frac{8 P_{55}}{3 (N-1) N^4 (N+1)^4 (N+2)^3}
                        +\frac{32 P_{43}}{3 (N-1) N^3 (N+1)^3 (N+2)^2} S_1
                \biggr) S_{-2}
\nonumber \\ &&
                -\frac{8 P_{57}}{9 (N-1)^2 N^3 (N+1)^3 (N+2)^2} \zeta_3
        \Biggr]
\Biggr\}.
\label{EQ:AQq}
\end{eqnarray}

The polynomials $P_i$ are given by
\begin{eqnarray}
P_{34} &=& 5 N^4+22 N^3+49 N^2+32 N+4 \\
P_{35} &=& 11 N^4+22 N^3-23 N^2-34 N-12 \\
P_{36} &=& 17 N^4-6 N^3+41 N^2-16 N-12 \\
P_{37} &=& 7 N^6+15 N^5+7 N^4-23 N^3-26 N^2-20 N-8 \\
P_{38} &=& 17 N^6+51 N^5+51 N^4+89 N^3+40 N^2-80 N-24 \\
P_{39} &=& 17 N^6+69 N^5+153 N^4+131 N^3-86 N^2-116 N-24 \\
P_{40} &=& 73 N^6+189 N^5+45 N^4+31 N^3-238 N^2-412 N-120 \\
P_{41} &=& 2 N^7+16 N^6+37 N^5+96 N^4+143 N^3+142 N^2+132 N+40 \\
P_{42} &=& 3 N^7-15 N^6-133 N^5-449 N^4-658 N^3-500 N^2-296 N-96 \\
P_{43} &=& 3 N^7+18 N^6+49 N^5+140 N^4+190 N^3+152 N^2+120 N+32 \\
P_{44} &=& 8 N^7+37 N^6+83 N^5+85 N^4+61 N^3+58 N^2-20 N-24 \\
P_{45} &=& 81 N^7+289 N^6+331 N^5+99 N^4-128 N^3-448 N^2-688 N-240 \\
P_{46} &=& 104 N^7+481 N^6+1064 N^5+1009 N^4+646 N^3+640 N^2-344 N-336 \\
P_{47} &=& 6 N^8+40 N^7+87 N^6+62 N^5+93 N^4+220 N^3+148 N^2+96 N+64 \\
_{48} &=& 269 N^8+1010 N^7+1558 N^6+2984 N^5+3633 N^4+1950 N^3-420 N^2
        \nonumber \\ &&
        -2632 N-864 \\
P_{49} &=& 6 N^9+24 N^8-6 N^7-138 N^6-191 N^5-422 N^4-927 N^3-526 N^2
        \nonumber \\ &&
        -132 N-136 \\
P_{50} &=& 6 N^9+39 N^8+89 N^7+136 N^6+85 N^5+183 N^4+358 N^3+344 N^2
        \nonumber \\ &&
        +440 N+112 
\end{eqnarray}

\begin{figure}[H]
\centering
\includegraphics[width=0.8\textwidth]{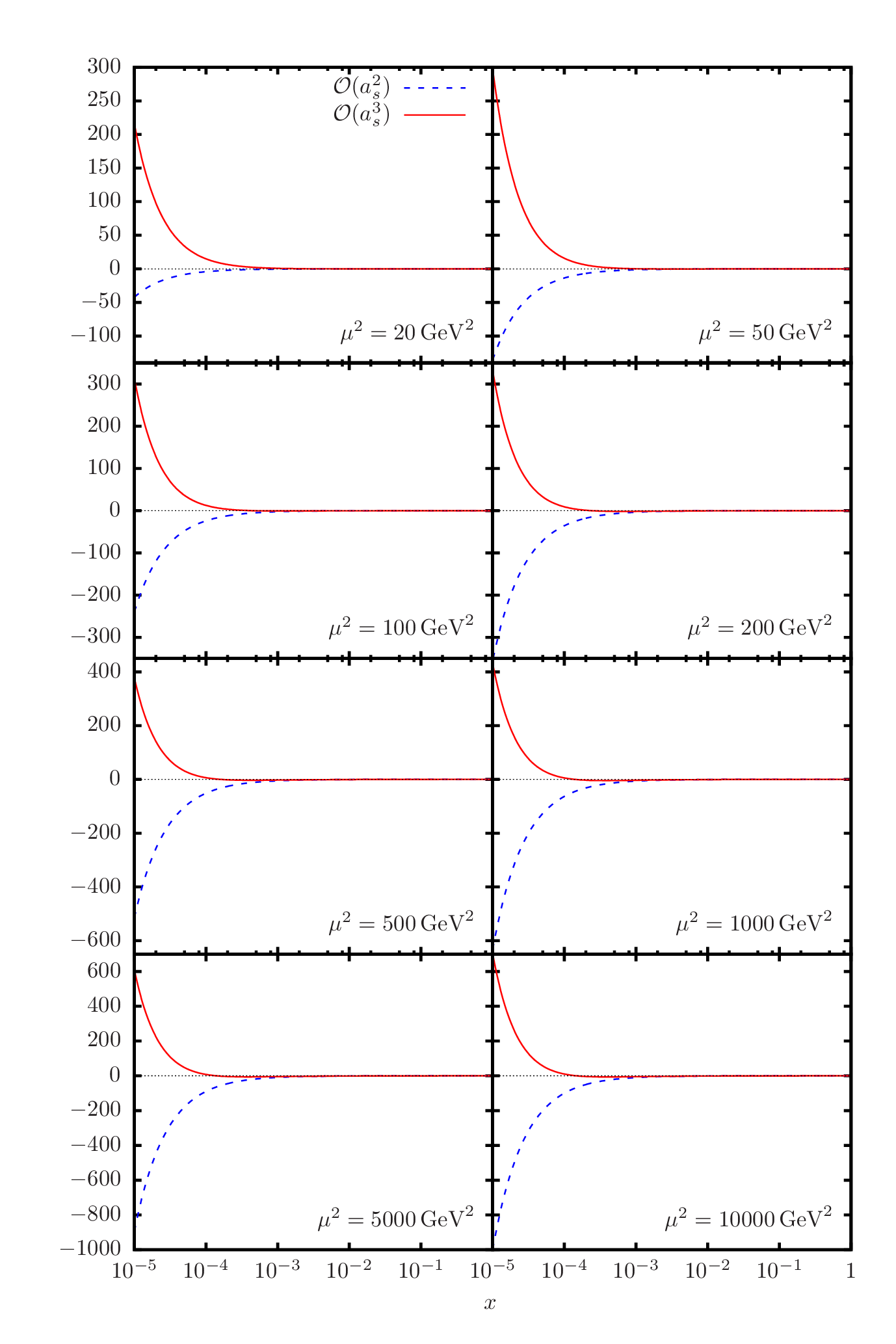}
\caption{\sf \small The OME $A_{Qq}^{\rm PS}$ up to  2- and  3-loop order in dependence of $x$ and $\mu^2$ for a 
heavy quark mass $m_c = 1.59~\GeV$ (on-shell scheme), cf.~\cite{Alekhin:2012vu}.}
\label{fig:A1}
\end{figure}
\begin{figure}[H]
\centering
\includegraphics[width=0.8\textwidth]{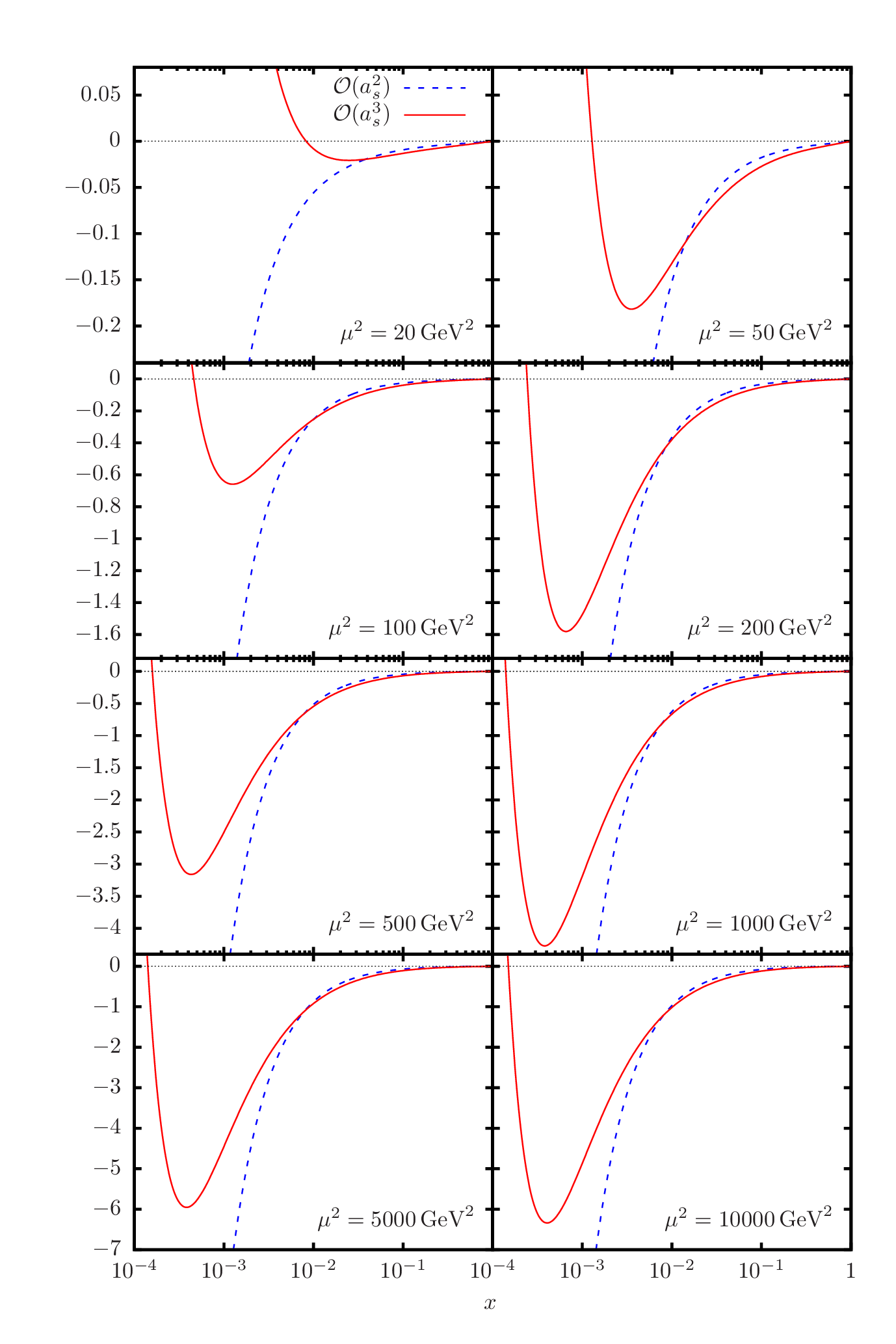}
\caption{\sf \small The OME $A_{Qq}^{\rm PS}$ up to  2- and  3-loop order in dependence of $x$ and $\mu^2$ for a 
heavy quark mass $m_c = 1.59~\GeV$ (on-shell scheme) at larger values of $x$.}
\label{fig:A2}
\end{figure}
\begin{eqnarray}
P_{51} &=& 6 N^9+39 N^8+105 N^7+88 N^6-91 N^5-329 N^4-410 N^3-344 N^2
        \nonumber \\ &&
        -264 N-80 \\
P_{52} &=& 72 N^9+432 N^8+965 N^7+757 N^6-729 N^5-3193 N^4-4848 N^3-1968 N^2
        \nonumber \\ &&
        +528 N+16 \\
P_{53} &=& N^{10}+8 N^9+29 N^8+49 N^7-11 N^6-131 N^5-161 N^4-160 N^3-168 N^2
        \nonumber \\ &&
        -80 N-16 \\
P_{54} &=& 3 N^{10}+39 N^9+111 N^8-27 N^7-692 N^6-1390 N^5-1232 N^4-636 N^3
        \nonumber \\ &&
        -248 N^2+80 N+96 \\
P_{55} &=& 5 N^{10}+32 N^9+46 N^8+82 N^7-137 N^6-658 N^5-1114 N^4-2576 N^3
        \nonumber \\ &&
        -3680 N^2-1952 N-416 \\
P_{56} &=& 8 N^{10}+133 N^9+564 N^8-720 N^7-9202 N^6-18333 N^5-13074 N^4
        \nonumber \\ &&
        -10744 N^3-5512 N^2+19440 N+14400 \\
P_{57} &=& 9 N^{10}-218 N^8-323 N^7+1211 N^6-398 N^5-5724 N^4-1035 N^3+810 N^2
        \nonumber \\ &&
        +76 N+984 \\
P_{58} &=& 19 N^{10}+143 N^9+427 N^8+567 N^7+454 N^6+822 N^5+1560 N^4+1784 N^3
        \nonumber \\ &&
        +1488 N^2+768 N+192 \\
P_{59} &=& 36 N^{10}+169 N^9+33 N^8-1407 N^7-4051 N^6-6392 N^5-8176 N^4-8212 N^3
        \nonumber \\ &&
        -5560 N^2-2736 N-736 \\
P_{60} &=& 43 N^{10}+320 N^9+939 N^8+912 N^7-218 N^6-510 N^5-654 N^4-1232 N^3
        \nonumber \\ &&
        +16 N^2+672 N+288 \\
P_{61} &=& 43 N^{10}+320 N^9+1059 N^8+1914 N^7+2431 N^6+2874 N^5+2379 N^4+820 N^3
        \nonumber \\ &&
        +352 N^2+336 N+144 \\
P_{62} &=& 104 N^{10}+1729 N^9+10752 N^8+31392 N^7+48422 N^6+57231 N^5+75450 N^4
        \nonumber \\ &&
        +59408 N^3+28136 N^2+47376 N+31680 \\
P_{63} &=& 135 N^{10}+702 N^9+1547 N^8+1319 N^7+553 N^6+2150 N^5-3213 N^4-6735 N^3
        \nonumber \\ &&
        -7854 N^2-7492 N-1272 \\
P_{64} &=& 136 N^{10}+647 N^9+1110 N^8-438 N^7-2555 N^6-2106 N^5-3105 N^4-3167 N^3
        \nonumber \\ &&
        +418 N^2+924 N+72 \\
P_{65} &=& 19 N^{11}-17 N^{10}+190 N^9+1350 N^8+1060 N^7-4480 N^6-12285 N^5-13625 N^4
        \nonumber \\ &&
        -5556 N^3+2768 N^2+4512 N+1872 \\
P_{66} &=& 118 N^{11}+793 N^{10}+2281 N^9+3402 N^8+2428 N^7+1457 N^6+1917 N^5
        \nonumber \\ &&
        +2476 N^4+4392 N^3+4976 N^2+2832 N+576 \\
P_{67} &=& 1669 N^{11}+10399 N^{10}+26752 N^9+36576 N^8+33436 N^7+39590 N^6
        \nonumber \\ &&
        +33039 N^5+8815 N^4+27708 N^3+47504 N^2+33312 N+8784 \\
P_{68} &=& 37 N^{12}+305 N^{11}+1107 N^{10}+2328 N^9+3520 N^8+5020 N^7
        \nonumber \\ &&
        +7642 N^6+10519 N^5+10938 N^4+8248 N^3+4656 N^2+1712 N+288 \\
P_{69} &=& 18 N^{13}+193 N^{12}+900 N^{11}+2378 N^{10}+3486 N^9+2817 N^8+2052 N^7+2256 N^6
        \nonumber \\ &&
        +2804 N^5+7272 N^4+12512 N^3+10304 N^2+4672 N+896 \\
P_{70} &=& 25 N^{13}+1016 N^{12}+11804 N^{11}+63190 N^{10}+184075 N^9+321474 N^8
        \nonumber \\ &&
        +375092 N^7+324832 N^6+221884 N^5+205760 N^4+302240 N^3+288576 N^2
        \nonumber \\ &&
        +153792 N+34560 \\
P_{71} &=& 158 N^{13}+1663 N^{12}+7309 N^{11}+17981 N^{10}+35774 N^9+59586 N^8+56374 N^7
        \nonumber \\ &&
        +23504 N^6+25457 N^5+30298 N^4-11384 N^3-30000 N^2-18864 N-4320 \\
P_{72} &=& 77 N^{14}+1046 N^{13}+7131 N^{12}+35512 N^{11}+87723 N^{10}+89530 N^9+46927 N^8
        \nonumber \\ &&
        +41002 N^7-194958 N^6-644698 N^5-589404 N^4-123376 N^3+61248 N^2
        \nonumber \\ &&
        +22752 N-1728 \\
P_{73} &=& 686 N^{14}+6560 N^{13}+25572 N^{12}+43489 N^{11}+9045 N^{10}-72944 N^9-125240 N^8
        \nonumber \\ &&
        -156761 N^7-206883 N^6-241600 N^5-250212 N^4-225808 N^3-150864 N^2
        \nonumber \\ &&
        -56448 N-8640 \\
P_{74} &=& 100 N^{15}+1170 N^{14}+6234 N^{13}+20518 N^{12}+49217 N^{11}+94274 N^{10}
        \nonumber \\ &&
        +145788 N^9+172682 N^8+139145 N^7+47068 N^6-50228 N^5-96416 N^4
        \nonumber \\ &&
        -82448 N^3-41536 N^2-11968 N-1536 \\
P_{75} &=& 158 N^{16}+6799 N^{15}+93011 N^{14}+633970 N^{13}+2547481 N^{12}+6605953 N^{11}
        \nonumber \\ &&
        +11841596 N^{10}+15808910 N^9+17140651 N^8+16081262 N^7+12756671 N^6
        \nonumber \\ &&
        +7253426 N^5+1318688 N^4-2323728 N^3-2738448 N^2-1334880 N-259200 \\
P_{76} &=& 2272 N^{17}+27343 N^{16}+135485 N^{15}+332260 N^{14}+398250 N^{13}+111012 N^{12}
        \nonumber \\ &&
        -530356 N^{11}-1134420 N^{10}-86378 N^9+3545573 N^8+7139427 N^7+8691144 N^6
        \nonumber \\ &&
        +9505284 N^5+9549872 N^4+7324752 N^3+3612672 N^2+1017792 N+124416
\end{eqnarray}

The difference between the OME in the $\overline{\rm MS}$-scheme and the on-shell scheme in $N$-space is given by
\begin{eqnarray}
\lefteqn{A_{Qq}^{\rm PS, \overline{\rm MS}}(N)
- A_{Qq}^{\rm PS, OMS}(N) =} \hspace*{2.3cm} \nonumber \\ &&
\textcolor{blue}{C_F^2 T_F} \Biggl\{
\frac{48 \big(N^2+N+2\big)^2}{(N-1) N^2 (N+1)^2 (N+2)} 
\ln^2\left(\frac{m^2}{\mu^2}\right)
\nonumber\\ &&
        -\frac{16   \big(
                4 N^7+20 N^6+37 N^5-4 N^4-43 N^3-34 N^2-52 N-24\big)}{(N-1) N^3 (N+1)^3 (N+2)^2}
\ln\left(\frac{m^2}{\mu^2}\right)
\nonumber\\ &&
        -\frac{64 \big(5 N^3+7 N^2+4 N+4\big)
\big(
                N^2+5 N+2\big)}{(N-1) N^3 (N+1)^3 (N+2)^2} 
\Biggr\}.
\end{eqnarray}
The corresponding expression in $x$-space reads
\begin{eqnarray}
A_{Qq}^{\rm PS, \overline{\rm MS}}(x)
- A_{Qq}^{\rm PS, OMS}(x)
&=& 
\textcolor{blue}{C_F^2 T_F} 
\Biggl\{
\Biggl[
                \frac{16 (1-x) \big(
                        4 x^2+7 x+4\big)}{x}
                +96 (x+1) H_0
        \Biggr] \ln^2\left(\frac{m^2}{\mu^2}\right)
\nonumber\\ &&
+ \Biggl[
                \frac{1}{x} \frac{32}{3} (1-x) (4 x-1) (5 x-2)
                +16 \big(
                        8 x^2+7 x-5\big) H_0
\nonumber\\ &&
                -48 (x+1) H_0^2
        \Biggr] \ln\left(\frac{m^2}{\mu^2}\right)
        -\frac{64}{3} \big(
                8 x^2+15 x+3\big) H_0
\nonumber\\ &&
        +64 (x+1) H_0^2
- \frac{128 (1-x) \big(28 x^2+x+10\big)}{9 x}
\Biggr\}.
\end{eqnarray}
Here we have set the heavy quark mass equal in both schemes to obtain a more compact expression. The corresponding representations
of the quark masses are given in \cite{MASS}.

The analytic continuation of the expressions in Mellin-$N$ space can be obtained using the asymptotic expressions, which can be derived 
in analytic form \cite{Blumlein:2009ta,Ablinger:2013cf}, and the recurrence relations. Alternatively, the Mellin-inversion can be performed
analytically and one may work in $x$-space. The corresponding relations for the OME are given in Appendix~\ref{appC}.

In the numerical representation of the following figures the harmonic polylogarithms were calculated using the code 
{\tt HPLOG5} \cite{Gehrmann:2001pz}.
In Figure~\ref{fig:A1} the massive OME $A_{Qq}^{\rm PS}$ is shown as a function of $x$ and $Q^2$ in the range of lower values of $x$.
Here and in the following we illustrate the corrections to $O(a_s^2)$ and up to $O(a_s^3)$ using next-to-next-to-leading order 
(NNLO) parton distribution functions and 
the value of $a_s$ at NNLO, i.e. the term $O(a_s^2)$ refers not to the next-to-leading order (NLO) correction, 
but to the $O(a_s^2)$ in the NNLO correction. We refer to the value of $a_s(\mu^2)$ as given in the parameterization 
\cite{Alekhin:2013nda} using the corresponding {\tt LHAPDF} library. 
While in the small-$x$ region the $O(a_s^2)$ term is negative, the NNLO correction is positive. 
Here and also in case of the heavy flavor Wilson 
coefficient in Section~\ref{sec:6}, one has to apply a Mellin convolution with the singlet quark densities at NNLO for a prediction
in the variable flavor number scheme and in calculating the contribution to the structure function $F_2(x,Q^2)$ in the fixed flavor number 
scheme. Since the singlet distribution is decreasing towards larger values of $x$, the medium-$x$ behaviour of the OME is important for 
the physical effect. Therefore, we enlarge the behaviour of the OME in the region of medium and large values of $x$
in Figure~\ref{fig:A2}. The $O(a_s^2)$ term remains negative and the NNLO correction turns to negative values approaching zero in the 
limit $x \rightarrow 1$ from below. At larger values of $x$ the NNLO correction becomes smaller than the $O(a_s^2)$  term. Unlike in the 
non-singlet case, we cannot present yet the matching in the VFNS to 3-loop order, since also the OME $A_{Qg}^{(3)}$ contributes here.
\section{The Pure Singlet Wilson Coefficient in the Asymptotic Region}
\label{sec:6}

\vspace*{1mm}
\noindent
The pure singlet Wilson coefficient in the asymptotic region is given by Eq.~(\ref{eq:WILS}). It also depends on the scale 
ratio $Q^2/\mu^2$ via
\begin{eqnarray}
L_Q = \ln\left(\frac{Q^2}{\mu^2}\right)~.
\end{eqnarray}
In Mellin--$N$ space it is given by


and the polynomials $P_i$ are given by
\begin{eqnarray}
P_{77}    &=& N^4+2 N^3+7 N^2+22 N+20 \\
P_{78}    &=& N^5+9 N^4+24 N^3+36 N^2+32 N+8 \\
P_{79}    &=& 11 N^5+26 N^4+57 N^3+142 N^2+84 N+88 \\
P_{80}    &=& 5 N^6+135 N^5+327 N^4+329 N^3+220 N^2-176 N-120 \\
P_{81} &=& 16 N^6+35 N^5+33 N^4-11 N^3-41 N^2-36 N-12 \\
P_{82} &=& 17 N^6-57 N^5-213 N^4-175 N^3-140 N^2+64 N+72 \\
P_{83} &=& N^7-15 N^5-58 N^4-92 N^3-76 N^2-48 N-16 \\
P_{84} &=& 2 N^7+14 N^6+37 N^5+102 N^4+155 N^3+158 N^2+132 N+40 \\
P_{85} &=& 3 N^7-15 N^6-153 N^5-577 N^4-854 N^3-652 N^2-408 N-128 \\
P_{86} &=& 5 N^7+19 N^6+61 N^5+197 N^4+266 N^3+212 N^2+136 N+32 \\
P_{87} &=& 7 N^7+21 N^6+5 N^5-117 N^4-244 N^3-232 N^2-192 N-80 \\
P_{88} &=& 9 N^7+15 N^6-103 N^5-575 N^4-998 N^3-948 N^2-696 N-256 \\
P_{89} &=& 11 N^7+37 N^6+53 N^5+7 N^4-68 N^3-56 N^2-80 N-48 \\
P_{90} &=& 25 N^7+91 N^6+101 N^5-195 N^4-546 N^3-556 N^2-520 N-224 \\
P_{91} &=& 99 N^7+379 N^6+553 N^5+465 N^4+232 N^3-256 N^2-688 N-336 \\
P_{92} &=& N^8+8 N^7+8 N^6-14 N^5-53 N^4-82 N^3+60 N^2+104 N+96 \\
P_{93} &=& 6 N^8-42 N^7-241 N^6-579 N^5-307 N^4+477 N^3+602 N^2+492 N+168 \\
P_{94} &=& 2 N^9+7 N^8+17 N^7+30 N^6+83 N^5+193 N^4+220 N^3+136 N^2+64 N+16 \\
P_{95} &=& 15 N^9+24 N^8-174 N^7-659 N^6-997 N^5-749 N^4-156 N^3+256 N^2
        \nonumber \\ &&
        +320 N+144 \\
P_{96} &=& 19 N^9+86 N^8+144 N^7-38 N^6-535 N^5-1016 N^4-1180 N^3-872 N^2
        \nonumber \\ &&
        -416 N-96 \\
P_{97} &=& 9 N^{10}-218 N^8-350 N^7+1238 N^6-317 N^5-5643 N^4-981 N^3+594 N^2
        \nonumber \\ &&
        +76 N+984 \\
P_{98} &=& 19 N^{10}+143 N^9+412 N^8+426 N^7-N^6+159 N^5+1066 N^4+1552 N^3
        \nonumber \\ &&
        +1456 N^2+848 N+224 \\
P_{99} &=& 20 N^{10}+111 N^9+219 N^8-3 N^7-331 N^6+920 N^5+3712 N^4+5080 N^3
        \nonumber \\ &&
        +4192 N^2+2272 N+576 \\
P_{100} &=& 47 N^{10}+823 N^9+5739 N^8+21510 N^7+53459 N^6+105381 N^5+160023 N^4
        \nonumber \\ &&
        +158774 N^3+104300 N^2+56664 N+18720 \\
P_{101} &=& 60 N^{10}+340 N^9+594 N^8-204 N^7-2167 N^6-4496 N^5-7339 N^4-8524 N^3
        \nonumber \\ &&
        -6112 N^2-3024 N-784 \\
P_{102} &=& 67 N^{10}+383 N^9+867 N^8+696 N^7-755 N^6-2391 N^5-3027 N^4-2744 N^3
        \nonumber \\ &&
        -1256 N^2-48 N+144 \\
P_{103} &=& 85 N^{10}+482 N^9+1146 N^8+1272 N^7+532 N^6+840 N^5+2427 N^4+2440 N^3
        \nonumber \\ &&
        +1768 N^2+1248 N+432 \\
P_{104} &=& 95 N^{10}+1621 N^9+10419 N^8+32166 N^7+55847 N^6+78615 N^5+111963 N^4
        \nonumber \\ &&
        +100934 N^3+57980 N^2+61560 N+36000 \\
P_{105} &=& 118 N^{10}+675 N^9+1588 N^8+1652 N^7+326 N^6+357 N^5+876 N^4+1672 N^3
        \nonumber \\ &&
        +3440 N^2+2544 N+576 \\
P_{106} &=& 127 N^{10}+644 N^9+1113 N^8-372 N^7-4016 N^6-4578 N^5-558 N^4+2008 N^3
        \nonumber \\ &&
        +2848 N^2+2496 N+864 \\
P_{107} &=& 151 N^{10}+708 N^9+1156 N^8+464 N^7-967 N^6+372 N^5+3672 N^4+5236 N^3
        \nonumber \\ &&
        +6152 N^2+3792 N+864 \\
P_{108} &=& 118 N^{11}+649 N^{10}+1996 N^9+5922 N^8+14389 N^7+26096 N^6+33057 N^5
        \nonumber \\ &&
        +29305 N^4+19668 N^3+8048 N^2+2016 N+432 \\
P_{109} &=& 37 N^{12}+305 N^{11}+1017 N^{10}+1462 N^9+592 N^8+408 N^7+4064 N^6+9645 N^5
        \nonumber \\ &&
        +12222 N^4+10280 N^3+6064 N^2+2192 N+352 \\
P_{110} &=& 45 N^{13}+485 N^{12}+2289 N^{11}+6064 N^{10}+8448 N^9+4398 N^8-1602 N^7
        \nonumber \\ &&
        -2715 N^6-584 N^5+9300 N^4+22624 N^3+21232 N^2+10112 N+1984 \\
P_{111} &=& 82 N^{13}+2471 N^{12}+27848 N^{11}+164605 N^{10}+597268 N^9+1483293 N^8
        \nonumber \\ &&
        +2732000 N^7+3846211 N^6+4059946 N^5+3144284 N^4+1798280 N^3
        \nonumber \\ &&
        +756000 N^2+222912 N+34560 \\
P_{112} &=& 83 N^{14}+636 N^{13}+1484 N^{12}-505 N^{11}-7588 N^{10}-8082 N^9+12896 N^8
        \nonumber \\ &&
        +30199 N^7-2799 N^6-73072 N^5-117444 N^4-105808 N^3-62992 N^2
        \nonumber \\ &&
        -23424 N-4032 \\
P_{113} &=& 1790 N^{14}+13034 N^{13}+34014 N^{12}+16729 N^{11}-108615 N^{10}-261746 N^9
        \nonumber \\ &&
        -246794 N^8-165593 N^7-316791 N^6-606160 N^5-724860 N^4-602224 N^3
        \nonumber \\ &&
        -352272 N^2-124416 N-19008 \\
P_{114} &=& 73 N^{15}+867 N^{14}+4698 N^{13}+16255 N^{12}+43958 N^{11}+97502 N^{10}+165558 N^9
        \nonumber \\ &&
        +200747 N^8+161729 N^7+60265 N^6-48800 N^5-106628 N^4-94640 N^3
        \nonumber \\ &&
        -48016 N^2-13696 N-1728 \\
P_{115} &=& 229 N^{16}-49 N^{15}-48956 N^{14}-530524 N^{13}-2816896 N^{12}-9419641 N^{11}
        \nonumber \\ &&
        -22464935 N^{10}-41400392 N^9-60928891 N^8-70644896 N^7-62314487 N^6
        \nonumber \\ &&
        -39968930 N^5-16753760 N^4-2474640 N^3+1995408 N^2+1334880 N
        \nonumber \\ &&
        +259200.
\end{eqnarray}

\noindent
In Appendix~\ref{appC} we present the corresponding expression in $x$-space. Again one may also represent the structure function
in Mellin-$N$ space, using the corresponding evolution operators, cf. e.g.~\cite{Blumlein:1997em} and Wilson coefficient. This requests 
the
representation of the corresponding (generalized) harmonic sums for $N \in \mathbb{C}$, which can be realized using their analytic
asymptotic representation to high accuracy and the known shift relations \cite{Blumlein:2009ta,Ablinger:2013cf}.

\begin{figure}[H]
\centering
\includegraphics[width=0.85\textwidth]{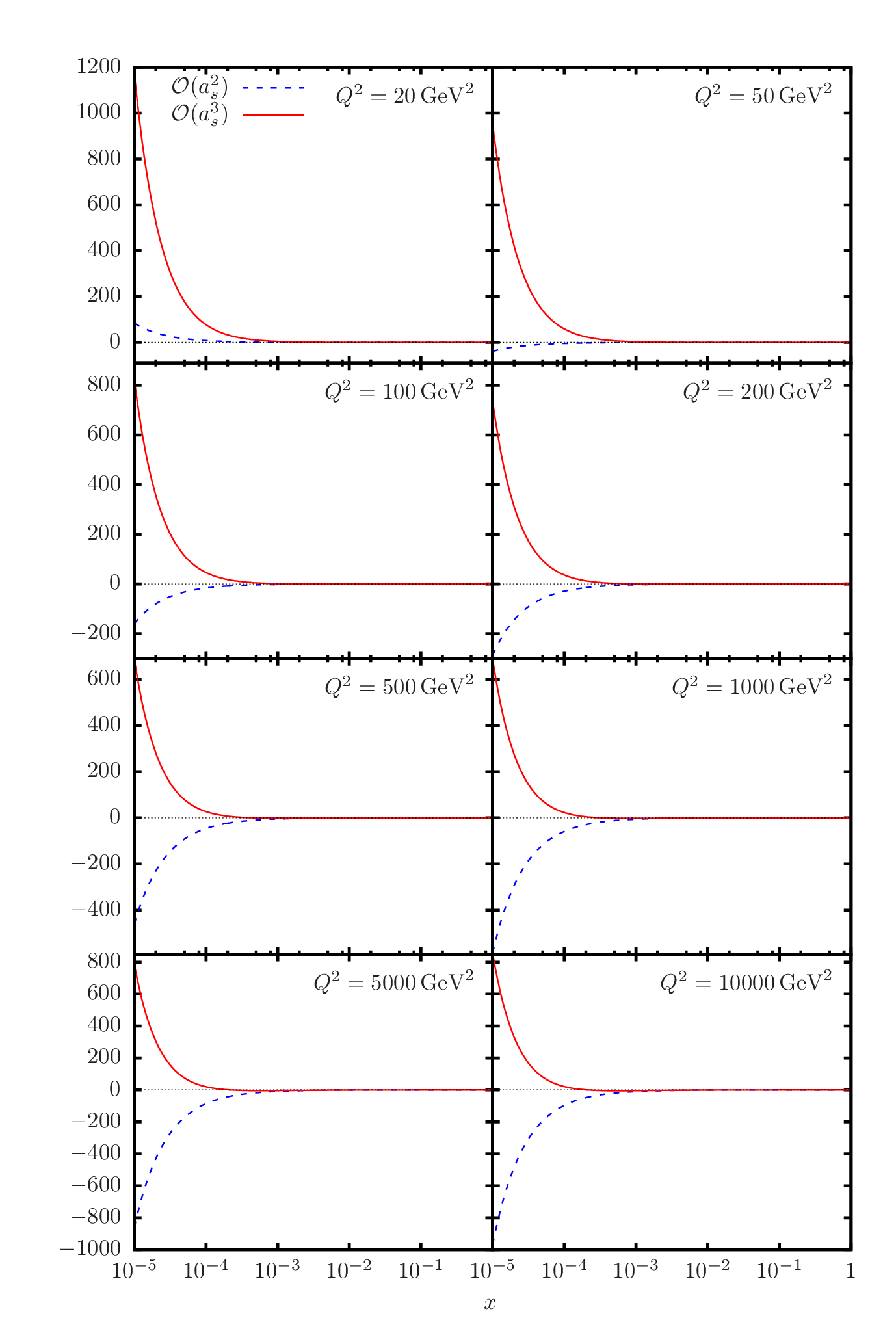}
\caption{\sf \small The Wilson coefficient $H_{Q,2}^{\rm PS}$ as a function of $x$ and $Q^2$ choosing $Q^2 = \mu^2$ and the heavy 
quark mass $m_c = 1.59~\GeV$ (on-shell scheme) using the parton distribution functions \cite{Alekhin:2013nda}.}
\label{FIG:WILS}
\end{figure}

\begin{figure}[H]
\centering
\includegraphics[width=0.95\textwidth]{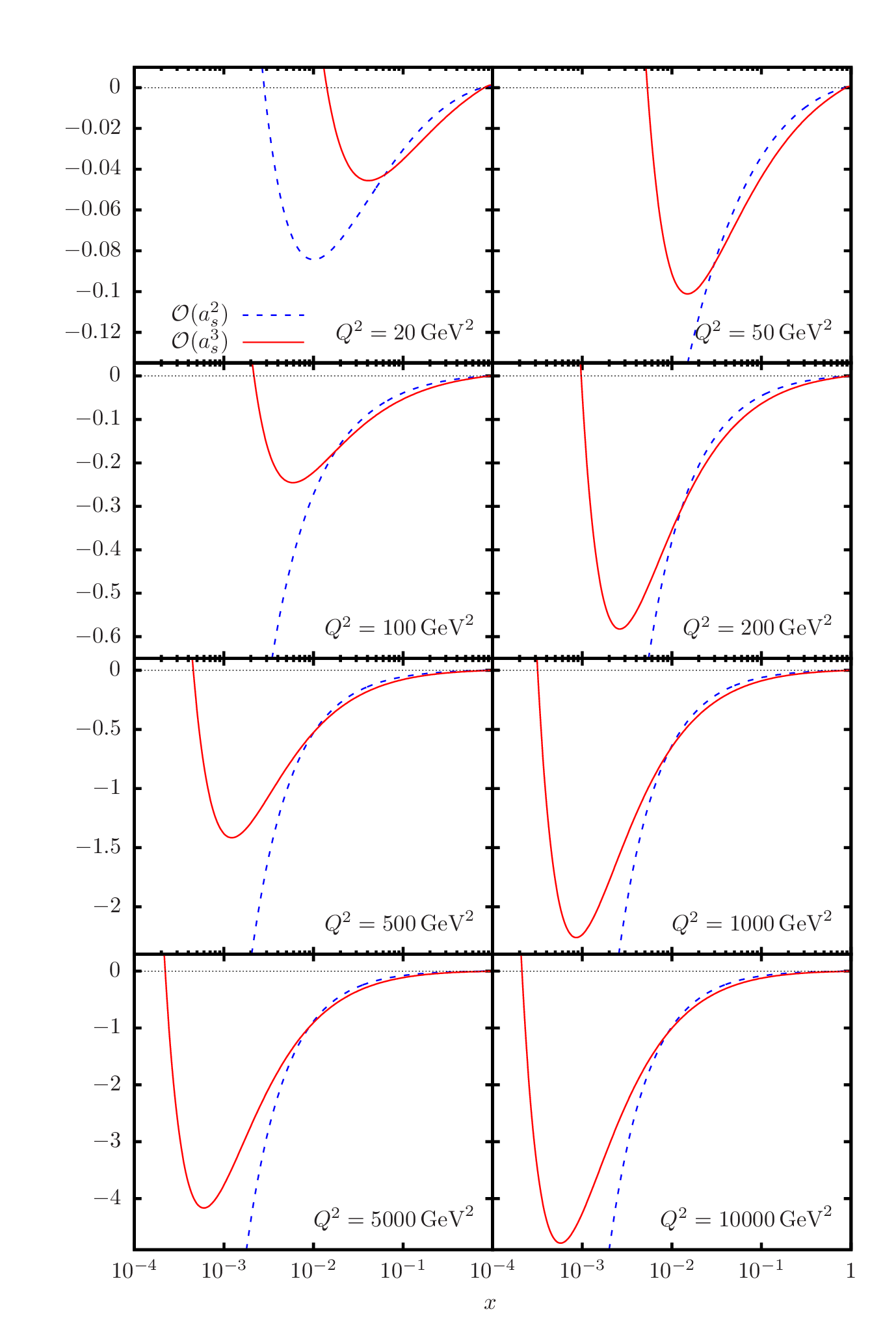}
\caption{\sf \small 
The Wilson coefficient $H_{Q,2}^{\rm PS}$ as a function of $x$ and $Q^2$ choosing $Q^2 = \mu^2$ and the heavy 
quark mass $m_c = 1.59~\GeV$ (on-shell scheme) at large values of $x$ using the parton distribution functions 
\cite{Alekhin:2013nda}.}
\label{FIG:WILS1}
\end{figure}

Figure~\ref{FIG:WILS} shows the $x$- and 
$Q^2$-dependence of the pure singlet Wilson coefficient up to 2- and 3-loop order at $Q^2 = \mu^2$ in the region of smaller values of $x$.
While at low scales $Q^2 \simeq 20 \GeV^2$ the $O(a_s^2)$ term is positive in this region, it turns negative at higher scales. On the 
other hand,
the corrections up to $O(a_s^3)$ are positive. At larger values of $x$, similar to the behaviour of the OME, also the corrections up to 
NNLO turn negative and, at even larger values of $x$ undershoot the $O(a_s^2)$ contributions turning to zero for $x \rightarrow 0$, as 
shown in Figure~\ref{FIG:WILS1}.

\begin{figure}[H]
\centering
\includegraphics[width=0.80\textwidth]{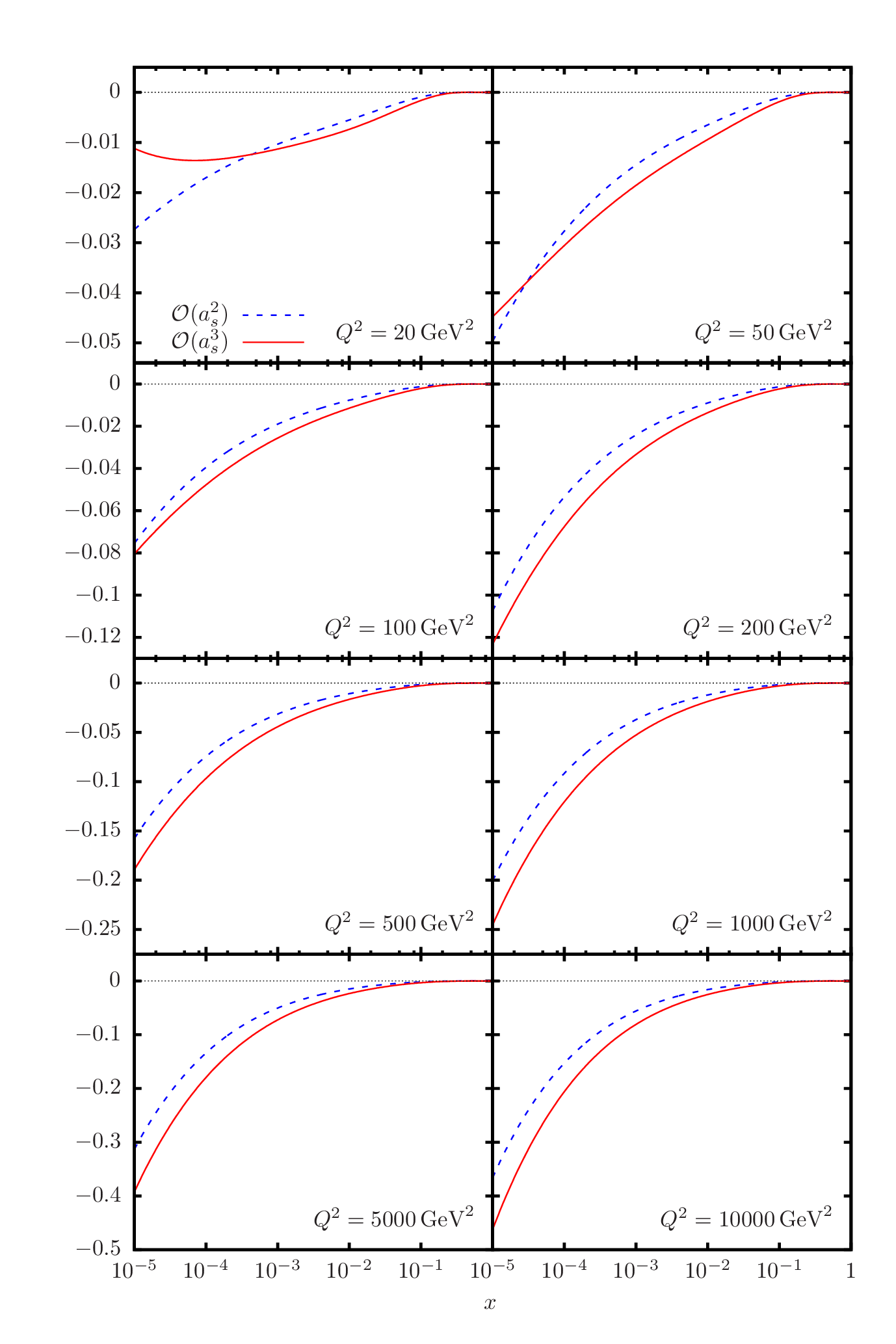}
\caption{\sf \small The charm contribution by the Wilson coefficient $H_{Q,2}^{\rm PS}$ to the structure function 
$F_2(x,Q^2)$ as a function of $x$ and $Q^2$ choosing $Q^2 = \mu^2, m_c = 1.59 \GeV$ (on-shell scheme) using the parton distribution 
functions \cite{Alekhin:2013nda}.}
\label{FIG:F2c}
\end{figure}

\begin{figure}[H]
\centering
\includegraphics[width=0.95\textwidth]{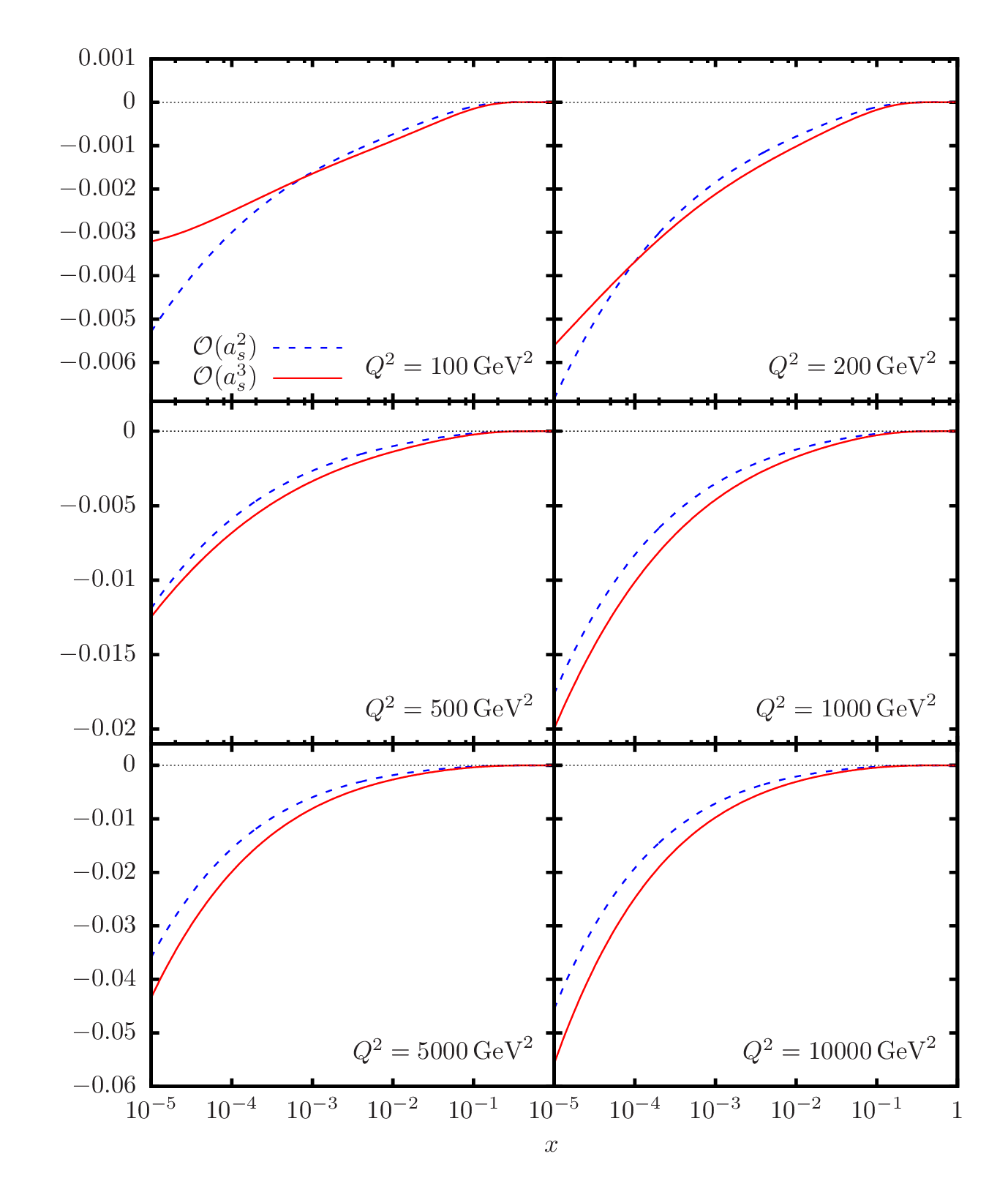}
\caption{\sf \small The bottom contribution by the Wilson coefficient $H_{Q,2}^{\rm PS}$ to the structure function 
$F_2(x,Q^2)$ as a function of $x$ and $Q^2$ choosing $Q^2 = \mu^2, m_b = 4.78 \GeV$ (on shell scheme \cite{PDG}) using the parton 
distribution 
functions \cite{Alekhin:2013nda}.}
\label{FIG:F2b}
\end{figure}

The latter behaviour is of importance for the Mellin convolution with the quark-singlet distribution function. This is displayed
in the contribution of the heavy flavor pure singlet contribution to the structure function $F_2(x,Q^2)$ in the case of the single 
heavy quark
contributions at $O(a_s^2)$ and up to $O(a_s^3)$ in Figures~\ref{FIG:F2c} and \ref{FIG:F2b}, respectively, again setting $\mu^2 = 
Q^2$ and referring to the on-shell masses\footnote{A thorough numerical study of $F_2(x,Q^2)$ for the scale dependence in 
$\mu^2$ requests to consider also the gluonic contributions due to the mixing in the singlet sector and will be carried out at a later 
stage of the present project.}. We apply the same setting for $a_s$ as in the case of the OME $A_{Qq}^{\rm PS}$.
Both corrections are negative. At lower scales $Q^2$, the $O(a_s^2)$ effects are larger than those at NNLO, which give larger negative
corrections at higher scales. To quote a few numbers, we obtain using the parton distribution functions \cite{Alekhin:2013nda} at
$x = 10^{-4}$ and $Q^2 = 100 \GeV^2$ a correction of $-0.026$ in case of charm and $-0.0015$ for bottom. At a given value of $Q^2$ and the 
lowest value of $x$ in the kinematic 

\noindent
range at HERA, $x_0 = Q^2/(s y)$ for $y = 1, s \simeq 10^5 \GeV^2$, one obtains the following 
corrections~: $-0.013~(x_0= 0.0001), -0.026~(x_0 = 0.005), -0.023~(x_0 = 0.01), -0.019~(x_0 = 0.02), -0.007~(x_0 = 0.05)$ and $0.003~(x_0 
= 0.1)$ for the contribution due to charm. In the whole kinematic region at HERA up to $Q^2 \simeq 10^4 \GeV^2$ the bottom quark 
contribution is about one order of magnitude smaller than that by the charm quark. 

\section{Conclusions}
\label{sec:7}

\vspace*{1mm}
\noindent
We have calculated the $O(a_s^3)$ heavy flavor contributions to the flavor pure singlet OME $A_{Qq}^{(3)}$, contributing to 
the matching relations in the VFNS  and the corresponding heavy flavor Wilson coefficient for single heavy quark flavors in 
the asymptotic region $Q^2 \gg m^2$. As a by-product of the calculation we computed the complete pure singlet anomalous dimension 
at 3-loop order in an independent way for the first time. Our result agrees with the result in Ref.~\cite{Vogt:2004mw}. 
On the technical side of the calculation, we used the integration by parts package {\tt Reduze2} to reduce the Feynman integrals 
carrying local operator insertions to master integrals. The master integrals were calculated using different techniques. 
Most notably, we used differential equations, turned them into difference equations and solved using the packages {\tt Sigma, 
EvaluateMultiSums, SumProduction, HarmonicSums} and {\tt OreSys}. Part of these packages were also used to compute the nested sums
obtained using representations by generalized hypergeometric functions and using Mellin-Barnes techniques. The massive OME and Wilson 
coefficient depend on new functions, which did not occur in the related 3-loop results having been dealt with previously in 
Refs.~\cite{Ablinger:2014lka,Ablinger:2014vwa}. In the present result generalized harmonic sums contribute to the final expression 
in $N$-space. In $x$-space their effect manifests in harmonic polylogarithms of argument $1 - 2x$. We studied the behaviour of the 
constant part of the unrenormalized massive OME, $a_{Qq}^{(3)}$. This quantity has been newly calculated beyond the terms contributing 
to $A_{Qq}^{(3)}$ and $H_{q,2}^{\rm PS}$ implied by renormalization \cite{Bierenbaum:2009mv}. The leading small-$x$ contribution  
of $O(a_s^3 \ln(x)/x)$ is not describing this quantity, not even at values of $x$ in the LHC region. To get physical values in the 
region of $x \simeq 10^{-4}$ one has to add the term $O(a_s^3/x)$, which we also obtained as a by-product of the present calculation. 
To describe the region of larger values of $x \simeq 10^{-2}$ quite a series of logarithmic corrections in $x$ have to be known.
The pure singlet corrections both to structure function $F_2(x,Q^2)$ are negative and are largest in the small $x$ region. The 
corrections due to bottom quarks are about one magnitude smaller than those by the charm quarks. We presented all quantities both in $N$ 
and $x$-space for the use in deep-inelastic data analyses. The relations presented in the present paper can be obtained in computer-readable 
form on request via e-mail to {\tt Johannes.Bluemlein@desy.de}. 
\appendix
\section{Integral families}
\label{K3amAppendix}

\vspace*{1mm}
\noindent
In this appendix, we show the integral families that were implemented 
in order to perform IBP reductions for the operator matrix elements in
the pure singlet case. 

\begin{figure}[H]
\begin{minipage}[c]{0.65\linewidth}
     \includegraphics[width=0.80\textwidth]{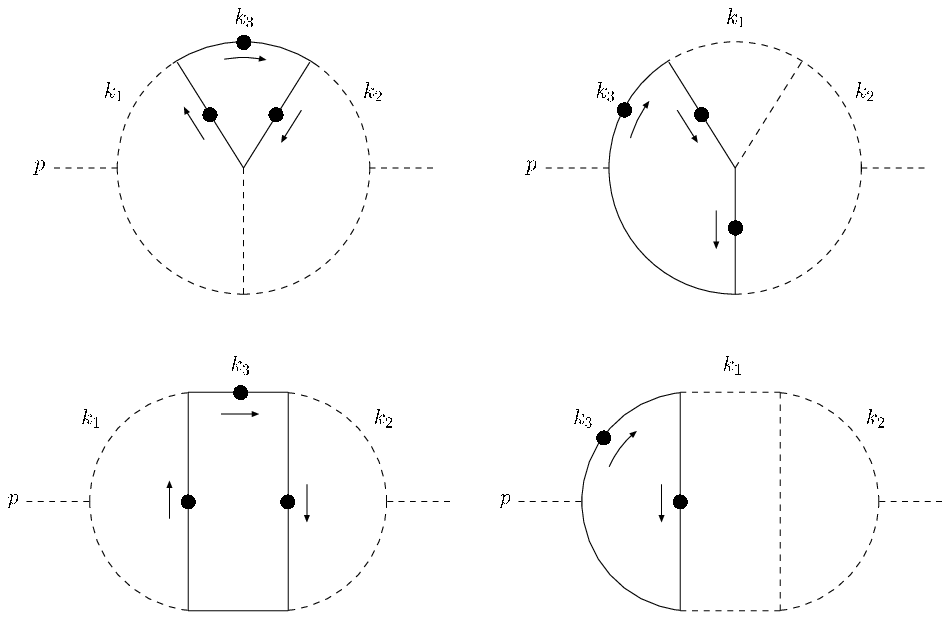}
\vspace*{-4mm}
\end{minipage}
\hspace*{1mm}
\begin{minipage}[c]{0.32\linewidth}
\small{
\begin{tabular}{c|l}
\multicolumn{2}{c}{} \\
\hline
$D_1$ & $k_1^2$ \\
$D_2$ & $(k_1-p)^2$ \\
$D_3$ & $k_2^2$ \\
$D_4$ & $(k_2-p)^2$ \\
$D_5$ & $k_3^2-m^2$ \\
$D_6$ & $(k_3-k_1)^2-m^2$ \\
$D_7$ & $(k_3-k_2)^2-m^2$ \\
$D_8$ & $(k_1-k_2)^2$ \\
$D_9$ & $(k_3-p)^2-m^2$ \\
$D_{10}$ & $1-x (\Delta.k3 - \Delta.k_1)$ \\
$D_{11}$ & $1-x \Delta.k_3$ \\
$D_{12}$ & $1-x (\Delta.k3 - \Delta.k_2)$ \\
\hline
\end{tabular}
}
\end{minipage}
\end{figure}

\vspace*{1mm}
\begin{center}
\small Family {\sf B1a} 
\end{center}

\noindent
We give the corresponding set of propagators and
depict the different topologies that are covered by each integral family,
although only a few of these topologies are actually related to
$A_{Qq}^{(3), \rm PS}$, the rest being related to other OMEs. In the same
way as for the diagrams on the right-hand side of Figure~\ref{Corresp}, a
given bilinear propagator is indicated by a large dot on the corresponding
line, and since the direction of the momentum is important in these
propagators, this is also indicated in the diagrams. A solid line in the
diagrams represents a massive particle (heavy quark), while a dashed
line is a massless particle (gluon or light quark).

The names we have given to the families are arbitrary, of course. There
is, however, a rationale behind the names we have chosen. The family
names shown here start with a {\sf B}, which indicates that these
families are associated with a Benz-like topology. 

\vspace*{3mm}
\begin{figure}[H]
\begin{minipage}[c]{0.65\linewidth}
     \includegraphics[width=0.80\textwidth]{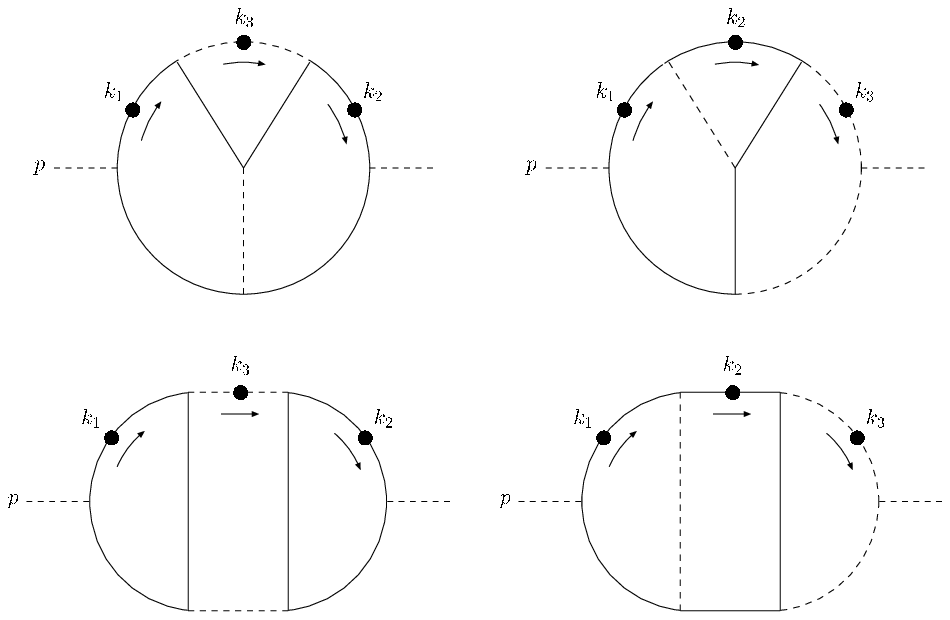}
\vspace*{-4mm}
\end{minipage}
\hspace*{1mm}
\begin{minipage}[c]{0.32\linewidth}
\small{   
\begin{tabular}{c|l}
\multicolumn{2}{c}{} \\
\hline
$D_1$ & $k_1^2-m^2$ \\
$D_2$ & $(k_1-p)^2-m^2$ \\
$D_3$ & $k_2^2-m^2$ \\
$D_4$ & $(k_2-p)^2-m^2$ \\
$D_5$ & $k_3^2$ \\
$D_6$ & $(k_3-k_1)^2-m^2$ \\
$D_7$ & $(k_3-k_2)^2-m^2$ \\
$D_8$ & $(k_1-k_2)^2$ \\
$D_9$ & $(k_3-p)^2$ \\
$D_{10}$ & $1-x \Delta.k_1$ \\
$D_{11}$ & $1-x \Delta.k_3$ \\
$D_{12}$ & $1-x \Delta.k_2$ \\
\hline
\end{tabular}
}
\end{minipage}

\vspace*{1mm}
\begin{center}
\small Family {\sf B5a} 
\end{center}
\end{figure}

\noindent
In addition, they are
constructed such that they also cover ladder topologies. Future calculations
will require also crossed-box (non-planar) topologies, which we have
labeled with names starting with a {\sf C}, and are not shown here.
The number that comes after the {\sf B} (or the {\sf C}) labels different routings
of the mass in the propagators, while the letter that comes after this
number labels the choice of three artificial propagators.

\vspace*{3mm}
\begin{minipage}[c]{0.65\linewidth}
     \includegraphics[width=0.80\textwidth]{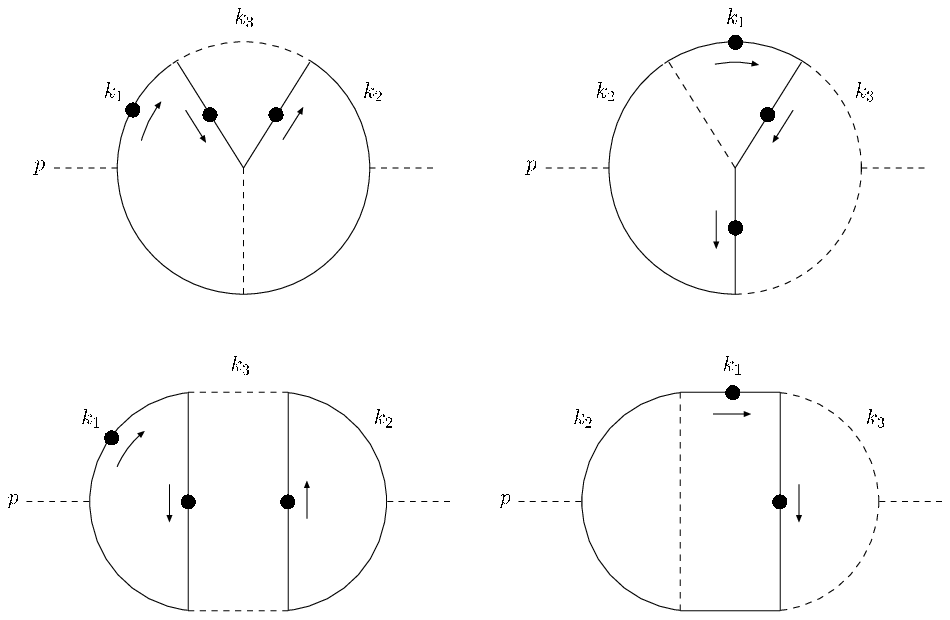}
\vspace*{-4mm}
\end{minipage}
\hspace*{1mm}
\begin{minipage}[c]{0.32\linewidth}
\small{
\begin{tabular}{c|l}
\multicolumn{2}{c}{} \\
\hline
$D_1$ & $k_1^2-m^2$ \\
$D_2$ & $(k_1-p)^2-m^2$ \\
$D_3$ & $k_2^2-m^2$ \\
$D_4$ & $(k_2-p)^2-m^2$ \\
$D_5$ & $k_3^2$ \\
$D_6$ & $(k_3-k_1)^2-m^2$ \\
$D_7$ & $(k_3-k_2)^2-m^2$ \\
$D_8$ & $(k_1-k_2)^2$ \\
$D_9$ & $(k_3-p)^2$ \\
$D_{10}$ & $1-x \Delta.k_1$ \\
$D_{11}$ & $1-x (\Delta.k_1 - \Delta.k_3)$ \\
$D_{12}$ & $1-x (\Delta.k_2 - \Delta.k_3)$ \\
\hline
\end{tabular}
}
\end{minipage}

\vspace*{1mm}
\begin{center}
\small Family {\sf B5c} ($A_{Qq}^{(3), \rm PS}$)
\end{center}  
\section{Integrals}
\label{app:B}

\vspace*{1mm}
\noindent
In the following we present representations of the generalized harmonic sums occurring in the
pure singlet OME as Mellin transforms and in intermediary steps of the calculation, partly with different support, of (generalized) 
harmonic polylogarithms~:
\begin{eqnarray} 
2^N S_1\left(\frac{1}{2},N\right) &=& \int_0^1 dx \frac{x^N}{(x-2)} + 2^N \ln(2) 
\\ 
2^N S_2\left(\frac{1}{2},N\right) &=& - \int_0^1 dx \frac{x^N}{x-2} H_0(x) 
+ 2^{N-1} \left[\zeta_2 - \ln^2(2)\right] 
\\ 
2^N S_3\left(\frac{1}{2}, N\right) &=& \int_0^1 dx \frac{x^N}{2 (x-2)} \ln^2(x) 
+ 2^{N-1} \left[\frac{1}{3} \ln^3(2) - \ln(2) \zeta_2  
+\frac{7}{4} \zeta_3 \right]
\\
S_3\left(2,N\right)                  &=&
2^N \int_0^1 dx \frac{x^N -(1/2)^N}{2 x - 1} H_0^2(x)
\\
S_{1,1}\left(\frac{1}{2},1,N\right)  &=&
2^{-N} \int_0^{1} dx x^N \frac{H_1(x)}{x-2}  + \frac{1}{2} \zeta_2
\\
S_{1,2}\left(\frac{1}{2},1,N\right)  &=&
2^{-N} \int_0^{1} dx x^N \frac{H_0(x) H_1(x) - H_{0,1}(x)}{2-x}  + \frac{5}{8} \zeta_3
\\
S_{2,1}\left(\frac{1}{2},1,N\right)  &=&
\int_0^{1} dx x^N \frac{\zeta_2 - H_{0,1}(x)}{x-2}  + \zeta_3 - 
\frac{1}{2} \ln(2) \zeta_2
\\
S_{1,1}\left(1,\frac{1}{2},N\right)  &=&
 2^{-N} \int_0^{1} dx x^N \frac{H_2(x) - \ln(2)}{x-2} 
+ \ln(2) S_1(N) - \frac{1}{2} \ln^2(2)
\\
S_{2,1}\left(1,\frac{1}{2},N\right)  &=&
 \int_0^{1/2} dx \frac{x^N -1}{x - 1} \left[-1 + \frac{1}{2}\left(\ln^2(2) + \zeta_2\right) - H_{0,1}(x)\right]
\nonumber\\ &&
+ \ln(2) S_1(N)
\\
S_{1,2}\left(1,\frac{1}{2},N\right)  &=&
2^{-N} \int_0^{1} dx x^N \frac{H_{0,2}(x) - H_2(x) H_0(x) + [\ln^2(2) - \zeta_2]/2}{x-2}  
\nonumber\\ &&
+ \frac{1}{2} \left( \zeta_2 - 
\ln^2(2)\right)S_1(N) + \frac{1}{6} \ln^3(2) - \frac{1}{8} \zeta_3
\\
S_{1,3}\left(2,\frac{1}{2},N\right)  &=&
\left(\frac{1}{6} \ln^3(2) - \frac{1}{2} \ln(2) \zeta_2 
+ \frac{7}{8} \zeta_3 \right) \int_1^2 \frac{x^N - 1}{x-1}
+ \int_0^1 dx \frac{x^N - 1}{x-1} 
\nonumber\\ && \times
\left[H_0^2(x) H_2(x)/2 
- H_0(x) H_{0,2}(x) + H_{0,0,2}(x) \right]
\\
S_{1,2}\left(2,1,N\right)            &=& 
2^N \int_0^1 dx (x^N -(1/2)^N)\frac{2(H_0(x)H_1(x) - H_{0,1}(x))}{1-2x}
\\
S_{2,1}\left(2,1,N\right)            &=&
\int_0^2 dx \frac{x^N-1}{x-1} \left[\zeta_2 - H_{0,2}(x)\right]
\\
S_{2,2}\left(2,\frac{1}{2},N\right)  &=&
\int_0^1 dx (x^N -1)\Bigg\{\frac{H_0(x) H_{0,2}(x)
-2 H_{0,0,2}(x)
}{x-1}
+\frac{H_0(x)}{2(x-1)} \big(
        \zeta_2
	-\ln^2(2)
\big)
\nonumber\\
&&+\frac{1}{x-1}  \Biggl[
        \frac{\ln^3(2)}{3}
        -\ln(2)\zeta_2
        +\frac{7 \zeta_3}{4}
\Biggr]\Biggr\}
\nonumber\\ 
&& +2^N \int_0^1 dx (x^N -(1/2)^N)\frac{H_0(x)}{2 x-1} \big(
        \ln^2(2)
        -\zeta_2
\big)
\\
S_{3,1}\left(2,\frac{1}{2},N\right)  &=& \int_0^1 dx (x^N-1) \Biggl\{
\frac{-\tfrac{\ln^3(2)}{6} 
+\tfrac{\ln(2) \zeta_2}{2}
-\tfrac{7 \zeta_3}{8}
- \tfrac{1}{2} \ln(2) H_0^2(x) + H_{0,0,2}(x)
}{x-1}
\nonumber
\end{eqnarray}
\begin{eqnarray}
&&
-\frac{H_0(x)}{x-1} \left(
        -\frac{\ln^2(2)}{2}
        +\frac{\zeta_2}{2}
\right)
\Biggr\}
\nonumber\\ &&
+2^N \int_0^1 dx (x^N-(1/2)^N)\frac{\ln(2) H_0^2(x)}{2 x-1}
\\
S_{1,1,1}\left(\frac{1}{2},1,1,N\right) &=&
2^{-N-1} \int_0^1 dx x^N \frac{H_1^2(x)}{x-2} 
+\frac{3}{4} \zeta_3
\\
S_{1,1,1}\left(1,\frac{1}{2},1,N\right) &=&
2^{-N} \int_0^1 dx x^N \frac{H_{2,1}(x) - \tfrac{\zeta_2}{2}}{x-2}
        -\frac{\ln(2) \zeta_2}{2}
        +\frac{\zeta_3}{8}
+{\zeta_2} S_1(N) 
\\
S_{1,1,1}\left(1,1,\frac{1}{2},N\right) &=& 
-2^{-N} \int_0^1 dx x^N \frac{\tfrac{1}{2}
H^2_{2}(x) - \ln(2) H_{2}(x) + \tfrac{\ln^2(2)}{2}}{2-x} 
\nonumber\\ &&
+ \int_0^1 dx \frac{x^N-1}{x-1} \left[\ln(2) H_1(x) - \frac{\ln^2(2)}{2}\right] + \frac{1}{6} \ln^3(2) 
\\
S_{1,1,1}\left(2,1,1,N\right)           &=&  
\frac{1}{2} \int_0^1 \frac{x^N-1}{x-1} H^2_2(x)
\\
S_{1,1,2}\left(2,\frac{1}{2},1,N\right) &=&
- 2^N \int_0^1 dx (x^N -(1/2)^N)\frac{5 \zeta_3}{4(1-2x)} 
\nonumber\\ &&
+ \int_0^1 dx (x^N-1) \frac{-\tfrac{5}{8} 
\zeta_3 - H_0(x) H_{2,1}(x) + H_{0,2,1}(x) 
+H_{2,0,1}(x)}{x-1} 
\\
S_{1,1,2}\left(2,1,\frac{1}{2},N\right) &=&
 2^N \int_0^1 dx (x^N-(1/2)^N) \frac{4 \ln^3(2) - 3 \zeta_3}{12 (-1 + 2 x)}
- \int_0^1 dx \frac{x^N-1}{24(1-x)} 
\nonumber\\ &&
\times \Biggl[-4 \ln^3(2) + 3 \zeta_3 
+ 12 [\ln^2(2) - \zeta_2] H_2(x)
- 12 H_0(x) H_2^2(x)
\nonumber\\ &&
+ 24 H_{0,2,2}(x)
+ 24 H_{2,0,2}(x)\Biggr]
- \frac{1}{2}[\ln^2(2) - \zeta_2] \int_0^2 \frac{x^N-1}{x-1} H_2(x) 
\\
S_{1,2,1}\left(2,\frac{1}{2},1,N\right) &=&
\frac{1}{2} 
\int_1^2 dx  \frac{x^N-1}{x-1} \left(2 \zeta_3 - \ln(2) \zeta_2\right)
+ \int_0^1 dx \frac{x^N-1}{x-1} \left[\zeta_2 H_2(x) - H_{2,0,1}(x) \right] 
\nonumber\\ && 
\\
S_{1,2,1}\left(2,1,\frac{1}{2},N\right) &=&
2^N \int_0^1 dx (x^N-(1/2)^N)
\nonumber\\ && \times
\frac{
 \tfrac{\ln^3(2)}{3} 
+\tfrac{\zeta_3}{2}
+ \ln(2)\left[
-2 H_0(x) H_1(x)
+2 H_{0,1}(x) - \zeta_2 \right]}{2 x-1}
\nonumber\\ &&
+\int_0^1 dx (x^N-1) \Biggl[\frac{
-\tfrac{\ln^3(2)}{6}-\tfrac{\zeta^3}{4}
-\tfrac{1}{2} \ln^2(2) H_2(x)
-H_{2,0,2}(x)
}{x-1}
\nonumber\\ &&
+ \frac{\ln(2)\left[H_0(x) H_2(x) - H_{0,2}(x) + \tfrac{\zeta_2}{2}\right]
}{x-1} \Biggr]
\\
S_{2,1,1}\left(2,\frac{1}{2},1,N\right) &=&
\left(\frac{13}{8} \zeta_3  - \ln(2) \zeta_2\right) S_1(N) 
+ \frac{1}{2} \int_1^2 dx \frac{x^N-1}{x-1} \left[\ln(2) - H_0(x) \right] \zeta_2
\nonumber\\ &&
- \int_0^1 dx \frac{x^N-1}{x-1} H_{0,2,1}(x) 
\\
S_{2,1,1}\left(2,1,\frac{1}{2},N\right) &=&
-2^N \int_0^1 dx x^N \frac{\ln^2(2) H_0(x)}{1-2x} - \int_0^1 dx \frac{x^N-1}{1-x} \Biggl[\frac{1}{3} \ln^3(2) 
- \frac{1}{2} \ln(2) 
\zeta_2
\nonumber\\ &&
+\frac{1}{8} \zeta_3 
- \frac{1}{2} \ln^2(2) H_0(x) + \ln(2) H_{0,2}(x) - H_{0,2,2}(x) \Biggr]
\nonumber\\ &&
+ \ln(2) \int_0^2 \frac{x^N-1}{x-1} \left[\zeta_2 - H_{0,2}(x)\right] 
\\
S_{1,1,1,1}\left(2,\frac{1}{2},1,1,N\right) &=&
\frac{3}{4} \zeta_3 \int_1^2 \frac{x^N-1}{x-1} + \int_0^1 \frac{x^N-1}{x-1} H_{2,1,1}(x)
\\
S_{1,1,1,1}\left(2,1,\frac{1}{2},1,N\right) &=&
\int_1^2 \frac{x^N-1}{x-1} \left[ \frac{\zeta_3}{8} - \frac{\zeta_2 \ln(2)}{2} + \frac{\zeta_2}{2} H_{2}(x)\right]
\nonumber\\ &&
+ \int_0^1 \frac{x^N-1}{x-1} H_{2,2,1}(x).
\end{eqnarray}
The generalized harmonic polylogarithms appearing above can be expressed in the following way
\begin{eqnarray}
H_2(x) &=& -\ln\left(1 - \frac{x}{2}\right)
\\
H_{0,2}(x) &=& \Li_2\left(\frac{x}{2}\right)
\\
H_{2,1}(x) &=& \Li_2(x-1) +\ln(1-x) \ln(2-x) + \frac{\zeta_2}{2}
\\
H_{0,0,2}(x) &=& \Li_3\left(\frac{x}{2}\right)
\\
H_{2,1,1}(x) &=& \frac{3}{4} \zeta_3 - \frac{1}{2} \ln^2(1-x) \ln(2-x) - \ln(1-x) \Li_2(x-1) + \Li_3(x-1)
\\
H_{2,2,1}(x) &=& -\frac{3 \zeta_2}{2} \ln(2-x) + i \frac{\pi}{2} \ln^2(2-x) + \Li_3(2-x) - \frac{7}{8} \zeta_3
\\
H_{0,2,2}(x) &=& 
\left[
        -\zeta_2
        +\frac{1}{2} \ln^2(2-x)
        +\ln(2-x) \ln(x)
\right] \ln(2)
-\Li_3\left(
        1-\frac{x}{2}\right)
\nonumber\\ &&
+(\ln(2)-\log (2-x)) \Li_2\left(
        \frac{x}{2}\right)
+\left(
        -\ln(2-x)
        -\frac{1}{2} \ln(x)
\right) \ln^2(2)
\nonumber\\ &&
-\frac{1}{2} \ln^2(2-x) \ln(x)
+\zeta_2 \ln(2-x)
+\zeta(3)
+\frac{1}{2} \ln^3(2) 
\\
H_{2,0,1}(x) &=&
(-\ln(2) - \ln(1-x) + \ln(2-x)) \Li_2(1-x)
\nonumber\\ &&
+(\log (2-x) - \ln(2) - \ln(1-x)) \left[\Li_2\left(\frac{2-x}{2 (1-x)}\right)-\Li_2\left(\frac{2-x}{1-x}\right) \right]
\nonumber\\ &&
+\left(
        -\frac{1}{2} \ln^2(1-x)
        +\ln(1-x) \ln(2-x)
        -\frac{\zeta_2}{2}
\right) \ln(2)
-\frac{15 \zeta_3}{8}
+\Li_3(1-x)
\nonumber\\ &&
+\Li_3\left(
        1-\frac{x}{2}\right)
-\Li_3\left(
        \frac{2-x}{2 (1-x)}\right)
+\Li_3\left(
        \frac{2-x}{1-x}\right)
+(\ln(2)-\ln(2-x)) \Li_2\left(
        1-\frac{x}{2}\right)
\nonumber\\ &&
-\ln(2-x) \zeta_2
+\ln^2(2) \left(
        \frac{i \pi }{2}-\log (1
        -x
        )\right)
\\
H_{0,2,1}(x) &=& 
( \ln(2)
+ \ln(1-x)
- \ln(2-x)) \left[\Li_2\left(\frac{2-x}{2 (1-x)}\right) 
                - \Li_2\left(\frac{2-x}{1-x}\right)\right]
\nonumber\\ &&
+\left(
        \frac{1}{2} \ln^2(1-x)
        -\ln(1-x) \ln(2-x)
        +\frac{\zeta_2}{2}
\right) \ln(2)
+\frac{11 \zeta_3}{4}
-\Li_3(1-x)
\nonumber\\ &&
-\Li_3(2-x)
-\Li_3\left(
        1-\frac{x}{2}\right)
+\Li_3\left(
        \frac{2-x}{2 (1-x)}\right)
-\Li_3\left(
        \frac{2-x}{1-x}\right)
-\Li_3(x)
\nonumber\\ &&
+\Li_3\left(
        \frac{x}{2-x}\right)
-\text{Li}_3\left(
        -\frac{x}{2-x}\right)
+(\ln(2) + \ln(1-x) - \ln(x)) \Li_2(1-x)
\nonumber\\ &&
+(\ln(2-x) -\ln(x)) \Li_2(2-x)
+(\ln(2-x)-\log (2)) \Li_2\left(1-\frac{x}{2}\right)
\nonumber\\ &&
+(\ln(2-x) - \ln(x)) \left[\Li_2\left(\frac{x}{2-x}\right) - \Li_2\left(-\frac{x}{2-x}\right)\right]
+\frac{5}{2} \ln(x) \zeta_2
\nonumber\\ &&
+\left[
        \ln(2-x) \ln(x)
        -\ln^2(x)
\right] \ln(1-x)
+\frac{1}{2} i \pi  \ln^2(2-x)
\nonumber\\ &&
+\ln^2(2) \left(
        \ln(1-x)-\frac{i \pi }{2}\right)
-i \pi  \ln(2-x) \ln(x)
\\
H_{2,0,2}(x) &=& 
\left[
        -\ln^2(2-x)
        -2 \ln(2-x) \ln(x)
        +2 \zeta_2
\right] \ln(2)
-2 \zeta_3
+2 \Li_3\left(
        1-\frac{x}{2}\right)
\nonumber\\ &&
+(\ln(2-x) - \ln(2)) \Li_2\left(
        \frac{x}{2}\right)
-2 \ln(2-x) \zeta_2
+(2 \ln(2-x)
+\ln(x)
) \ln^2(2)
\nonumber\\ &&
+\ln^2(2-x) \ln(x)
-\ln^3(2).
\end{eqnarray}
\noindent
On the expense of more complicated arguments they can be represented in terms 
of classical polylogarithms up to weight {\sf w = 3}. This is generally expected for 3-letter
alphabets, see \cite{Blumlein:1998if}.
\section{Expressions in \boldmath $x$-Space} 
\label{appC}
The massive OME $A_{Qq}^{\rm PS}$ to 3-loop order in $x$-space is given by

The massive Wilson coefficient  $H_{2,q}^{\rm PS}$ in the asymptotic region to 3-loop order in $x$-space is given by
%

\vspace{5mm}
\noindent
{\bf Acknowledgment.}~
We would like to thank I. Bierenbaum, S. Klein, C.G.~Raab, M.~Round, A. Vogt, and F.~Wi\ss{}brock for discussions, 
and M.~Steinhauser for providing the code {\tt MATAD 3.0}. The graphs have been drawn using {\tt Axodraw}~\cite{Vermaseren:1994je}. 
This work was supported in part by DFG Sonderforschungsbereich Transregio 9, Computergest\"utzte Theoretische Teilchenphysik, 
the Austrian Science Fund (FWF) grants P20347-N18 and SFB F50 (F5009-N15), the European Commission through contract PITN-GA-2010-264564 
({LHCPhenoNet}) and PITN-GA-2012-316704 ({HIGGSTOOLS}), and by the Research Center ``Elementary Forces and Mathematical Foundations (EMG)'' 
of J. Gutenberg University Mainz and DFG.

\end{document}